\documentclass[aps,prx,twocolumn,superscriptaddress,longbibliography]{revtex4-2}
\usepackage{lineno}
\usepackage{graphicx}  
\usepackage{dcolumn}   
\usepackage{bm}        
\usepackage{amssymb}   
\usepackage{amsmath}
\usepackage{blkarray, multirow, graphicx, diagbox, color, xcolor, colortbl}
\usepackage{bbm, bbold}
\usepackage{glossaries}
\usepackage{hyphenat}
\usepackage{ifthen}
\usepackage{xkeyval}
\usepackage{moreverb}
\usepackage{rotating}
\usepackage{wrapfig}
\usepackage{slashbox}
\usepackage{xspace}
\usepackage{nicefrac}
\usepackage[]{units}
\usepackage{physics}
\usepackage{braket}
\usepackage[inline]{enumitem}
\usepackage{tabto}
\usepackage{listings}
\usepackage{xstring}
\def\ReplaceStr#1{%
	\IfSubStr{#1}{p}{%
		\StrSubstitute{#1}{p}{.}}{#1}}

\usepackage[caption=false]{subfig}
\newcommand\subfigref[1]{\protect\subref{#1}}

\usepackage[customcolors]{hf-tikz} 
\usepackage{tikz}
\usepackage{calc}
\usepackage{pgffor}
\usepackage{pgfplots}
\pgfplotsset{compat=1.13}
\usepackage{pgfplotstable}
\usepgfplotslibrary{groupplots}
\usepgfplotslibrary{fillbetween}

\tikzstyle{n} = [draw,shape=ellipse,minimum size=1.5em,inner sep=0pt,fill=white!20, minimum width=2.5em]
\tikzstyle{Init} = [n,color=green,fill=green!20,text=black]
\tikzstyle{Fin} = [n,color=red,fill=red!20,text=black]
\tikzstyle{Ghost} = [minimum size=1.5em,inner sep=0pt,color=white,text=black]
\tikzstyle{Multiple} = [draw,shape=rect,minimum size=2em,inner sep=0pt]

\tikzstyle{ghostA} = [text=red!70,thick, minimum size=2*(5pt-\pgflinewidth), inner sep=0pt, outer sep=0pt]
\tikzstyle{ghostB} = [text=blue!70,thick, minimum size=2*(5pt-\pgflinewidth), inner sep=0pt, outer sep=0pt]
\tikzstyle{siteA} = [regular polygon, regular polygon sides=3, shape border rotate= 30, draw=red!50,fill=red!20,thick,inner sep=0pt,minimum width=1.5em,font=\footnotesize]
\tikzstyle{siteB} = [regular polygon, regular polygon sides=3, shape border rotate= -30, draw=green!50,fill=green!20,thick,inner sep=0pt,minimum width=1.5em,font=\footnotesize]
\tikzstyle{op} = [regular polygon, regular polygon sides=4, draw=orange!50, fill=orange!20, thick, inner sep=0.2pt, minimum width=1.25em, minimum height=1.5em,font=\footnotesize]
\tikzstyle{opghost} = [regular polygon, regular polygon sides=4, thick, inner sep=0.2pt, minimum width=1.25em, minimum height=1.5em,font=\footnotesize]
\tikzstyle{site} = [circle,draw=blue!50,fill=blue!20,thick,inner sep=0.2pt,minimum width=1.25em,font=\footnotesize]
\tikzstyle{hiddensite} = [circle,draw=white!50,fill=white!20,thick,inner sep=0.2pt,minimum width=1.25em,font=\footnotesize]
\tikzstyle{nosite} = [circle,draw=white,fill=white,thick,inner sep=0.1pt,minimum width=1.5em]
\tikzstyle{ghost} = [font=\footnotesize]
\tikzstyle{intersite} = [regular polygon, regular polygon sides=4, shape border rotate= 45, draw=black!50,fill=black!20,thick,inner sep=0pt,minimum width=1.5em]
\tikzstyle{ld} = [inner sep=1pt, font=\small]
\tikzstyle{unsite} = [circle, outer sep=0pt,inner sep=0.2pt,minimum width=1.25em]

\definecolor{colorA}{rgb} {0.58,0,0.8275}
\definecolor{colorB}{rgb} {0.11,0.663,0.51}
\definecolor{colorC}{rgb} {0.3373,0.7059,0.9137}
\definecolor{colorD}{rgb} {0.902,0.6235,0}
\definecolor{colorE}{rgb} {0.9451,0.902,0.3255}
\definecolor{colorF}{rgb} {0.3373,0.3255,0.902}
\definecolor{colorG}{rgb} {0.9451,0.3255,0.3373}

\def\gpmarkers{{"+","x","star","square","square*","o","*","triangle","triangle*"}}
\def\gpcolors{{"colorA","colorB","colorC","colorD","colorE"}}

\usetikzlibrary
{
	calc,
	decorations,
	plotmarks,
	patterns,
	positioning,
	petri,
	arrows,
	decorations.markings,
	backgrounds,
	fit,
	graphs,
	shapes.geometric,
	decorations.pathmorphing,
	shapes.misc,
	shapes,
	tikzmark,
	pgfplots.colorbrewer,
	fpu,
}

\usepackage{ocgx}

\usetikzlibrary{ocgx}

\pgfplotsset{
        cycle from colormap manual style/.style={
            x=3cm,y=10pt,ytick=\empty,
            stack plots=y,
            every axis plot/.style={line width=2pt},
        },
}

\tikzset{->-/.style={decoration={
			markings,
			mark=at position .5 with {\arrow{>}}},postaction={decorate}}}

\tikzset{-<-/.style={decoration={
			markings,
			mark=at position .5 with {\arrow{<}}},postaction={decorate}}}

\tikzstyle{orientedsnake} = [
decorate, 
decoration={snake},
->
]  
\tikzstyle{orientedshortarrow} = [
decoration={markings,
	mark=at position .33 with {\arrow{>}}},
postaction={decorate}
]  
\tikzstyle{orientedlongarrow} = [
decoration={markings,
	mark=at position .67 with {\arrow{>}}},
postaction={decorate}
]
\tikzset{dbl/.style={double,
		double equal sign distance,
		-implies,
		shorten >=10pt,
		shorten <=10pt}}
\tikzset{
	between/.style args={#1 and #2}{
		at = ($(#1)!0.5!(#2)$)
	}
}

\pgfmathdeclarefunction{peierlspotential}{6}%
{
	\pgfmathparse
	{
		#2 / (#3 * #5) 
		* sin(deg(#1 * #3 - #3 * #5 * (x)))
		* exp(-( ( #1 - #5 * ((x) - #4) ) )^2.0 / ( #6^2.0 ) )
	}%
}

\pgfmathdeclarefunction{linearFct}{2}%
{%
	\pgfmathparse{#1*x+#2}%
}
\pgfmathdeclarefunction{quadFct}{3}%
{%
	\pgfmathparse{#1*x*x+#2*x+#3}%
}
\pgfmathdeclarefunction{logFct}{2}%
{%
	\pgfmathparse{#1*log10(x)+#2}%
}
\pgfmathdeclarefunction{expFct}{2}%
{%
	\pgfmathparse{#1*exp(-x/#2)}%
}

\newboolean{buildCurrentBlockInline}
\setboolean{buildCurrentBlockInline}{false}
\newboolean{rebuildCurrentBlock}
\setboolean{rebuildCurrentBlock}{false}

\newboolean{rebuildData}
\setboolean{rebuildData}{false}

\newboolean{removeData}
\setboolean{removeData}{false}

\newboolean{buildtikzpics}
\setboolean{buildtikzpics}{false}
\newif\ifrebuildtikz
\newif\ifChangeMode
\ChangeModetrue
\ChangeModefalse
\ifthenelse{\boolean{buildtikzpics}}
{
	\rebuildtikztrue
	\usetikzlibrary{external}
	\tikzexternalize[optimize=false,prefix=Figures/autogen/]
}
{
	\rebuildtikzfalse
}
\usepackage[colorlinks,bookmarks=false,citecolor=blue,linkcolor=black,urlcolor=blue]{hyperref}
\usepackage{cleveref}
\Crefname{appendix}{Appendix}{Appendices}
\Crefname{equation}{Equation}{Equations}
\Crefname{figure}{Figure}{Figures}
\Crefname{section}{Section}{Sections}
\Crefname{tabular}{Tabular}{Tabulars}
\crefname{appendix}{App.}{Apps.}
\crefname{equation}{Eq.}{Eqs.}
\crefname{figure}{Fig.}{Figs.}
\crefname{section}{Sec.}{Secs.}
\crefname{tabular}{Tab.}{Tabs.}

\newcommand{\ingoing}[1]{#1}

\ExplSyntaxOn
\DeclareExpandableDocumentCommand \eval { m } { \fp_eval:n { #1 } }
\ExplSyntaxOff

\makeatletter
\def\pgfplotsutil@decstringcounter#1{%
 \begingroup
  \c@pgf@counta=#1\relax
  \advance\c@pgf@counta by -1
  \edef#1{\the\c@pgf@counta}%
  \pgfmath@smuggleone#1%
 \endgroup
}%

\pgfplotsset{
/pgfplots/each nth point*/.style 2 args={%
/pgfplots/x filter/.append code={%
 \ifnum\coordindex=0
  \def\c@pgfplots@eachnthpoint@xfilter{0}%
  \edef\c@pgfplots@eachnthpoint@xfilter@cmp{#1}%
 \else
  \ifnum\coordindex>#2\relax
   \pgfplotsutil@advancestringcounter\c@pgfplots@eachnthpoint@xfilter
   \ifx\c@pgfplots@eachnthpoint@xfilter@cmp\c@pgfplots@eachnthpoint@xfilter
    \def\c@pgfplots@eachnthpoint@xfilter{0}%
   \else
    \let\pgfmathresult\pgfutil@empty
   \fi
  \fi
 \fi
}%
},
}
\makeatother
\usepackage{braket}
\usepackage{amsmath, amssymb}
\usepackage{graphicx}
\usepackage{dcolumn}
\usepackage{bm}
\usepackage{float}
\usepackage{csvsimple}
\usepackage{nicefrac}
\usepackage{glossaries}
\usepackage{placeins}
\graphicspath{{Figures/}}
\usepackage{tikz}
\newacronym[shortplural={MPS}]{MPS}{MPS}{matrix\hyp product state}
\newacronym{ACA}{ACA}{adaptive cross approximation}
\newacronym{1D}{1D}{one\hyp dimensional}
\newacronym{2D}{2D}{two\hyp dimensional}
\newacronym{MPO}{MPO}{matrix\hyp product operator}
\newacronym{SVD}{SVD}{singular\hyp value decomposition}
\newacronym{QCS}{QCS}{quantum\hyp computer simulator}
\newacronym{QC}{QC}{quantum computer}
\newacronym{FSM}{FSM}{finite\hyp state machine}
\newacronym{CDW}{CDW}{charge\hyp density wave}
\newacronym{SDW}{SDW}{spin\hyp density wave}
\newacronym{ARPES}{ARPES}{angle-resolved photoemission spectroscopy}
\newacronym{OBC}{OBC}{open-boundary conditions}
\newacronym{PBC}{PBC}{periodic-boundary conditions}
\newacronym{TEBD}{TEBD}{time-evolution block-decimation}
\newacronym{TDVP}{TDVP}{time\hyp dependent variational principle}
\newacronym{iff}{iff}{if and only if}
\newacronym{DFT}{DFT}{density\hyp functional theory}
\newacronym{DMFT}{DMFT}{dynamical mean\hyp field theory}
\newacronym{DMRG}{DMRG}{density\hyp matrix renormalization group}
\newacronym{1DMRG}{1DMRG}{single-site density\hyp matrix renormalization group}
\newacronym{2DMRG}{2DMRG}{two-site density\hyp matrix renormalization group}
\newacronym{DMRG3S}{DMRG3S}{strictly single-site density\hyp matrix renormalization group}
\newacronym{iDMRG}{iDMRG}{inifinite\hyp size density\hyp matrix renormalization group}
\newacronym{tDMRG}{tDMRG}{time\hyp dependent density\hyp matrix renormalization group}
\newacronym{PP-DMRG}{PP-DMRG}{projected purified density\hyp matrix renormalization group}
\newacronym{QMC}{QMC}{quantum Monte Carlo}
\newacronym{AIM}{AIM}{Anderson impurity model}
\newacronym{SIAM}{SIAM}{single impurity Anderson model}
\newacronym{LDA}{LDA}{local\hyp density approximation}
\newacronym{VQE}{VQE}{variational\hyp quantum eigensolver}
\newacronym{ED}{ED}{exact diagonalization}
\newacronym{QPT}{QPT}{quantum phase transition}
\newacronym{QCP}{QCP}{quantum critical point}
\newacronym{ETH}{ETH}{eigenstate thermalization hypothesis}
\newacronym{EHM}{EHM}{extended Hubbard model}
\newacronym{AKLT}{AKLT}{Affleck\hyp Lieb\hyp Kennedy\hyp Tasaki}
\newglossaryentry{QR}{name={QR},description={QR decomposition}}
\newacronym{TNS}{TNS}{tensor\hyp network state}
\newacronym{SM}{SM}{supplemental material}
\newacronym{NOO}{NOO}{natural orbital occupation}
\newacronym{NO}{NO}{natural orbital}
\newacronym{LRO}{LRO}{long\hyp range order}
\newacronym{qLRO}{qLRO}{quasi\hyp long\hyp range order}
\newacronym{SC}{SC}{Superconductivity}
\newacronym{VBGS}{VBGS}{valence bond ground-state}
\newacronym{PEPS}{PEPS}{projected entangled pair\hyp states}
\newacronym{ALS}{ALS}{alternating least squares}
\newacronym{BdG}{BdG}{Bogoljubov de-Gennes}
\newacronym{TFIM}{TFI}{transverse field Ising model}
\newacronym{PP}{PP}{projected purification}
\newacronym{BEC}{BEC}{Bose\hyp Einstein condensate}
\newacronym{JWT}{JWT}{Jordan\hyp Wigner transformation}
\newacronym{NISQ}{NISQ}{noisy intermediate scale quantum}
\newacronym{NN}{NN}{nearest\hyp neighbor}
\newacronym{NNN}{NNN}{next\hyp nearest\hyp neighbor}
\newacronym{SPDM}{SPDM}{single\hyp particle density matrix} 
\newacronym{HCB}{HCB}{hardcore bosons}
\newacronym{SF}{SF}{spinless fermions}
\newacronym{fRG}{fRG}{functional renormalization group}
\newacronym{LE}{LE}{Luther\hyp Emery}
\newacronym{TLL}{TLL}{Tomonaga-Luttinger liquid}
\newacronym{PLL}{PLL}{pair Luttinger liquid}
\newacronym{FB}{FB}{flat band}
\newacronym{SU}{SU}{superfluid}
\newacronym{QO}{QO}{quasi-order}
\newacronym{LL}{LL}{Luttinger liquid}
\newacronym{CFT}{CFT}{conformal field theory}
\usetikzlibrary{external}
\usepackage{color}
\definecolor{blue}{rgb}{0.1, 0.1, 1}
\newcommand{\tb}[1]{\ensuremath{\textbf{#1}}}
\newcommand{\R}{\ensuremath{\textbf{R}}}
\renewcommand{\rm}{\mathrm}

\usepackage{ifthen}
\usepackage{pgf}
\usepackage{tikz}
\tikzset{>=stealth}
\usepackage{pgffor}
\usepackage{pgfplots}
\pgfplotsset{compat=newest}
\usepackage{pgfplotstable}
\usepgfplotslibrary{groupplots}
\usetikzlibrary
{
	calc,
	decorations,
	calligraphy,
	pgfplots.fillbetween,
	pgfplots.patchplots,
	plotmarks,
	patterns,
	positioning,
	petri,
	arrows,
	intersections,
	decorations.markings,
	backgrounds,
	fit,
	matrix,
	graphs,
	shapes.geometric,
	decorations.pathreplacing, 
	decorations.pathmorphing,
	decorations,
	shapes.misc,
	shapes.multipart,
	shapes,
	through,
	tikzmark
}

\newcommand{\Systems}[7]%
{%
	\pgfmathparse{int((#2*#5-1))}%
	\pgfmathsetmacro{\ymax}{\pgfmathresult}%
	\pgfmathparse{int(#1-1)}%
	\pgfmathsetmacro{\xmax}{\pgfmathresult}%
	\tikzset{external/export next=false}%
	\begin{tikzpicture}[baseline]%
		\foreach \x in {0,...,\xmax}%
		{%
			\foreach \y in {0,...,\ymax}%
			{%
				\node[circle, draw, inner sep = 0pt, outer sep = 0pt] at ($(\x*1.5*#3,\y*1.25*#3)$) (\x\y){\pgfmathparse{int(10*#3)}\fontsize{\pgfmathresult pt}{10pt}#4};%
			}%
		}%
		\foreach \x [remember=\x as \lastx (initially 0)] in {0,...,\xmax}%
		{%
			\foreach \y [remember=\y as \lasty (initially 1)] in {0,...,\ymax}%
			{%
				\ifthenelse{\y>0}%
				{%
					\pgfmathparse{int(mod(\y,#2))}%
					\ifthenelse{\pgfmathresult=0}%
					{%
						\draw[densely dotted] (\x\lasty) -- node[left, inner ysep = 0.5em] {} (\x\y);%
					}%
					{%
						\draw (\x\lasty) -- node[left, inner ysep = 0.5em] {} (\x\y);%
					}%
				}{}%
				\ifthenelse{\ymax>0}%
				{%
					\draw[densely dotted, opacity=#7] (\lastx\lasty) -- node[midway] (m) {} (\x\y);%
				}%
				{}%
				\ifthenelse{\x>0}%
				{%
					\pgfmathparse{int(mod(\x,#2))}%
					\ifthenelse{\pgfmathresult=0}%
					{%
						\begin{scope}[on background layer]
							\draw[opacity=#6] (\lastx\y.east) -- node[above, inner xsep = 0.5em, inner ysep = 0.1em] {} (\x\y.west);%
						\end{scope}
					}%
					{%
						\draw (\lastx\y) -- node[above, inner xsep = 0.5em, inner ysep = 0.1em] {} (\x\y);%
					}%
				}{}%
			}%
		}%
		\pgfmathparse{int((#5-1)/2*(#2))}%
		\pgfmathsetmacro{\yPhysMin}{\pgfmathresult}%
		\pgfmathparse{int(\yPhysMin+#2-1)}%
		\pgfmathsetmacro{\yPhysMax}{\pgfmathresult}%
		\pgfmathparse{int(#1-1)}%
		\pgfmathsetmacro{\xmax}{\pgfmathresult}%
		\node[draw, fit = (0\yPhysMin) (\xmax\yPhysMax)] (Phys) {};
		\node[anchor=west] at (Phys.north east) {$H_0$};
	\end{tikzpicture}%
}%
\newcommand{\Fermions}[5]%
{%
	\pgfmathparse{rnd}%
	\pgfmathsetmacro{\uprnd}{\pgfmathresult}%
	\ifthenelse{\lengthtest{\uprnd pt > #1pt }}%
	{%
		{\color{#2}$\uparrow$}%
	}%
	{%
		\color{#3}$\uparrow$%
	}%
	\pgfmathparse{rnd}%
	\pgfmathsetmacro{\downrnd}{\pgfmathresult}%
	\ifthenelse{\lengthtest{\downrnd pt > #1pt }}%
	{%
		{\color{#4}$\downarrow$}%
	}%
	{%
		{\color{#5}$\downarrow$}%
	}%
}%
\newcommand{\Bosons}[3]%
{%
	\pgfmathparse{rnd}%
	\pgfmathsetmacro{\rnd}{\pgfmathresult}%
	\ifthenelse{\lengthtest{\rnd pt > #1pt }}%
	{%
		{\color{#2}\textbullet}%
	}%
	{%
		{\color{#3}\textbullet}%
	}%
}%

\definecolor{colorA}{rgb} {0.58,0,0.8275}
\definecolor{colorB}{rgb} {0.11,0.663,0.51}
\definecolor{colorC}{rgb} {0.3373,0.7059,0.9137}
\definecolor{colorD}{rgb} {0.902,0.6235,0}
\definecolor{colorE}{rgb} {0.9451,0.902,0.3255}
\definecolor{colorF}{rgb} {0.3373,0.3255,0.902}
\definecolor{colorG}{rgb} {0.9451,0.3255,0.3373}
\definecolor{colorH}{rgb} {0.11,0.3255,0.3373}

\pgfmathdeclarefunction{critExponentFct}{3}%
{%
	\pgfmathparse{(x<#2)?(#1*abs(x-#2)^(#3)):(0)}%
}

\pgfmathdeclarefunction{exponentialInFct}{3}%
{%
	\pgfmathparse{#1+#2*exp(-(#3)/x)}%
}

\newcommand{\printpgfnumberwithouterrorInMath}[3][0]%
{%
	\pgfmathfloatparsenumber{#2}%
	\pgfmathfloattomacro{\pgfmathresult}{\Fn}{\Mn}{\En}%
	\pgfmathparse{#1==2 ? (\Fn==2 ? "+" : "-") : (\Fn==2 ? "-" : (#1==1 ? "+" : ""))}%
	\edef\Sn{\pgfmathresult}%
	\pgfmathfloatparsenumber{#3}%
	\pgfmathfloattomacro{\pgfmathresult}{\Fe}{\Me}{\Ee}%
	\pgfmathparse{int(sqrt((\Ee-\En)^2))}%
	\edef\precisionAbsEe{\pgfmathresult}%
	\pgfmathparse{int(\Ee-\En)}%
	\edef\precisionE{\pgfmathresult}%
	\pgfmathparse{\eval{\Me*10^(\precisionE)}}%
	\ifthenelse{\En=0}%
	{%
		\Sn\pgfmathprintnumber[fixed, precision=\precisionAbsEe, zerofill]{\Mn}%
	}%
	{%
		\ifthenelse{\En=1}%
		{%
			\pgfmathparse{\eval{\Mn*10}}%
			\edef\Mn{\pgfmathresult}%
			\pgfmathparse{int(\precisionAbsEe+1)}%
			\edef\precisionAbsEe{\pgfmathresult}%
			\Sn\pgfmathprintnumber[fixed, precision=\precisionAbsEe, zerofill]{\Mn}%
		}%
		{%
			\ifthenelse{\En=2}%
			{%
				\pgfmathparse{\eval{\Mn*100}}%
				\edef\Mn{\pgfmathresult}%
				\pgfmathparse{int(\precisionAbsEe+1)}%
				\edef\precisionAbsEe{\pgfmathresult}%
				\Sn\pgfmathprintnumber[fixed, precision=\precisionAbsEe, zerofill]{\Mn}%
			}%
			{%
				\ifthenelse{\En=-1}%
				{%
					\pgfmathparse{\eval{\Mn/10}}%
					\edef\Mn{\pgfmathresult}%
					\pgfmathparse{int(\precisionAbsEe+1)}%
					\edef\precisionAbsEe{\pgfmathresult}%
					\Sn\pgfmathprintnumber[fixed, precision=\precisionAbsEe, zerofill]{\Mn}%
				}%
				{%
					\ifthenelse{\En=-2}%
					{%
						\pgfmathparse{\eval{\Mn/10}}%
						\edef\Mn{\pgfmathresult}%
						\pgfmathparse{int(\precisionAbsEe+1)}%
						\edef\precisionAbsEe{\pgfmathresult}%
						\Sn\pgfmathprintnumber[fixed, precision=\precisionAbsEe, zerofill]{\Mn}%
					}%
					{%
						\Sn\pgfmathprintnumber[std, precision=\precisionAbsEe, zerofill]{\Mn} \cdot10^{\En}%
					}%
				}%
			}%
		}%
	}%
}

\tikzstyle{ghostA} = [text=red!70,thick, minimum size=2*(5pt-\pgflinewidth), inner sep=0pt, outer sep=0pt]
\tikzstyle{ghostB} = [text=blue!70,thick, minimum size=2*(5pt-\pgflinewidth), inner sep=0pt, outer sep=0pt]
\tikzstyle{siteA} = [regular polygon, regular polygon sides=3, shape border rotate= 30, draw=red!50,fill=red!20,thick,inner sep=0pt,minimum width=1.5em,font=\footnotesize]
\tikzstyle{siteB} = [regular polygon, regular polygon sides=3, shape border rotate= -30, draw=green!50,fill=green!20,thick,inner sep=0pt,minimum width=1.5em,font=\footnotesize]
\tikzstyle{op} = [regular polygon, regular polygon sides=4, draw=orange!50, fill=orange!20, thick, inner sep=0.2pt, minimum width=1.25em, minimum height=1.5em,font=\footnotesize]
\tikzstyle{opghost} = [regular polygon, regular polygon sides=4, thick, inner sep=0.2pt, minimum width=1.25em, minimum height=1.5em,font=\footnotesize]
\tikzstyle{site} = [circle,draw=blue!50,fill=blue!20,thick,inner sep=0.2pt,minimum width=0.75em,font=\footnotesize]
\tikzstyle{hiddensite} = [circle,draw=white!50,fill=white!20,thick,inner sep=0.2pt,minimum width=1.25em,font=\footnotesize]
\tikzstyle{nosite} = [circle,draw=white,fill=white,thick,inner sep=0.2pt,minimum width=1.25em]
\tikzstyle{ghost} = [font=\footnotesize]
\tikzstyle{intersite} = [regular polygon, regular polygon sides=4, shape border rotate= 45, draw=black!50,fill=black!20,thick,inner sep=0pt,minimum width=1.5em]
\tikzstyle{ld} = [inner sep=1pt, font=\small]
\tikzstyle{unsite} = [circle, outer sep=0pt,inner sep=0.2pt,minimum width=0.75em]

\begin{document}
	\title{Solving 2D and 3D lattice models of correlated fermions - combining matrix product states with mean field theory}
	\author{Gunnar Bollmark}
	\affiliation{Department of Physics and Astronomy, Uppsala University, Box 516, S-751 20, Uppsala, Sweden}
	\author{Thomas K\"ohler}
	\affiliation{Department of Physics and Astronomy, Uppsala University, Box 516, S-751 20, Uppsala, Sweden}
	\author{Lorenzo Pizzino}
	\affiliation{DQMP, University of Geneva, 24 Quai Ernest-Ansermet, 1211 Geneva, Switzerland}
	\author{Yiqi Yang}
	\affiliation{Department of Physics, College of William and Mary, Williamsburg, Virginia 23187, USA}
	\author{Johannes S. Hofmann}
	\affiliation{Department of Condensed Matter Physics, Weizmann Institute of Science, Rehovot 76100, Israel}
	\author{Hao Shi}
	\affiliation{Department of Physics and Astronomy, University of Delaware, Newark, Delaware 19716, USA}
	\author{Shiwei Zhang}
	\affiliation{Center for Computational Quantum Physics, Flatiron Institute, 162 Fifth Avenue, New York, New York, 10010, USA}
	\author{Thierry Giamarchi}
	\affiliation{DQMP, University of Geneva, 24 Quai Ernest-Ansermet, 1211 Geneva, Switzerland}
	\author{Adrian Kantian}
	\affiliation{Department of Physics and Astronomy, Uppsala University, Box 516, S-751 20, Uppsala, Sweden}
	\affiliation{SUPA, Institute of Photonics and Quantum Sciences, Heriot-Watt University, Edinburgh EH14 4AS, United Kingdom}

	\begin{abstract}
	Correlated electron states are at the root of many important phenomena including unconventional superconductivity (USC), where electron-pairing arises from repulsive interactions. Computing the properties of  correlated electrons, such as the critical temperature $T_c$ for the onset of USC, efficiently and reliably from the microscopic physics with quantitative methods remains a major challenge for almost all models and materials. In this theoretical work we combine matrix product states (MPS) with static mean field (MF) to provide a solution to this challenge for quasi-one-dimensional (Q1D) systems: Two- and three-dimensional (2D/3D) materials comprised of weakly coupled correlated 1D fermions. This MPS+MF framework for the ground state and thermal equilibrium properties of Q1D fermions is developed and validated for attractive Hubbard systems first, and further enhanced via analytical field theory. We then deploy it to compute $T_c$ for superconductivity in 3D arrays of weakly coupled, doped and repulsive Hubbard ladders. The MPS+MF framework thus enables the reliable, quantitative and unbiased study of USC and high-$T_c$ superconductivity - and potentially many more correlated phases - in fermionic Q1D systems from microscopic parameters, in ways inaccessible to previous methods. It opens the possibility of designing deliberately optimized Q1D superconductors, from experiments in ultracold gases to synthesizing new materials.

	\end{abstract}

	\maketitle
	
	\section{\label{Sec::Introduction}Introduction}
	Obtaining quantitative and reliable solutions to strongly correlated fermionic models from first principles is one of the greatest challenges to the theory of quantum matter, arising in many different areas, from solid state physics to ultra cold atomic gases. Its prominence derives, to a large degree, from the  fundamental and technological importance of many of the collective quantum states emerging from electronic correlations. 
	
	Unconventional superconductivity (USC) epitomizes this challenge: Repulsive interactions lead to electrons forming correlated pairs which attain macroscopic phase coherence and enter a superconducting state at the critical temperature $T_c$. Crucially, the high-$T_c$ superconducting models and materials belong to this group~\cite{BookAnderson1997,Orenstein2000,Stewart2017}.
    These can be divided into two classes: the quasi-two-dimensional (Q2D) models and materials, and the quasi-one-dimensional (Q1D) ones. In both cases, the relevant three-dimensional (3D) system is comprised of weakly coupled lower-dimensional sub-units, 2D and 1D ones for Q2D and Q1D respectively.
    
	The present work introduces a quantitative, efficient and reliable theoretical framework for Q1D systems, capable of calculating properties of unconventional and high-$T_c$ superconductivity as well as other, competing correlated phases of fermions from microscopic parameters. It thus allows to deliberately engineer high-$T_c$ superconductivity from the bottom up in such systems.
	
	This is achieved by leveraging two unique advantages with respect to USC that the 1D sub-units within a Q1D system can have compared to the known 2D sub-units in the Q2D ones: (1) The microscopic mechanism of pair formation from repulsive interactions can typically be worked out~\cite{Giamarchi2003}. (2) The pairing energy resulting from it can be computed accurately from microscopic parameters~\cite{Karakonstantakis2011,Dolfi2015b}. These pairing energies can moreover be very high. For doped two-leg Hubbard ladders, pairing energies up to about 15\% of electron tunneling have already been demonstrated~\cite{Karakonstantakis2011}, as the 1D nature of these ladders naturally enhances the repulsively mediated effective attraction. As fluctuations preclude superconductivity in isolated 1D sub-units, the coupling of these into a three-dimensional Q1D array is essential to enable macroscopic ordering of pair-phases.

	Analysis of the resulting arrays has so far been limited to Tomonaga-Luttinger liquid (TLL) field theory~\cite{Giamarchi2003} combined with static mean-field (MF) techniques~\cite{Essler2002,Karakonstantakis2011}. The TLL approach has many uses, such as explaining repulsively-mediated pairing, or mapping the low-energy theory of Hubbard-ladders to that of the simpler negative-$U$ Hubbard chains. However, it allows neither to compute $T_c$ of a Q1D array quantitatively, nor to determine from microscopic parameters whether it realizes USC or another correlated phase. Furthermore, this approach also disregards exchange processes between 1D sub-units. Thus, it cannot inform the search for USC and high-$T_c$ systems in the Q1D-space, be it for synthesizing candidate materials or guiding quantum simulations towards analogue states of USC within current experimental capabilities. Both are intensely sought goals that remain highly challenging for the theory of Q2D superconductors with repulsively-mediated pairing~\cite{BookAnderson1997,Orenstein2000,Scalapino2012a,LeBlanc2015,Zheng2017,Qin2020,Wietek2021,Bohrdt2019,Kantian2019a}. The latter issue is acute for finding USC-like states in highly controlled ultracold gas experiments with fermions, i.e., analog quantum simulators, for the 2D Hubbard models~\cite{Bloch2008b,Cheuk2016,Boll2016,Parsons2016,Mazurenko2017,Bohrdt2021a}. As current theory cannot efficiently determine where and whether these realize a USC state within the achievable entropies, the impasse on classical computational theory for 2D fermions still impedes progress on the quantum simulation front.

\begin{figure*}[!t]%
		\centering%
		\subfloat[\label{fig:geometry:chains}]%
		{%
			\includegraphics[width=0.5\textwidth]{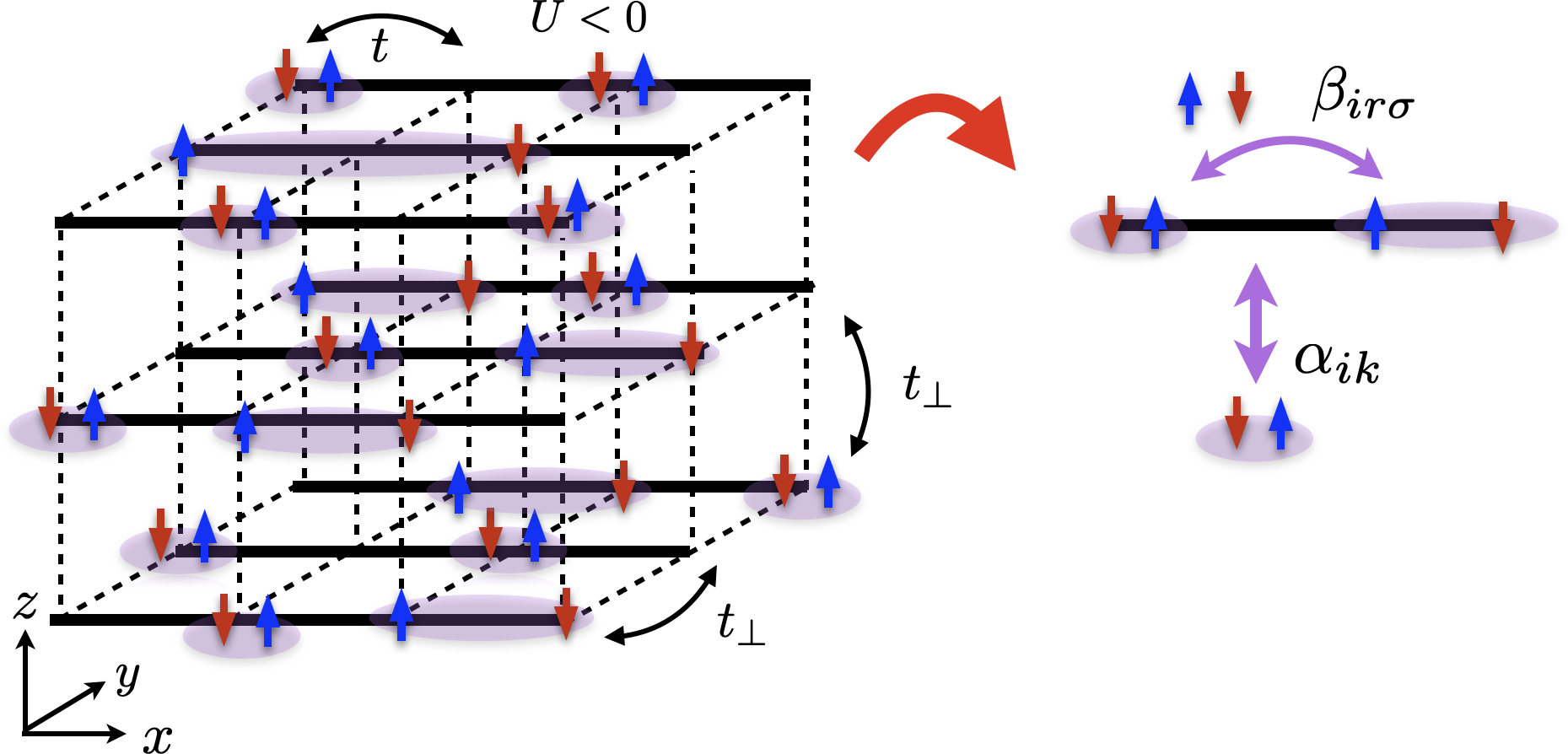}%
		}%
		\subfloat[\label{fig:geometry:ladders}]%
		{%
			\includegraphics[width=0.5\textwidth]{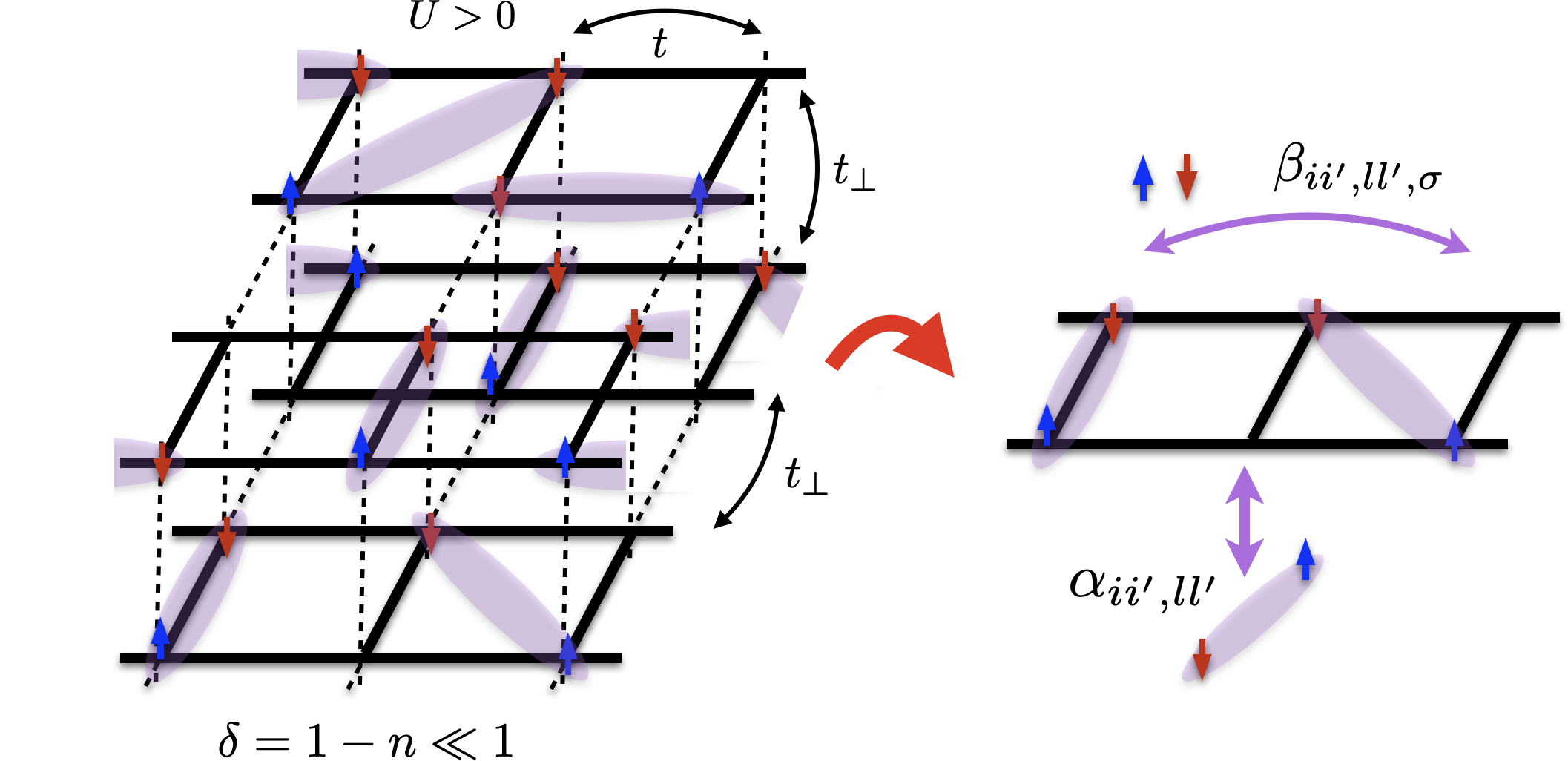}%
		}%
		\caption%
		{%
			\label{fig:schem_model}%
			Overview of MPS+MF for fermions. 
			A 3D array composed of weakly coupled 1D sub-units is mapped onto a self-consistent 1D problem with mean-field amplitudes $\alpha_{ik}$ and $\beta_{ir\sigma}$.
			\subfigref{fig:geometry:chains} Array of negative-$U$ Hubbard chains, with onsite attraction $U$, tunneling $t$ along the chain and inter-chain-tunneling $t_\perp$.
			\subfigref{fig:geometry:ladders} Array of repulsive-$U$ doped Hubbard ladders. Doping levels are $\delta$, n.n.-tunneling inside the ladders is $t$ and $t_\perp$ in-between them.
			%
		}%
		\label{fig:geometry}
	\end{figure*}

	Shifting focus to Q1D systems instead, this work establishes a numerical theory framework capable of meeting these aims. We further enhance the framework via novel TLL+MF theory that allows to compute $T_c$ from easier-to-obtain zero-temperature numerics. Briefly summarized, the framework exploits that in the weak coupling regime between the 1D sub-units we can apply perturbation theory in the ratio of the single-fermion tunneling in-between 1D sub-units to the pairing energy. We show how this ratio naturally controls the possible superconducting $T_c$. In this perturbative regime, which is different from USC and high-$T_c$ superconductivity in Q2D system such as the cuprates, the full 3D array can be decoupled via static MF theory. This maps the problem to that of a single 1D sub-unit with multiple MF-amplitudes, which is then solved self-consistently at the microscopic level. As summarized in Fig.~\ref{fig:schem_model}, these represent the various possible injections/ejections of fermion pairs into/out of the 1D sub-unit stemming from the 3D array, as well as the previously neglected exchange processes. Single-fermion tunneling in-between 1D sub-units is naturally suppressed in this perturbative regime.

	This framework is not limited to superconductivity. It can be applied to any fermionic Q1D setup in which the 1D sub-units have some type of gap sufficient for inter-system tunneling to be relevant first at second order. Such a regime is realized for instance in the various insulating phases of the Bechgaard and Fabre salts~\cite{Giamarchi2004,Bourbonnais2007}. Alternatively, such regimes may be found, with minor modifications, in the recently proposed and partially realized mixed-dimensional large-$J$ systems in ultracold fermionic lattice gases ~\cite{Bohrdt2021,Hirthe2022}. At the same time, the quantitative study of USC regimes of the same compounds, as well as of other Q1D materials such as Chromium pnictide~\cite{Bao2015,Watson2017} and the telephone-number compounds~\cite{Nagata1998,Dagotto1999}, would require including single-electron tunneling in-between 1D sub-units. This would be beyond the MPS+MF framework described here, necessitating an approach akin to chain-dynamical mean field theory~\cite{Biermann2001,Berthod2006}.
	
	The present work hinges on an efficient, reliable numerical method at the microscopic level for correlated-fermion 1D sub-units including MF-amplitudes. Algorithms using matrix product states (MPS)~\cite{Schollwock2011}, such as the density matrix renormalization group (DMRG)~\cite{White1992,White1993} are uniquely suited here. Frameworks using MPS+MF for Q1D spin~\cite{Klanjsek2008,Bouillot2011} and bosonic systems~\cite{Bollmark2020} have been successfully employed, e.g., on experiments in spin-ladder materials. They compare well to Quantum Monte Carlo (QMC) algorithms, despite the MF approximation in the weak-coupling directions, at greatly increased efficiency. 
	
	Applying the MPS+MF approach to fermions is more demanding than for spins or bosons, as the MF-amplitudes are more numerous than for those systems, which typically have one such amplitude. Moreover, these amplitudes range over multiple sites, raising the bond dimension of the matrix product operator~(MPO) representing the total Hamiltonian. When the 1D sub-units have internal structure, as the two-leg Hubbard ladder does, even higher performance is required. Modern MPS implementations can treat even such complex models with many long-range coupling terms, despite the potentially large bipartite entanglement, which sets computational complexity in MPS-methods. Work on DMRG for ground states of 2D Hubbard models demonstrate this fact~\cite{Stoudenmire2012,LeBlanc2015,Zheng2017,Qin2020,Wietek2021,Bohrdt2019}. A recent DMRG-implementation for distributed-memory architectures, parallel DMRG (pDMRG), can treat very large clusters by MPS standards, of e.g., the 2D U-V Hubbard models of the Bechgaard and Fabre salts~\cite{Kantian2019a}.
	Yet, calculations for correlated fermions in Q1D-systems cannot be handled in this way, the bipartite entanglement in 3D would be far too large.
	Auxiliary-field QMC (AFQMC) methods~\cite{HaoPRE2016,Shiwei2019,Assaad2021} are applied in this work for the cases with attractive interactions, but they are typically computationally intensive, and in general the fermionic sign problem prevents exact solutions.

	The MF-approximation within the MPS+MF scheme for fermions thus allows the present work to study the ground state and thermal properties of much larger correlated Q1D systems, including those with repulsive interactions, than either brute-force MPS-methods or AFQMC. Being primarily MPS-based, the framework can also be extended to non-equilibrium real-time problems, such as the study of dynamically induced superconductivity in a Q1D system~\cite{Martens2022}.

	The present work is structured as follows: \cref{Sec::Model} describes a basic Q1D model of superconductivity, an array of negative-$U$ Hubbard-chains, and how to obtain a 1D problem with self-consistent MF-amplitudes from it. \cref{Sec::MPSMF} defines the MPS+MF algorithm used to solve this effective 1D model, and motivates various optimizations. The MF-order treated here is a superconducting one, but we briefly discuss how the framework can be extended to address multiple competing orders simultaneously. Then, \cref{Sec::Bosonization} features an analytical treatment of the same model. In \cref{Sec::Results}, results of the MPS+MF framework are shown. These are compared to analytical methods, and we show how these can be used to obtain $T_c$ from ground-state calculations. Furthermore, we compare our results for a 2D version of our basic Q1D model against AFQMC. Separately, \cref{Sec::HubbardLadder} features all the developed tools being used to solve an array of weakly coupled, doped, repulsive-$U$ Hubbard ladders in 3D, whose low-energy properties are analogous to the array of negative-$U$ Hubbard chains, and the $T_c$ for USC is obtained. In \cref{Sec::Resource_requirements} examples of the required resources for running the different presented MPS+MF use-cases are presented. Finally, in \cref{Sec::Conclusion} the results are discussed and an outlook is given on future work.
	
	\section{\label{Sec::Model}Model}
	The two related models that are studied in this paper are illustrated in \cref{fig:geometry}. Both are Q1D arrays in 2D and 3D, comprised of many identical 1D sub-units in parallel (e.g., chains and ladders). They are described by a Hamiltonian $H_0$, and connected by a perturbing Hamiltonian $H_\perp$ with the general form 
	\begin{equation}\label{fullham}
	H=H_0+t_\perp H_\perp.
	\end{equation}
	The first model we consider is shown in \cref{fig:geometry:chains} and permits extensive testing and validation and is treated in sections \cref{Sec::Model,Sec::MPSMF,Sec::Bosonization,Sec::Results}. It features a 3D Hamiltonian with anisotropic tunneling and interactions where $H_0$ and $H_\perp$ take the following forms:
	\begin{flalign}\label{3dHam}
	\nonumber H_0 = &\sum_{\lbrace\textbf{R}_i\rbrace}H_0(\textbf{R}_i) = \\ \nonumber -t&\sum_n^{L-1}\sum_{\lbrace\R_i\rbrace,\sigma}c^\dagger_{n+1,\R_i,\sigma}c_{n,\R_i\sigma} + \text{h.c.}\\
	\nonumber -\mu&\sum_n^L\sum_{\lbrace\R_i\rbrace,\sigma}c^\dagger_{n,\R_i,\sigma} c_{n,\R_i,\sigma}\\
	+U&\sum_n^L\sum_{\lbrace\R_i\rbrace} n_{n,\R_i,\uparrow}n_{n,\R_i,\downarrow},
	\end{flalign}
	and
	\begin{equation}\label{perpham}
	H_\perp = -\sum_n^L\sum_{\lbrace\R_i\rbrace} \sum_{\hat{\tb{a}}\in\lbrace \hat{\tb{y}},\hat{\tb{z}}\rbrace} c^\dagger_{n,\R_i+\hat{\tb{a}},\sigma}c_{n,\R_i\sigma} + \text{h.c.}.
	\end{equation}
	Here, $\lbrace\R_i\rbrace$ indicates a 2D subspace spanned by ${\hat{\tb{y}},\hat{\tb{z}}}$. The operator of $c_{n,\R_i,\sigma}^\dagger$ creates a fermion on site $n$ in a chain at $\R_i$ with spin $\sigma$. The parameters $t$, $\mu$ and $U$ denote hopping, chemical potential and on-site interaction respectively. The only higher-dimensional effect in this model is given by $H_\perp$. 
	
	The second model we consider is a generalization of the first. It is made up of weakly coupled doped Hubbard ladder systems with repulsive interactions shown in \cref{fig:geometry:ladders}. The definition of $H_0$ and $H_\perp$ differ somewhat in appearance from the first and is specified further in \cref{Sec::HubbardLadder} and \cref{App::Ladder_SMF_model}. The methods developed to treat the first model in \cref{Sec::Model,Sec::MPSMF,Sec::Bosonization,Sec::Results} will be used to treat this latter, more complicated case in \cref{Sec::HubbardLadder}.
	
	For both models, there are two fundamental energy scales: the spin gap and the pairing energy. The former measures the cost of energy to flip a spin in a single 1D system at any position $\R_i$. It is defined by
	\begin{equation}\label{spingap}
	\Delta E_s = E(N,S=1)-E(N,S=0),
	\end{equation}
	where $E(N,S)$ is the ground state energy of ${H_0(\R_i)}$ (i.e., a single 1D sub-unit) at charge and spin quantum numbers $N$ and $S$ respectively.  Conversely, the pairing energy is the cost to move one spin from a ${S=0}$ chain to a neighbouring 1D sub-unit, also at ${S=0}$, creating two chains at ${S=1/2}$
	\begin{equation}\label{pairingenergy}
		\Delta E_p = 2E\left(N+1,\frac{1}{2}\right)-E(N,0)-E(N+2,0).
	\end{equation}
	As will be seen in the following section these energy scales and in particular $\Delta E_p$, for which generally $\Delta E_s \leq \Delta E_p$ (see for instance~\cite{Karakonstantakis2011}), will govern the strength of higher dimensional effects.

	\subsection{\label{Sec::Model/sub::pert_theo}Perturbation theory}
	\begin{figure}
		\ifthenelse{\boolean{buildtikzpics}}%
	    {%
		\tikzsetnextfilename{SchematicSpectrum}
		\tikzset{external/export next=true}
		\begin{tikzpicture}[rotate=90]
		\foreach \i in {0,...,5}
		\draw[thick] (\i/10,0) -- (\i/10,3);
		\foreach \i in {0,...,5}
		\draw[thick] (3+\i/10,0) -- (3+\i/10,3);
		\foreach \i in {0,...,5}
		\draw[thick] (4.0+\i/10,0) -- (4.0+\i/10,3);
		\draw[decorate, decoration={brace}, thick] (3+0.25,-0.1) -- (0.25,-0.1)
		node[pos=0.5,right=3pt]{$\Delta E_s$};
		\draw[decorate, decoration={brace}, thick] (4+0.25,-1) -- (0.25,-1)
		node[pos=0.5,right=3pt]{$\Delta E_p$};
		\draw[decorate, decoration={brace}, thick] (0,3.1) -- (0.5,3.1)
		node[pos=0.5,left=3pt]{$E_{i0}$};
		\draw[decorate, decoration={brace}, thick] (3,3.1) -- (3.5,3.1)
		node[pos=0.5,left=3pt]{$E_{i1}$};
		\draw[decorate, decoration={brace}, thick] (4,3.1) -- (4.5,3.1)
		node[pos=0.5,left=3pt]{$E_{i2}$};
		\end{tikzpicture}
		}%
    	{%
    		\includegraphics{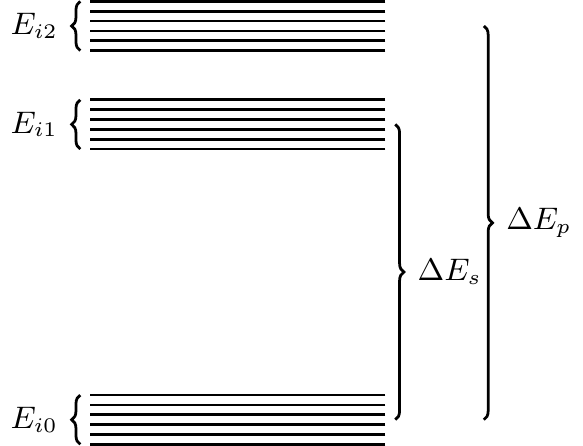}
    	}
		\caption{Schematic representation of the spectrum for Hamiltonians of form \cref{fullham} which are considered. The gaps highlighted are the spin gap $\Delta E_s$ and pairing energy $\Delta E_p$. Notably, the spin gap and pairing energy are approximated as independent of small variations in energy.}
		\label{fig:schematic_spectrum}
	\end{figure}
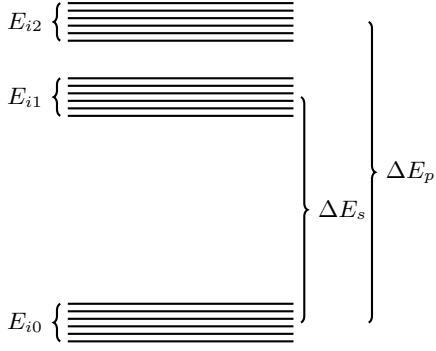
	The models presented in the previous section are typically a challenge to solve numerically and analytically. In particular, the doped Hubbard ladder with repulsive interactions would be impossible to treat in a 3D array of any meaningful size. However, restricting ourselves to the case of the gaps in \cref{spingap,pairingenergy} being large in comparison to the strength of the perturbing Hamiltonian $H_\perp$ we can construct a perturbation theory in the ratio ${t_\perp/\Delta E_p}$, that is, in order to solve Hamiltonians of the form \cref{fullham} we specialize to the case where ${t_\perp\ll \Delta E_p}$.
	
	Specifically, when ${U<0}$ and ${t_\perp=0}$ this model describes a set of disconnected 1D Hamiltonians $H_0(\R_i)$. Each such Hamiltonian exhibits a spectrum of the form~\cite{Giamarchi2003}
	\begin{equation}\label{basic_spectrum}
	|E_{i\alpha}-E_{j\alpha}| \ll |E_{i\alpha} - E_{j\beta}|,~ \alpha \not= \beta,
	\end{equation}
	where $\alpha,\beta$ indicate changes to the state that induce a large energy shift and ${i,j}$ small shifts. The labeling is used to distinguish energy states ${i,j}$ which lie in the same manifold and $\alpha,\beta$ which denote what manifold the state is in. In the case of our model this is related to spin singlet pair-forming by the parameter ${U<0}$. The sub-sequent break-up of such pairs by spin-flip costs ${\Delta E_s}$ in the isolated 1D system, giving rise to~(\ref{basic_spectrum}) which is illustrated in \cref{fig:schematic_spectrum}. Conversely, when a pair is broken by a process like \cref{perpham} two well-separated and unpaired spins are left which changes this cost to ${\Delta E_p}$.
	
	When $t_\perp$ is much smaller than the pairing energy it is possible to produce an effective Hamiltonian with perturbation theory acting in manifold $\alpha$ with matrix elements \cite{CohenTannoudji1998}
	\begin{flalign}
	\nonumber&\braket{i|H_{\mathrm{eff}}^\alpha|j} = E_{i\alpha}\delta_{ij} + \braket{i,\alpha|t_\perp H_\perp|j,\alpha} \\ + \nonumber&\frac{1}{2}\left<i,\alpha\left|t_\perp H_\perp\left(\frac{1}{E_{i\alpha}-H_0}+\frac{1}{E_{j\alpha}-H_0}\right)t_\perp H_\perp\right|j,\alpha\right> \\ + &\mathcal{O}(t_\perp^3).
	\end{flalign}
	The action of $H_\perp$ on $\ket{j,\alpha}$ changes the manifold of the state. We assume that the energy differences between states inside manifolds are much smaller than that of states in different manifolds:
	\begin{equation}
	E_{i0}-E_{j2} \approx -\Delta E_p,
	\end{equation}
	where we let ${\alpha=0}$ indicate the lowest manifold and ${\alpha=2}$ the one reached by $H_\perp$. In other words, we approximate the energy difference between states of different manifolds to be independent of the exact state in that manifold. This yields the effective Hamiltonian
	\begin{equation}\label{eff_ham_nufhc}
	H_{\mathrm{eff}}^0 = H_0P_0 + t_\perp P_0H_\perp P_0 - \frac{t_\perp^2}{\Delta E_p}P_0H_\perp^2P_0,
	\end{equation}
	where $P_\alpha = \sum_{i}\ket{i,\alpha}\bra{i,\alpha}$ is a projector onto manifold $\alpha$. By design, $H_\perp$ puts a state in manifold $0$ into manifold $2$. This means that	
	\begin{equation}
	\braket{i,0|t_\perp P_0H_\perp P_0|j,\beta} = \braket{i,0|t_\perp P_0|k,2} = 0.
	\end{equation}
	Thus, the first order and odd order contribution of the effective Hamiltonian disappears and we are left with
	\begin{equation}
	H_{\mathrm{eff}}^0 = H_0P_0 - \frac{t_\perp^2}{\Delta E_p}P_0H_\perp^2P_0 + \mathcal{O}(t_\perp^4).
	\end{equation}
	In order to understand how $H_\perp^2$ acts on a state we expand:
	\begin{flalign}
	\nonumber H_\perp^2 = &\sum_{n,m}^{L}\sum_{\sigma\sigma'}\sum_{\lbrace\R_i,\R_j\rbrace}\sum_{\hat{\tb{a}},\hat{\tb{b}}\in\lbrace \hat{\tb{y}},\hat{\tb{z}}\rbrace}\\
	\nonumber \bigl(&c^\dagger_{n,\R_i+\hat{\tb{a}},\sigma}c_{n,\R_i\sigma}c^\dagger_{m,\R_j+\hat{\tb{b}},\sigma'}c_{m,\R_j\sigma'} \\+ \nonumber &c^\dagger_{n,\R_i+\hat{\tb{a}},\sigma}c_{n,\R_i\sigma}c^\dagger_{m,\R_j\sigma'}c_{m,\R_j+\hat{\tb{b}},\sigma'} \\+ \nonumber &c^\dagger_{n,\R_i\sigma}c_{n,\R_i+\hat{\tb{a}},\sigma}c^\dagger_{m,\R_j+\hat{\tb{b}},\sigma'}c_{m,\R_j\sigma'} \\+ &c^\dagger_{n,\R_i\sigma}c_{n,\R_i+\hat{\tb{a}},\sigma}c^\dagger_{m,\R_j\sigma'}c_{m,\R_j+\hat{\tb{b}},\sigma'}\bigr),
	\end{flalign}
	obtaining a Hamiltonian describing two-particle tunneling events. What characterizes the $\alpha=0$ manifold is pair formation of opposite spins due to attractive interaction. Some terms within $H_\perp^2$ will put the state in a manifold $\beta>\alpha$ and will subsequently be projected out. Most importantly, any terms flipping spins in two chains simultaneously, such as $c^\dagger_{n,\R_i,\uparrow}c_{n,\R_i+\hat{\tb{y}},\uparrow}c^\dagger_{n,\R_i+\hat{\tb{y}},\downarrow}c_{n,\R_i,\downarrow}$, which would move two chains initially in their $S=0$ manifolds into their $S=\pm 1$ manifolds, will be projected out due to describing a state with at least energy $2\Delta E_s$ above the lowest-energy manifold. In this manner, it is clear that each chain in our model has to conserve spin and in particular pair all spins such that total spin $S=0$.
	
	The parts of $H_\perp^2$ which remain after projecting to $\alpha=0$ must either move particles as pairs (simultaneous tunneling of an up-spin and down-spin particle) or exchange them between chains (an up/down-spin particle is moved out of the chain and another of the same spin is moved in). Any of these processes involve at most two chains in order to conserve spin. While $H_\perp^2$ allows for processes involving chains at arbitrary distance, these would concern four separate chains, with a final state at least $2\Delta E_p$ above the low-energy manifold, and thus removed by projection.
	
	With these restrictions $H_\perp^2$ becomes heavily reduced:
	\begin{flalign} \label{Htp_sq}
	\nonumber H_\perp^2 = &\sum_\sigma\sum_{n,m}^L\sum_{\lbrace\R_i\rbrace} \sum_{\hat{\tb{a}}\in\lbrace \hat{\tb{y}},\hat{\tb{z}}\rbrace} \\ \nonumber\bigl(&c^\dagger_{n,\R_i+\hat{\tb{a}},\sigma}c_{n,\R_i\sigma}c^\dagger_{m,\R_i+\hat{\tb{a}},-\sigma}c_{m,\R_i,-\sigma} \\\nonumber+
	&c^\dagger_{n,\R_i\sigma}c_{n,\R_i+\hat{\tb{a}},\sigma}c^\dagger_{m,\R_i,-\sigma}c_{m,\R_i+\hat{\tb{a}},-\sigma}\bigr) \\\nonumber + 
	&\sum_\sigma\sum_{n,m}^L\sum_{\lbrace\R_i\rbrace} \sum_{\hat{\tb{a}}\in\lbrace \hat{\tb{y}},\hat{\tb{z}}\rbrace} \\\nonumber\bigl( &c^\dagger_{n,\R_i+\hat{\tb{a}},\sigma}c_{n,\R_i\sigma}c^\dagger_{m,\R_i\sigma}c_{m,\R_i+\hat{\tb{a}},\sigma} \\\nonumber + &c^\dagger_{n,\R_i\sigma}c_{n,\R_i+\hat{\tb{a}},\sigma}c^\dagger_{m,\R_i+\hat{\tb{a}},\sigma}c_{m,\R_i,\sigma}\bigr) \\= &H_{\mathrm{pair}} + H_{\mathrm{exc}}.
	\end{flalign}
	The first set of operators, $H_{\mathrm{pair}}$, correspond to pairs of fermions jumping from one chain to another. This conserves spin but not number of particles within a chain. The second set of operators, $H_{\mathrm{exc}}$, correspond to fermions of like spin switching chains (not necessarily at the same site within a chain).
	
	Within this Hamiltonian there is still the degree of freedom to pick how large $|n-m|$ we include. This is something that cannot be restricted by symmetry arguments or the smallness of $t_\perp/\Delta E_p$.
	\subsection{Mean-field theory}
	Notably, the perturbation theory has produced an effective quartic interaction from the single-particle tunneling Hamiltonian. This can be further simplified by casting \cref{Htp_sq} in a form where there is no explicit mention of other chains. This allows us to solve an effective 1D model instead of the full 3D system. We use an ansatz of quasi-free states:
	\begin{equation}
	\braket{c^\dagger_ic^\dagger_jc_kc_l} = \braket{c^\dagger_ic^\dagger_j}\braket{c_kc_l} + \braket{c^\dagger_ic_l}\braket{c^\dagger_jc_k} - \braket{c^\dagger_ic_k}\braket{c^\dagger_jc_l},
	\end{equation}
	which produces a mean-field Hamiltonian from the quartic operators in \cref{Htp_sq}. We further approximate the expectation value of any operator creating/annihilating particles on two different chains to zero, i.e., pair constituents cannot live on different chains. This is motivated by choosing small enough $t_\perp/\Delta E_p$, amounting to the standard mean-field approximation.

	\subsubsection{Pairing terms}
	For each $\R_i$ of $H_{\mathrm{pair}}$ we obtain
	\begin{flalign}
	\nonumber H_{\mathrm{pair},\mathrm{MF}}&=-\sum_{n,m}\sum_\sigma\sum_{\hat{\tb{a}}\in\lbrace \hat{\tb{y}},\hat{\tb{z}}\rbrace}\\\nonumber&\bigl(\braket{c^\dagger_{n\R_i+\hat{\tb{a}},\sigma}c^\dagger_{m\R_i+\hat{\tb{a}},-\sigma}}c_{n,\R_i\sigma}c_{m,\R_i,-\sigma} \\\nonumber&+ c^\dagger_{n\R_i,\sigma}c^\dagger_{m\R_i,-\sigma}\braket{c_{n,\R_i+\hat{\tb{a}},\sigma}c_{m,\R_i+\hat{\tb{a}},-\sigma}} \\\nonumber&+ \braket{c^\dagger_{n\R_i-\hat{\tb{a}},\sigma}c^\dagger_{m\R_i-\hat{\tb{a}},-\sigma}}c_{n,\R_i\sigma}c_{m,\R_i,-\sigma} \\\nonumber&+ c^\dagger_{n\R_i,\sigma}c^\dagger_{m\R_i,-\sigma}\braket{c_{n,\R_i-\hat{\tb{a}},\sigma}c_{m,\R_i-\hat{\tb{a}},-\sigma}}\bigr) \\ = 2z_c \sum_{n,m}&\braket{c_{n\uparrow} c_{m,\downarrow}}c_{n\uparrow}c_{m,\downarrow} + \braket{c_{n\uparrow}c_{m,\downarrow}}c^\dagger_{m\downarrow} c^\dagger_{n,\uparrow},
	\end{flalign}
	where $z_c=4$ is the coordination number for three dimensions. Notably the mean-field is performed on two dimensions instead of the full three. We have assumed that
	\begin{multline} \label{pair_ident}
	\braket{c_{n\uparrow}c_{m,\downarrow}} = 	\braket{c^\dagger_{m\downarrow}c^\dagger_{n,\uparrow}} =  \braket{c_{n\R_i,\uparrow}c_{m\R_i,\downarrow}} \\ =
	\braket{c_{n\R_i+p\hat{\tb{a}},\uparrow}c_{m\R_i+p\hat{\tb{a}},\downarrow}}
	,~\forall p\in\mathbb{N},~\hat{\tb{a}}\in\lbrace\hat{\tb{y}},\hat{\tb{z}}\rbrace,
	\end{multline}
	i.e., that all chains are identical and the mean-field is real-valued.
	
	\subsubsection{Exchange terms}
	For each $\R_i$ of $H_{\mathrm{exc}}$ we obtain
	\begin{flalign}
	\nonumber H_{\mathrm{exc},\mathrm{MF}} = -\sum_\sigma\sum_{n,m}^L &\sum_{\hat{\tb{a}}\in\lbrace \hat{\tb{y}},\hat{\tb{z}}\rbrace} \\\nonumber\braket{c^\dagger_{n,\R_i+\hat{\tb{a}},\sigma}c_{m,\R_i+\hat{\tb{a}},\sigma}}&c^\dagger_{m,\R_i\sigma}c_{n,\R_i\sigma} \\\nonumber+\braket{c^\dagger_{n,\R_i-\hat{\tb{a}},\sigma}c_{m,\R_i-\hat{\tb{a}},\sigma}}&c^\dagger_{m,\R_i\sigma}c_{n,\R_i\sigma} \\\nonumber+\braket{c^\dagger_{m,\R_i+\hat{\tb{a}},\sigma}c_{n,\R_i+\hat{\tb{a}},\sigma}}&c^\dagger_{n,\R_i\sigma}c_{m,\R_i,\sigma} \\\nonumber+\braket{c^\dagger_{m,\R_i-\hat{\tb{a}},\sigma}c_{n,\R_i-\hat{\tb{a}},\sigma}}&c^\dagger_{n,\R_i\sigma}c_{m,\R_i,\sigma}
	\\\nonumber = -z_c\sum_\sigma\sum_{n,m}^{L} &\braket{c^\dagger_{n,\sigma}c_{m,\sigma}}c^\dagger_{m,\sigma}c_{n,\sigma} \\\nonumber+&\braket{c^\dagger_{m,\sigma}c_{n,\sigma}}c^\dagger_{n,\sigma}c_{m,\sigma} \\\nonumber=-2z_c\sum_{n,\sigma}\sum_{r=1}^{L-n} &\braket{c^\dagger_{n,\sigma}c_{n+r,\sigma}}c^\dagger_{n+r,\sigma}c_{n,\sigma} \\+&\braket{c^\dagger_{n+r,\sigma}c_{n,\sigma}}c^\dagger_{n,\sigma}c_{n+r,\sigma},
	\end{flalign}
	where we have done similarly as in \cref{pair_ident}.
	\subsection{Effective 1D Hamiltonian}
	To summarize: within the MPS+MF approach, the total Hamiltonian is described by its one-dimensional subsets in a mean-field sense. It is sufficient to consider the effectively 1D Hamiltonian
	\begin{multline}\label{mf_ham}
	H_{\mathrm{MF}} = H_0(\textbf{R}_i) - \sum_{ik}\alpha_{ik}\left(c_{i\uparrow}c_{k\downarrow} + c^\dagger_{k\downarrow}c^\dagger_{i\uparrow}\right) \\+ \sum_{i\sigma}\sum_{r=1}^{L-n} \beta_{i,r,\sigma}\Big(c^\dagger_{i+r,\sigma}c_{i,\sigma} + c^\dagger_{i,\sigma}c_{i+r,\sigma}\Big),
	\end{multline}
	where
	\begin{equation}
	\alpha_{ik} = \frac{2z_ct_\perp^2}{\Delta E_p}\braket{c_{i\uparrow}c_{k\downarrow}},
	\end{equation}
	are the pair-MF parameters, describing pair-tunneling into/out of the 1D system mediated by $H_\rm{pair}$, and
	\begin{equation}
	\beta_{ir\sigma} = \frac{2z_ct_\perp^2}{\Delta E_p}\braket{c^\dagger_{i+r,\sigma}c_{i,\sigma}},
	\end{equation}
	are the exchange processes mediated by $H_\rm{exc}$ (c.f. \cref{Htp_sq}). Notably, $\alpha_{ik}$ is proportional to the bound state pair-wavefunction. In practice, keeping all ranges will make the problem intractable and limits must be introduced. Fortunately, the existence of a spin gap causes $\alpha$ and $\beta$ to decay exponentially with distance controlled by the spin correlation length~\cite{Giamarchi2003}. This allows us to include a cut-off for which amplitudes to keep. The choice of this value ultimately depends on microscopic parameters (in particular interaction strength) and will be exemplified in \cref{Sec::Results}.
	
	In addition, it is important to note that the zero-range terms ${\beta_{i0\sigma}\propto \langle n_{i\sigma}\rangle}$ are absent in $H_{\mathrm{MF}}$ as we assume constant density in the superconducting order and exclude potential insulating orders. We have already extended the MPS+MF framework to study the competition of insulating charge-orders with superconducting ones, based on adding $\beta_{i0\sigma}$-parameters, and are applying this to the Hubbard-ladder arrays in forthcoming work.
	
	Finally, as discussed in \cref{Sec::Model/sub::pert_theo}, MF-terms such as $\langle S^+_i\rangle S^-_i$ cannot occur in $H_\rm{MF}$ by construction, as such magnetic exchange-terms are suppressed through the condition that $t_\perp\ll \Delta E_s$.
	\section{MPS+MF - Numerical solutions to Q1D systems using mean-field theory}\label{Sec::MPSMF}
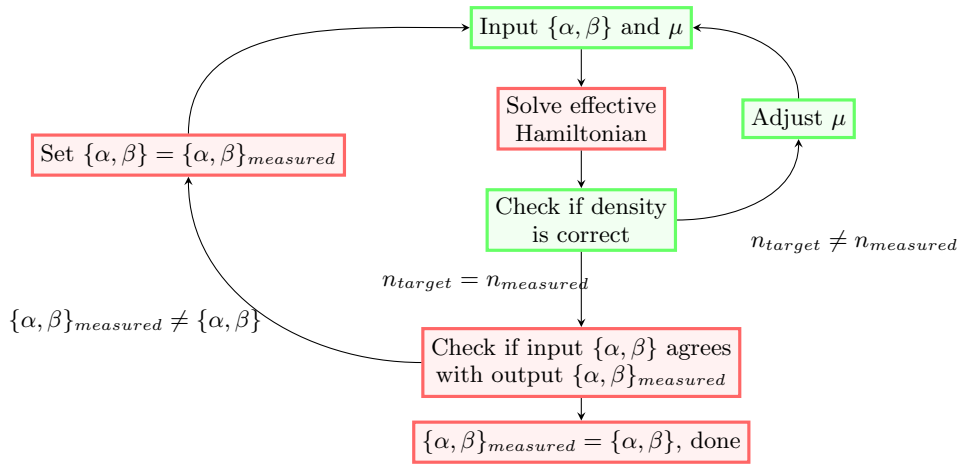
\begin{figure*}[!t]
	    \ifthenelse{\boolean{buildtikzpics}}%
    	{%
		\tikzsetnextfilename{MPSMFSchematic}
		\tikzset{external/export next=true}
		\begin{tikzpicture}[recnode/.style={rectangle, draw=red!60, fill=red!5, very thick, minimum size=5mm}, chemnode/.style={rectangle, draw=green!60, fill=green!5, very thick, minimum size=5mm}]
			\node[chemnode] (input) {Input $\{\alpha, \beta\}$ and $\mu$};
			\node[recnode, align=center] (solve) [below=0.5cm of input] {Solve effective \\ Hamiltonian};
			\node[chemnode, align=center] (dens) [below=0.5cm of solve] {Check if density \\ is correct};
			\node[chemnode, align=center] (densloop) [right=of solve] {Adjust $\mu$};
			\node[recnode, align=center] (measure) [below=of dens] {Check if input \{$\alpha, \beta\}$ agrees\\ with output $\{\alpha, \beta\}_{measured}$};
			\node[recnode, align=center] (loop) [above left=2cm and 1cm of measure] {Set $\{\alpha,\beta\}=\{\alpha,\beta\}_{measured}$};
			\node[recnode, align=center] (exit) [below= 0.3cm of measure] {$\{\alpha,\beta\}_{measured} = \{\alpha,\beta\}$, done};
			
			\node[below left=1.1cm and -3cm of densloop] {$n_{target}\not=n_{measured}$};
			\node[below left=0.2cm and -1.5cm of dens] {$n_{target}=n_{measured}$};
			\node[above left=-0.2cm and 2cm of measure] {$\{\alpha,\beta\}_{measured}\not=\{\alpha,\beta\}$};
			
			\draw[->] (input.south) -- (solve.north);
			\draw[->] (solve.south) -- (dens.north);
			\draw[->] (dens.south) -- (measure.north);
			\draw[->] (measure.west) to[out=180, in=270] (loop.south);
			\draw[->] (loop.north) to[out=90, in=180] (input.west);
			\draw[->] (measure.south) -- (exit.north);
			\draw[->] (dens.east) to[out=0, in=270] (densloop.south);
			\draw[->] (densloop.north) to[out=90, in=0] (input.east);
		\end{tikzpicture}
		}%
	    {%
		    \includegraphics{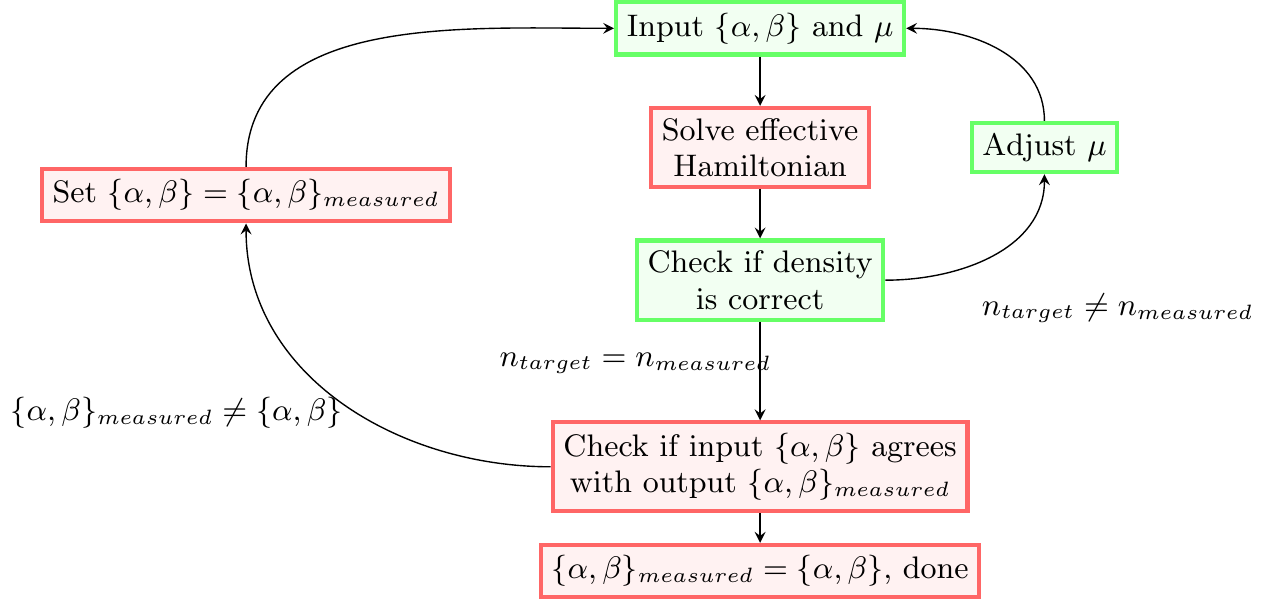}%
		}%
		\caption{Schematic representation of the MPS+MF framework. Green boxes highlight the density fixation part of the routine. Red boxes denote the mean-field amplitude fixation part. The mean-field amplitude set is denoted by $\{\alpha,\beta\}$ where $\{\alpha\}$ represent pairing amplitudes and $\{\beta\}$ the exchange amplitudes.}
		\label{fig:mpsmf_schem}
	\end{figure*}

	MPS+MF has been developed to solve Q1D models by relying on mean-field approximations of the full Hamiltonian which converts it to an effectively 1D system. The produced Hamiltonian in \cref{mf_ham} may be solved iteratively until self-consistency is reached for the mean-field amplitudes as shown schematically in \cref{fig:mpsmf_schem}.
	This framework has been developed as an approach to anisotropic systems in two or more dimensions for which the 1D correlations are most important and where single-particle tunneling and spin-flipping processes between chains can be neglected on account of $t_\perp\ll \Delta E_p$. The primary cost of the routine comes from the repeated solutions of effective 1D mean-field Hamiltonian such as~(\ref{mf_ham}) using MPS-based methods. Notably, any MPS-method may be used to iterate mean-field amplitudes, for example original DMRG, MPS-based DMRG or imaginary time evolution on purified states to obtain thermal states. In practice, the only requirement is that the mean-field amplitudes converge. Thus, the framework scales as the utilized DMRG method does with bond dimension and system size~\cite{Schollwock2011}. In this paper we utilize MPS-based DMRG to solve for ground states~\cite{Schollwock2011}. For thermal states we utilize both trotterized imaginary time evolution and the time-dependent variational principle (TDVP)~\cite{Paeckel2019}.
	\begin{figure}
        \ifthenelse{\boolean{buildtikzpics}}%
	    {%
		\tikzsetnextfilename{OrpTruncExtrap}
		\tikzset{external/export next=true}
		\begin{tikzpicture}
			\begin{axis}
				[
					xlabel	=	{$\epsilon_\psi$},
					ylabel	=	{$\delta\alpha$},
					width	=	0.24\textwidth,
					height	=	0.2\textheight,
				]
				\pgfplotstableread[header=false]{Data/observable_trunc_extrap_fig_1_fit.dat}{\fittingtable}
				\pgfplotstableread[header=true]{Data/observable_trunc_extrap_fig_1_dps.dat}{\datatable}
				\pgfplotstablegetelem{0}{[index]0}\of\fittingtable
				\edef\slope{\pgfplotsretval}
				\pgfplotstablegetelem{0}{[index]1}\of\fittingtable
				\edef\infval{\pgfplotsretval}
				\pgfplotstablegetelem{0}{[index]0}\of\datatable
				\edef\maxtrunc{\pgfplotsretval}
				\addplot
				[	
					colorD,
					only marks,
					thick,
					mark=o,
				]
				table
				[
					y expr = (\thisrowno{1}-(\infval)),
					x expr = \thisrowno{0},
				]
				{Data/observable_trunc_extrap_fig_1_dps.dat};
				\addplot
				[
					colorD,
					no marks,
					thick,
					domain = 0:\maxtrunc
				]
				{linearFct(\slope, 0)};
				\addplot
				[
				red,
				thick,
				mark=x
				]
				coordinates {(0,0)};
			\end{axis}
		\end{tikzpicture}
		}%
    	{%
    		\includegraphics{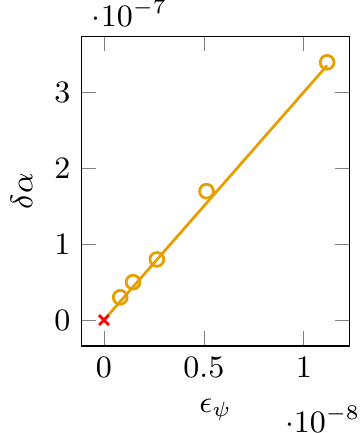}%
    	}%
    	\ifthenelse{\boolean{buildtikzpics}}%
	    {%
		\tikzsetnextfilename{EnerTruncExtrap}
		\tikzset{external/export next=true}
		\begin{tikzpicture}
		\begin{axis}
		[
		xlabel	=	{$\epsilon_\psi$},
		ylabel	=	{$\delta E$},
		width	=	0.24\textwidth,
		height	=	0.2\textheight
		]
		\pgfplotstableread[header=false]{Data/observable_trunc_extrap_fig_2_fit.dat}{\fittingtable}
		\pgfplotstableread[header=true]{Data/observable_trunc_extrap_fig_2_dps.dat}{\datatable}
		\pgfplotstablegetelem{0}{[index]0}\of\fittingtable
		\edef\slope{\pgfplotsretval}
		\pgfplotstablegetelem{0}{[index]1}\of\fittingtable
		\edef\infval{\pgfplotsretval}
		\pgfplotstablegetelem{0}{[index]0}\of\datatable
		\edef\maxtrunc{\pgfplotsretval}
		\addplot
		[	
		colorD,
		thick,
		only marks,
		mark=o,
		]
		table
		[
		y expr = {\eval{\thisrowno{1}-(\infval)}},
		x expr = \thisrowno{0},
		]
		{Data/observable_trunc_extrap_fig_2_dps.dat};
		\addplot
		[
			colorD,
			thick,
			no marks,
			domain = 0:\maxtrunc
		]
		{linearFct(\slope,0)};
		\addplot
		[
		red,
		thick,
		mark=x
		]
		coordinates {(0,0)};
		
		\end{axis}
		\end{tikzpicture}
		}%
	    {%
		    \includegraphics{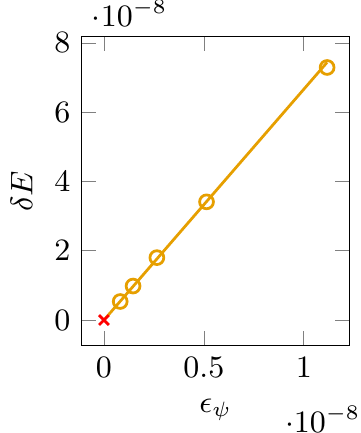}%
		}%
		\caption{Example of truncation error extrapolation of order parameter and ground state energy for a chain of length $L=100$ at attractive interaction $U=-4t$ and density $n=0.5$ and $t_\perp=0.05t$. The y-axes represent difference of order parameter $\delta\alpha=\alpha(\epsilon_\psi)-\alpha(0)$ and similarly for energy. The range of MF terms taken into account is $r=4$.}
		\label{fig:trunc_err_extrap}
	\end{figure}
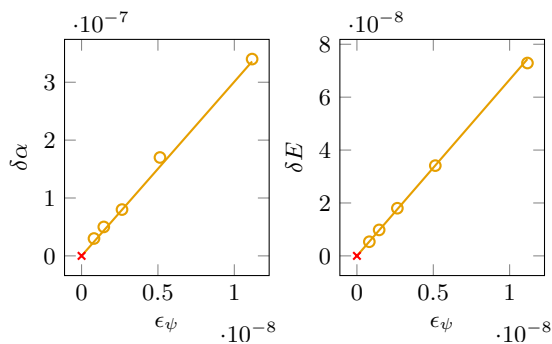

One common feature of states obtained with DMRG ground state algorithms is that the error of local quantities scales linearly with the discarded weight~\cite{Schollwock2005}. Given that the system under study is 1D, albeit representing a 3D system, we might expect that such linear scaling holds for MPS+MF as well. In \cref{fig:trunc_err_extrap} we show an example of our general finding that this holds true for quantities such as energy and order parameter.
	
	\subsection{Achieving self-consistent convergence}
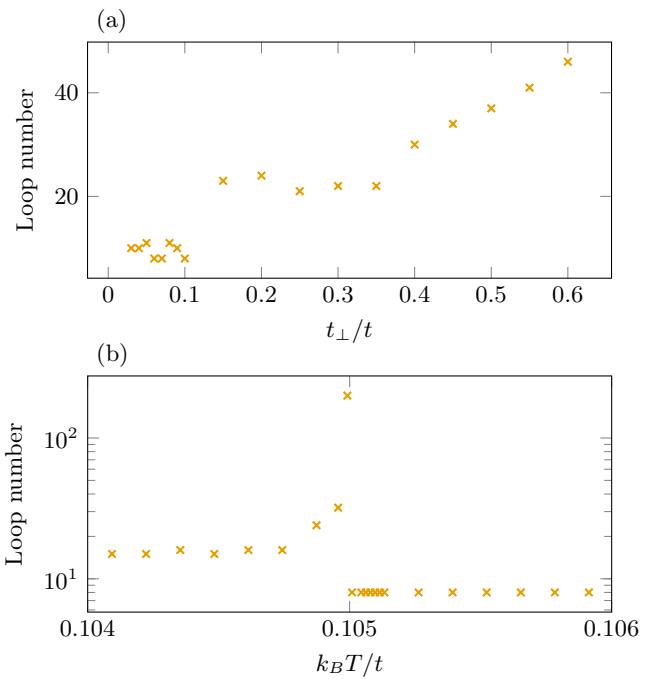
\begin{figure}
	    \ifthenelse{\boolean{buildtikzpics}}%
	    {%
		\tikzsetnextfilename{GSTimeCostInLoopNumber}
		\tikzset{external/export next=true}
		\begin{tikzpicture}
			\begin{groupplot}
			[
				group style=
				{
					group size= 1 by 2,
					vertical sep		=	4em,				
				},
				width	=	\columnwidth-2.8pt,
				height	=	0.2\textheight,
			]
				\nextgroupplot
				[
					xlabel	=	{$t_\perp/t$},
					ylabel	=	{Loop number},
					title = {(a)},
					every axis title/.style={above right, at={(0,1)}},
				]
					\addplot
					[
						colorD,
						thick,
						only marks,
						mark=x
					]
						table
						[
							y expr = \thisrowno{1},
							x expr = \thisrowno{0}
						]
							{Data/time_cost_in_loop_count_subfig_1.dat};

				\nextgroupplot
				[
					xlabel	=	{$k_BT/t$},
					ylabel	=	{Loop number},
					xmin = 0.104,
					xmax = 0.106,
					ymode = log,
					xticklabels={0.104,0.105,0.106},
					xtick={0.104,0.105,0.106},
					title = {(b)},
					every axis title/.style={above right, at={(0,1)}},
				]
					\addplot
					[
						colorD,
						thick,
						only marks,
						mark=x,
					]
						table
						[
							y expr = \thisrowno{1},
							x expr = \thisrowno{0},
						]
							{Data/time_cost_in_loop_count_subfig_2.dat};
			\end{groupplot}
		\end{tikzpicture}
		}%
    	{%
    		\includegraphics{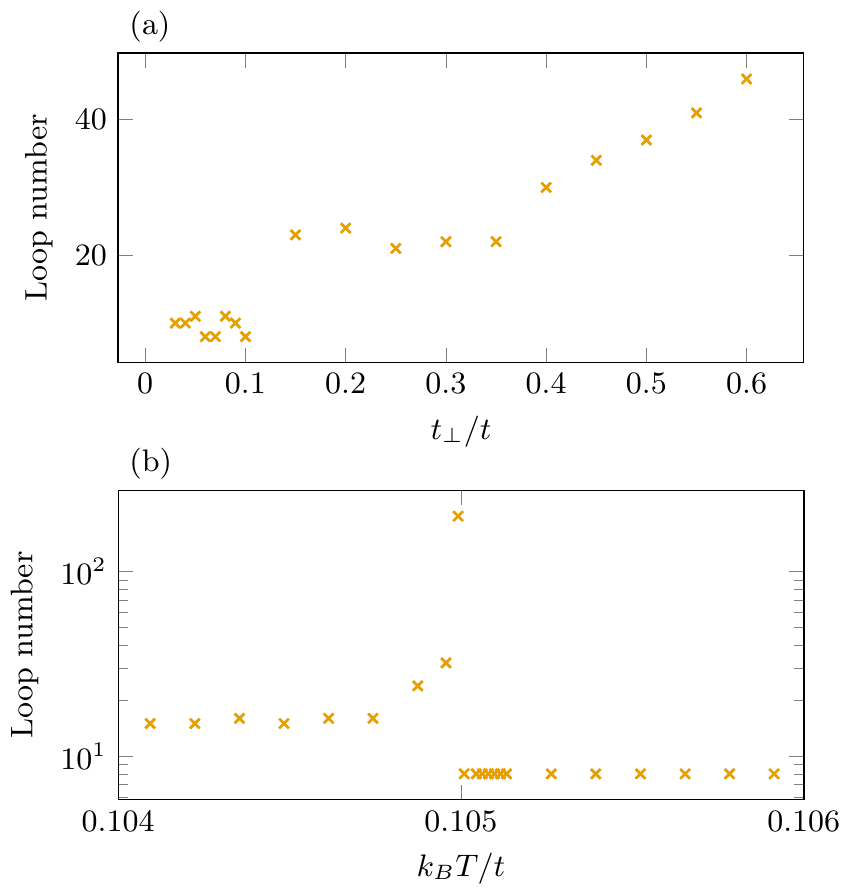}%
    	}%
		\caption{ Number of self-consistent loops required for convergence for (a) A ground state calculation for different values $t_\perp$, at $L=100$, $n=0.5$, $U=-4t$, $\chi=300$ and MF term range $r=4$ (b) A finite temperature calculation for different temperatures at $L=60$, $n=0.5$, $U=-10t$, $t_\perp=0.3t$, $\chi=200$ and MF terms with range $r=1$.}
		\label{fig:loop_number}
	\end{figure}

	When iterating a MF-theory to self-consistency, the required number of iterations becomes crucial as each one may have significant cost. The number of solutions required to achieve a self-consistent state is not constant over parameter space as shown in \cref{fig:loop_number}a. However, it only varies modestly with Hamiltonian parameters, such as $t_\perp$. The notable increase in loop number with larger $t_\perp$ is primarily due to increased difficulty fixing the density~(see \cref{Sec::MPSMF}D).
	
	Furthermore, we find that the number of required iterations peaks strongly around the phase transition from superconducting to normal phase as shown in \cref{fig:loop_number}b where the transition occurs around $T=0.105$~\cite{Ho2004,Cazalilla2006a,Bollmark2020}. When the superconductor is close to its transition to a metal the MF-coupling between chains is weak but non-zero. We find this slows down convergence rate explaining the increase in the number of required loops for convergence.

	In both this work and the primary uses to which the MPS+MF framework would be applied, the interest in phase transitions is central and we shall have to resolve the points which are difficult to obtain. This has prompted the development of several heuristics in this method to produce a faster convergence.
	
		\subsection{Extrapolation}
\begin{figure}
        \ifthenelse{\boolean{buildtikzpics}}%
	    {%
		\tikzsetnextfilename{LoopExtrapSample}
		\tikzset{external/export next=true}
		\begin{tikzpicture}
			\begin{semilogyaxis}
			[
				width = \columnwidth,
				height = 0.3\textheight,
				xlabel = {Loop number},
				ylabel = {$\alpha_n-\alpha_\infty$}
			]
			\pgfplotstableread[header=true]{Data/extrap_routine_sample.dat}{\datatable}
			\pgfplotstablegetrowsof{\datatable}
			\pgfmathsetmacro{\rowcount}{\pgfplotsretval-1}
			\pgfplotstablegetelem{\rowcount}{1}\of\datatable
			\edef\infval{\pgfplotsretval}
			\pgfplotstablegetelem{\rowcount}{0}\of\datatable
			\edef\lastloop{\pgfplotsretval+2}
			\pgfplotstableread[header=true]{Data/extrap_routine_sample_fit_1.dat}{\fit1}
			\pgfplotstablegetelem{0}{[index]0}\of\fit1
			\pgfmathsetmacro{\caa}{\pgfplotsretval}
			\pgfplotstablegetelem{0}{[index]1}\of\fit1
			\pgfmathsetmacro{\cab}{\pgfplotsretval}
			\pgfplotstablegetelem{0}{[index]2}\of\fit1
			\pgfmathsetmacro{\cac}{\pgfplotsretval}
			\pgfplotstablegetelem{0}{[index]3}\of\fit1
			\pgfmathsetmacro{\cad}{\pgfplotsretval}
			\pgfplotstableread[header=true]{Data/extrap_routine_sample_fit_2.dat}{\fit2}
			\pgfplotstablegetelem{0}{[index]0}\of\fit2
			\pgfmathsetmacro{\cba}{\pgfplotsretval}
			\pgfplotstablegetelem{0}{[index]1}\of\fit2
			\pgfmathsetmacro{\cbb}{\pgfplotsretval}
			\pgfplotstablegetelem{0}{[index]2}\of\fit2
			\pgfmathsetmacro{\cbc}{\pgfplotsretval}
			\pgfplotstablegetelem{0}{[index]3}\of\fit2
			\pgfmathsetmacro{\cbd}{\pgfplotsretval}
			\pgfplotstableread[header=true]{Data/extrap_routine_sample_fit_3.dat}{\fit3}
			\pgfplotstablegetelem{0}{[index]0}\of\fit3
			\pgfmathsetmacro{\cca}{\pgfplotsretval}
			\pgfplotstablegetelem{0}{[index]1}\of\fit3
			\pgfmathsetmacro{\ccb}{\pgfplotsretval}
			\pgfplotstablegetelem{0}{[index]2}\of\fit3
			\pgfmathsetmacro{\ccc}{\pgfplotsretval}
			\pgfplotstablegetelem{0}{[index]3}\of\fit3
			\pgfmathsetmacro{\ccd}{\pgfplotsretval}
			\addplot
			[
				blue,
				only marks,
				thick,
				mark=o,
				restrict y to domain=\eval{ln(1e-4)}:0
			]
			table
			[
				y expr = {\eval{\thisrowno{1}-(\infval)}},
				x expr = \thisrowno{0}
			]
			{\datatable};
			\addplot
			[
				orange,
				no marks,
				thick,
				domain=\caa:\lastloop
			]
			{\cab*exp(-(\cac)*(x-\caa))+\cad-(\infval)};
			\addplot
			[
			green,
			no marks,
			thick,
			domain=\cba:\lastloop
			]
			{\cbb*exp(-(\cbc)*(x-\cba))+\cbd-(\infval)};
			\addplot
			[
				orange,
				no marks,
				thick,
				dashed,
				domain=0:\lastloop
			]
			{linearFct(0,\cad-(\infval))};
			\addplot
			[
				green,
				no marks,
				thick,
				dashed,
				domain=0:\lastloop
			]
			{linearFct(0,\cbd-(\infval))};
			\addplot
			[
				red,
				no marks,
				thick,
				domain=\cca:\lastloop
			]
			{\ccb*exp(-(\ccc)*(x-\cca))+\ccd-(\infval)};
			\end{semilogyaxis}
		\end{tikzpicture}
		}%
    	{%
    		\includegraphics{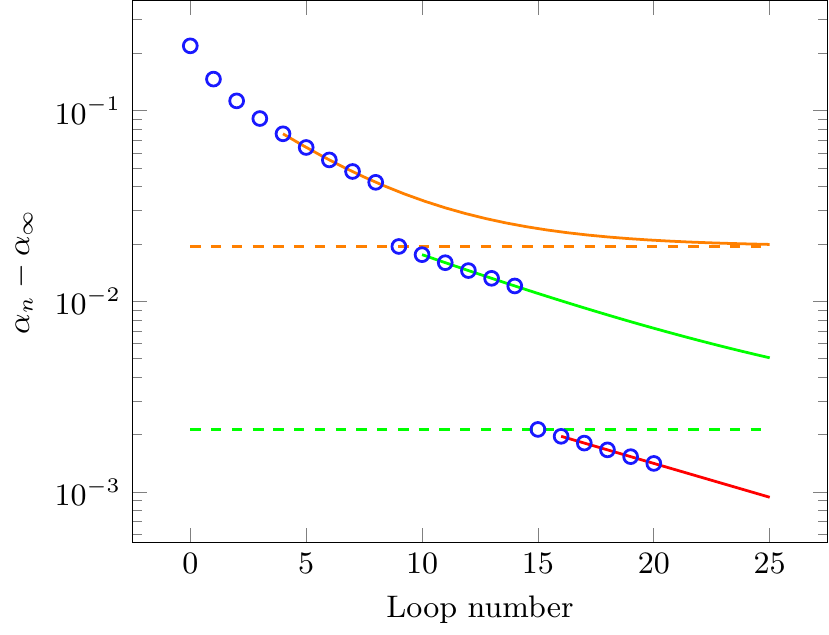}%
    	}%
		\caption{ Example of the exponential trend when iterating to self-consistency. The mean-field amplitude over loop number $n$ is denoted by $\alpha_n$ and converged amplitude by $\alpha_\infty$. The system parameters are $L=60$, $U=-10t$, $t_\perp=0.4t$, $\beta t=6.4286$, $\chi=200$ and MF terms with range $r=1$.}
		\label{fig:expdecay}
	\end{figure}
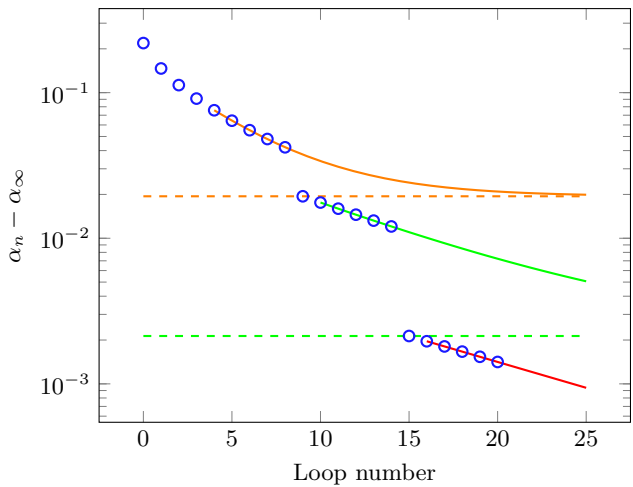

	Convergence implies that the mean-field amplitudes approach a set which no longer changes with further iteration, i.e., the change of these amplitudes with each iteration decreases. As can be seen from \cref{fig:expdecay} we find such exponential behaviour after an initial fluctuation (related to the initial guess). Given the clear trend an extrapolation can be performed to attempt a prediction of the converged amplitudes.
	
	In practice, such fits seldom give precise results but typically lead to a more converged amplitude compared to simply iterating one more loop. As such, we employ a strategy of repeated extrapolation in an attempt to speed up the self-consistent iteration as shown in \cref{fig:expdecay}. The result of 5 consecutive iterations is checked for sufficient exponential behaviour and is fitted to an exponential function
	\begin{equation}
	f(x) = Ae^{-|B|x} + C,
	\end{equation}
	where the fitting parameters $A,B,C$ are determined by a least-squares fit. In this manner, $C$ is the fitted result when approaching infinite iterations and is chosen as the amplitude to use for the next iteration.
	
	Typically, we find that this scheme reduces the number of necessary iterations to varying degree. In particular, close to transitions the slow convergence can be greatly aided by skipping ahead using extrapolations. In effect, the scheme is a method of breaking off the iterative procedure in order to generate an improved initial guess using exponential extrapolation. The amount of speedup depends on how evenly and slowly the mean-field amplitudes converge. If the convergence is quick, fitting to an exponential form is less faithful. Conversely, if convergence is too slow the extrapolated result makes predictions far outside any region of reliability (e.g., using 10 iterations to predict what would happen after 1000 iterations). In the best case scenarios we find the number of iterations reduced by up to a factor of $5$ and in the worst case it only executes once or not at all leading to small or no speedup.
	\subsection{Initial guesses\label{subsec::init_guess}}
	As shown in \cref{fig:mpsmf_schem} the routine has to be started with some input mean-field amplitudes and chemical potential. This indicates that the number of DMRG solutions is dependent on the quality of the initial guess. In certain cases bosonization can be used to make estimations of what the converged amplitudes would be, in particular where analytical control is good. However, in general such estimations will differ from the DMRG result and are often harder to obtain than simple heuristic guesses.
	
	What is always available for initial values is that of previous solutions using the framework. Indeed, the best guess is the set of converged amplitudes leading to an immediate solution. This becomes useful when considering that mean-field amplitudes should often only change marginally when subject to a marginal shift in parameter space (notable exception is that of first-order transitions). Hence, if we seek a point in parameter space that is close to one which is already computed the converged amplitudes of the computed point serve as a guess which should be close to the converged result.
	
\begin{figure}
	    \ifthenelse{\boolean{buildtikzpics}}%
	    {%
		\tikzsetnextfilename{LowBDinHighBDSample}
		\tikzset{external/export next=true}
		\pgfdeclarelayer{background}
		\pgfdeclarelayer{foreground}
		\pgfsetlayers{background,main,foreground}
		\begin{tikzpicture}[plot box/.style={thick}]
			\begin{pgfonlayer}{background}
				\begin{axis}
				[
					width = \columnwidth,
					height = 0.3\textheight,
					xlabel = {Loop number},
					ylabel = {$\alpha_n$},
					yticklabel style = {/pgf/number format/precision=3,
						/pgf/number format/fixed},
					legend pos = south west
				]
				\addplot
				[
					blue,
					thick,
					mark=o,
					only marks
				]
				table
				[
					y expr = \thisrowno{1},
					x expr = \thisrowno{0}
				]
				{Data/low_bd_in_high_bd_low_chi.dat};
				\addlegendentry{Guess}
				\addplot
				[
					thick,
					orange,
					mark=x,
					only marks
				]
				table
				[
					y expr = \thisrowno{1},
					x expr = \thisrowno{0}
				]
				{Data/low_bd_in_high_bd_high_chi.dat};
				\addlegendentry{Low-$\chi$ guess}
				\coordinate (inset) at (axis description cs:0.95,0.9);
				\draw[plot box, name=cool] (24.5,0.045) rectangle (40,0.07);
				\coordinate (zoomboxcornerleft) at (axis cs:24.5,0.07);
				\coordinate (zoomboxcornerright) at (axis cs:40,0.07);
				\end{axis}
			\end{pgfonlayer}
			\begin{pgfonlayer}{foreground}
				\begin{axis}
				[
					at = {(inset)},
					anchor = north east,
					width = 0.5\columnwidth,
					height = 0.15\textheight,
					xticklabels={},
				]
				\addplot
				[
					blue,
					thick,
					mark=o,
					only marks,
					restrict x to domain=25:40
				]
				table
				[
					y expr = \thisrowno{1},
					x expr = \thisrowno{0}
				]
				{Data/low_bd_in_high_bd_low_chi.dat};
				\addplot
				[
					thick,
					orange,
					mark=x,
					only marks,
					restrict x to domain=25:40
				]
				table
				[
					y expr = \thisrowno{1},
					x expr = \thisrowno{0}
				]
				{Data/low_bd_in_high_bd_high_chi.dat};
				\coordinate (lowleft) at (axis description cs:0,0);
				\coordinate (lowright) at (axis description cs:1,0);
				\end{axis}
			\end{pgfonlayer}
			\draw (lowleft) -- (zoomboxcornerleft);
			\draw (lowright) -- (zoomboxcornerright);
		\end{tikzpicture}
		}%
    	{%
    		\includegraphics{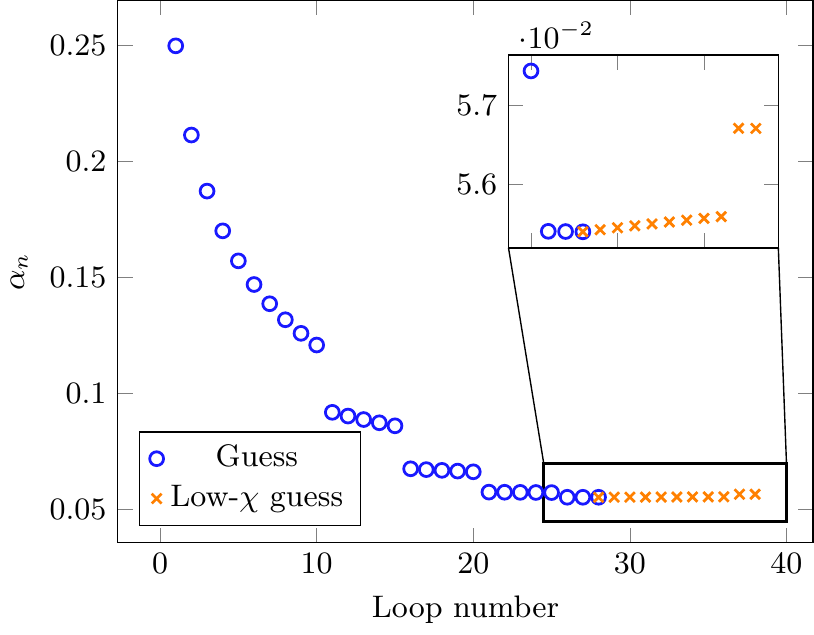}%
    	}
		\caption{Comparison of loop count $n$ and mean-field amplitude $\alpha_n$ when using a simple, unguided, guess at bond dimension $\chi=100$ and when using the $\chi=100$ converged amplitudes as a guess for $\chi=200$. The system parameters are $L=60$, $U=-10t$, $t_\perp=0.3t$, $\beta t=9.5948$ and $\chi=100$ (blue circles) and $\chi=200$ (orange crosses). The MF terms in both cases have range $r=1$.}
		\label{fig:low_bd_in_high_bd}
	\end{figure}
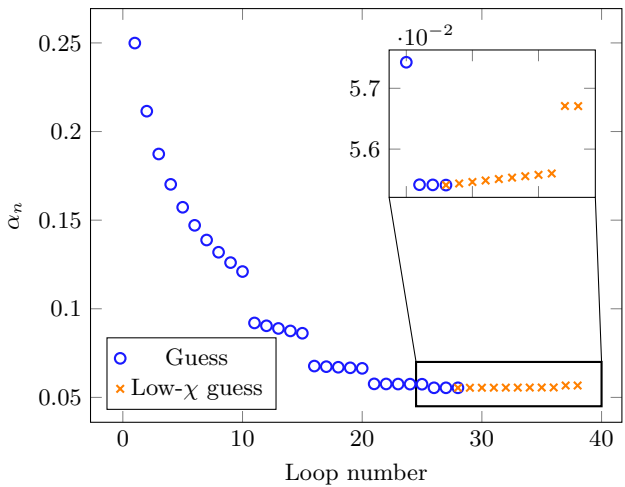

	This observation is of particular use when considering the bond dimension. DMRG scales cubically in this quantity~\cite{Schollwock2011} (which may be improved using conserved quantum numbers). It is imperative to keep this parameter large enough in order for the obtained state to reasonably approximate the targeted state. With the MPS+MF framework it is possible to compute the self-consistent amplitudes at a lower bond dimension starting with unguided guesses at low cost. When high precision is desired we use the cheap results to compute amplitudes at higher bond dimension. We find that the number of required iterations drops significantly with this strategy, as exemplified in~\cref{fig:low_bd_in_high_bd}.

	Further, this strategy may be applied to the issue of phase transitions. For such cases a dense grid of data is often necessary with amplitudes varying only modestly between points. Once a grid of some sparsity has been generated a tighter one is cheaper due to the possibility of interpolating between existing points, providing good initial guesses.

	\subsection{Density-fixing}
	While the original Hamiltonian \cref{fullham} is explicitly particle number conserving the derived effective 1D Hamiltonian \cref{mf_ham} loses this property. Since carrier density has been shown to affect superconductivity it is important to consider what density the converged solution obtains. In practice, being able to fix the density is important, since there is no guarantee that converged solutions at different parameters have the same density thereby making comparison difficult.
	
	In \cref{fig:mpsmf_schem} two classes of amplitudes are considered: $\{\alpha,\beta\}$, which represent the set of mean-field amplitudes, and $\mu$, the chemical potential, which may be used to control the density of the system. However, there is no way to determine which chemical potential to use for a given density. In previous applications this issue was resolved by sampling a grid of exact data from exact diagonalization on small systems in order to obtain a range of $\mu$ in which interpolation to high precision was possible~\cite{Bollmark2020}.
	
	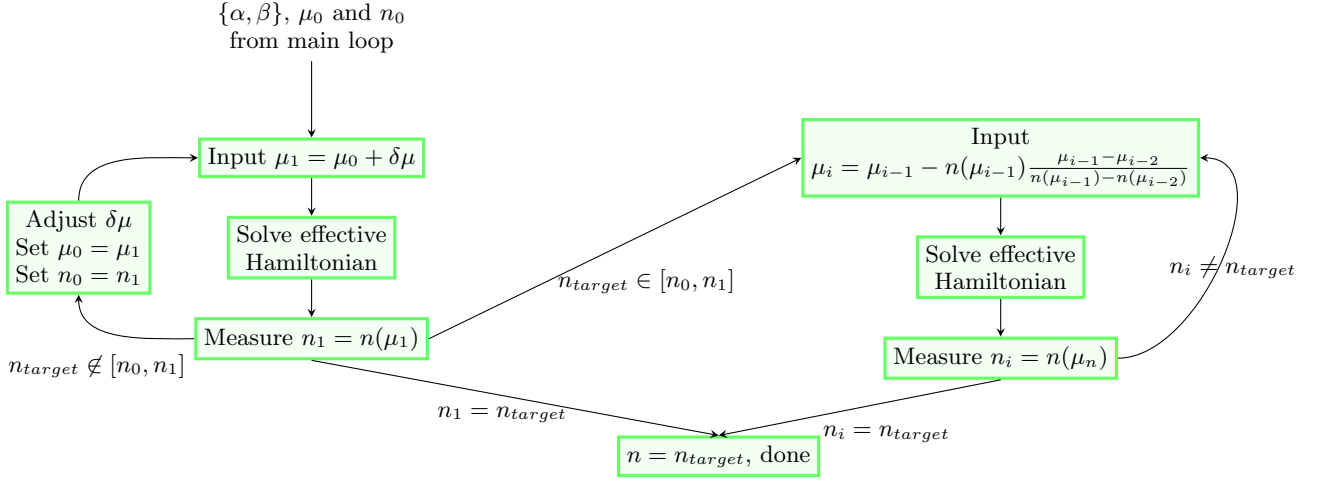
\begin{figure*}
    	\ifthenelse{\boolean{buildtikzpics}}%
    	{%
		\tikzsetnextfilename{DensityFixRoutineSchem}
		\tikzset{external/export next=true}
		\begin{tikzpicture}[recnode/.style={rectangle, draw=red!60, fill=red!5, very thick, minimum size=5mm}, chemnode/.style={rectangle, draw=green!60, fill=green!5, very thick, minimum size=5mm}]
			\node[above=of input, align=center](frommain){$\{\alpha,\beta\}$, $\mu_0$ and $n_0$ \\ from main loop};
			\node[chemnode, align=center] (input) {Input $\mu_1=\mu_0+\delta\mu$};
			\node[chemnode, align=center] (solve) [below=0.5cm of input] {Solve effective \\ Hamiltonian};
			\node[chemnode, align=center] (densmeas) [below=0.5cm of solve] {Measure $n_1=n(\mu_1)$};
			\node[chemnode, align=center] (increm) [left=of solve] {Adjust $\delta\mu$ \\ Set $\mu_0=\mu_1$ \\ Set $n_0=n_1$};

			\node[chemnode, align=center] (secant) [right=5cm of input] {Input \\ $\mu_i=\mu_{i-1}-n(\mu_{i-1})\frac{\mu_{i-1}-\mu_{i-2}}{n(\mu_{i-1})-n(\mu_{i-2})}$};
			\node[chemnode, align=center] (secsolve) [below=0.5cm of secant] {Solve effective \\ Hamiltonian};
			\node[chemnode, align=center] (secmeas) [below=0.5cm of secsolve] {Measure $n_i=n(\mu_n)$};
			\node[chemnode, align=center] (finish) [below right=1cm and 2.5cm of densmeas] {$n=n_{target}$, done};
			
			\node[below left=-0.2cm and 0cm of densmeas]{$n_{target}\not\in\lbrack n_0,n_1\rbrack$};
			\node[above right=0.2cm and 1.6cm of densmeas]{$n_{target}\in\lbrack n_0,n_1\rbrack$};
			\node[right=of secsolve]{$n_i\not=n_{target}$};
			\node[below right=0.5cm and 0cm of densmeas]{$n_1=n_{target}$};
			\node[below left=0.5cm and -1cm of secmeas]{$n_i=n_{target}$};

			\draw[->] (frommain.south) -- (input.north);
			\draw[->] (input.south) -- (solve.north);
			\draw[->] (solve.south) -- (densmeas.north);
			\draw[->] (densmeas.west) to[out=180, in=270] (increm.south);
			\draw[->] (increm.north) to[out=90, in=180] (input.west);
			\draw[->] (densmeas.east) -- (secant.west);
			\draw[->] (secant.south) -- (secsolve.north);
			\draw[->] (secsolve.south) -- (secmeas.north);
			\draw[->] (secmeas.east) to[out=0, in=0] (secant.east);
			\draw[->] (densmeas.south) -- (finish.north);
			\draw[->] (secmeas.south) -- (finish.north);
		\end{tikzpicture}
    	}%
    	{%
    		\includegraphics{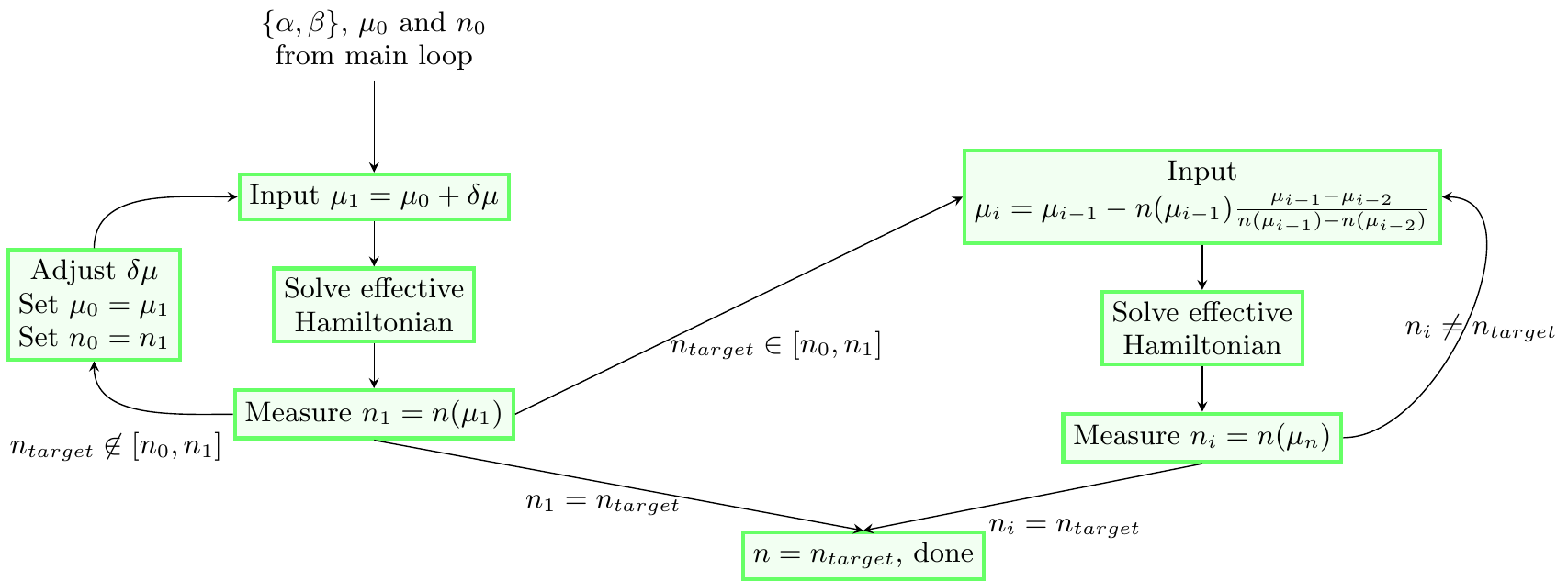}%
    	}%
		\caption{Schematic representation of the density fixing routine. Chemical potential at different stages of the loop are named $\mu_i$ and density $n_i$.}
		\label{fig:densityfix}
	\end{figure*}
	For models considered in this work the number of mean-field amplitudes is too great to attempt such a solution. Instead, a heuristic algorithm has been designed to obtain the appropriate density as shown in~\cref{fig:densityfix}. In essence, the algorithm attempts to find a chemical potential $\mu_{target}$ such that
	\begin{equation}
	n(\mu_{\mathrm{target}}) = n_{\mathrm{target}},
	\end{equation}
	where computing the density $n(\mu)$ for different $\mu$ necessitates a full DMRG solution leading to a significant time cost. To alleviate this issue we assume that $n(\mu)$ is a monotonic function of $\mu$ and look for a range in which $n(\mu)=n_{target}$. Once obtained we use the secant method to narrow the range until precision is achieved.
	
	It is notable that the density-fixing routine is most important when the mean-field amplitudes are strongly varying with each iteration. When approaching the self-consistently converged solution the density typically changes modestly, allowing for loops without running the density-fixing subroutine.

	\subsection{Self-consistent excited states}
In a superconducting system, much information can be obtained from the energy gap between the ground and the first excited state. This is especially true in Q1D systems, as is shown in \cref{Sec::Bosonization} analytically and confirmed numerically in \cref{Sec::Results/sub::hybrid}. Strikingly, finite-temperature properties such as superconducting $T_c$ can be obtained from this gap with minimal computational effort, which is readily exploited in \cref{Sec::HubbardLadder}.
Thus, we employ the ability of DMRG to compute the lowest-lying eigenstates orthogonal to the ground state. This allows the computation of the first excited state energy
	\begin{equation}\label{first_exc}
		E_{\mathrm{exc}} = \braket{\psi_1|H|\psi_1},~\braket{\psi_0|\psi_1}=0,
	\end{equation}
	where $|\psi_1\rangle$ is the first excited state which minimizes the energy $E_{\mathrm{exc}}$ with the constraint that it is orthogonal to $|\psi_0\rangle$, the ground state. The excitation gap may then be defined as
	\begin{equation}\label{first_exc_gap}
		\Delta = E_{exc}-\braket{\psi_0|H|\psi_0}.
	\end{equation}
	The excited state will not generally obey the self-consistency constraint, i.e., mean-field amplitudes measured in the excited state
	\begin{equation}
		\alpha_{\mathrm{measured}} \not=\alpha,
	\end{equation}
	where $\alpha$ is the self-consistent amplitude obtained for the ground state. This means our numerics indicate a depletion of the ground state of condensed pairs, in line with earlier analytical theory~\cite{Cazalilla2006a}.
	
\begin{figure}
        \ifthenelse{\boolean{buildtikzpics}}%
    	{%
		\tikzsetnextfilename{ExcDensityExtrap}
		\tikzset{external/export next=true}
		\begin{tikzpicture}
			\begin{axis}
			[
				xlabel = $1/L$,
				ylabel = $n-n_{\mathrm{target}}$,
			]
			\pgfplotstableread[header=true]{Data/exc_dens_fsize_fit_U_-2.0_tp_0.08_params.dat}{\datatable}
			\pgfplotstablegetelem{0}{[index]0}\of\datatable
			\pgfmathsetmacro{\a}{\pgfplotsretval}
			\pgfplotstablegetelem{0}{[index]1}\of\datatable
			\pgfmathsetmacro{\b}{\pgfplotsretval}
			\pgfplotstablegetelem{0}{[index]2}\of\datatable
			\pgfmathsetmacro{\c}{\pgfplotsretval}
			\addplot
			[
				only marks,
				thick,
				mark=o,
			]
			table
			[
				y expr = \thisrowno{1}-0.5,
				x expr = \thisrowno{0}
			]
			{Data/exc_dens_fsize_U_-2.0_tp_0.08.dat};
			\addplot
			[
				no marks,
				thick,
				domain = 0:0.05
			]
			{quadFct((\a),(\b),(\c-0.5))};
			\end{axis}
		\end{tikzpicture}
    	}%
    	{%
    		\includegraphics{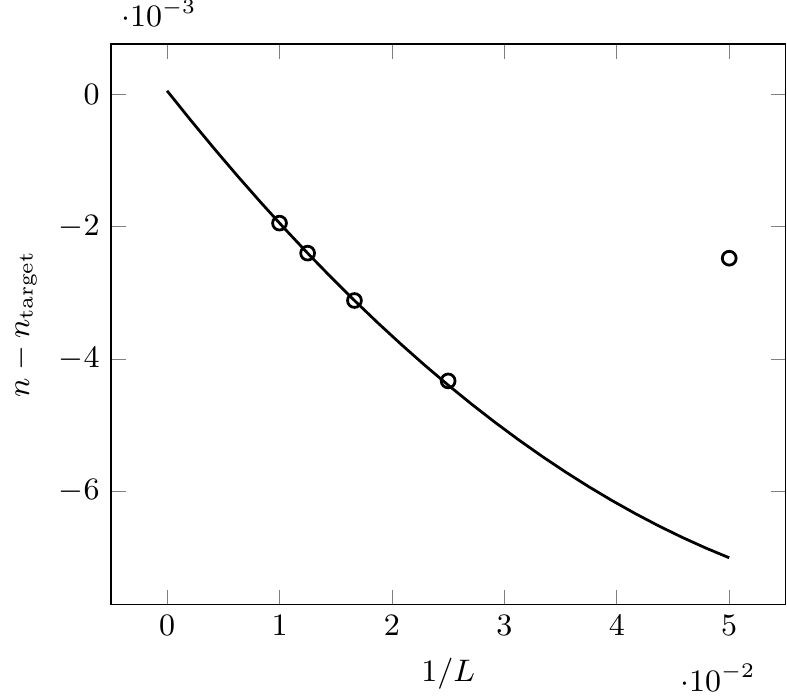}%
		}%
		\caption{Example of excited state density extrapolated to infinite system size for $U=-2t$, $t_\perp=0.08t$ and density $n_{\mathrm{target}}=0.5$.}
		\label{fig:exc_dens_extrap}
	\end{figure}
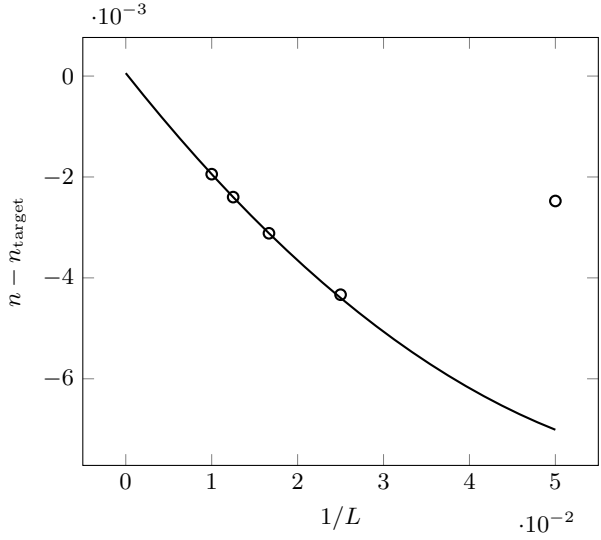

	Depletion of pairing can be a concern as the Hamiltonian~\cref{mf_ham} does not conserve density. Using the above procedure will yield whatever state of lowest energy that is orthogonal to the ground state and could well be at another density, as is exemplified in \cref{fig:exc_dens_extrap}.
	
	At the same time, the data also shows the deviation decreasing with system size. An extrapolation may thus be performed using a general form of finite size behavior
	\begin{equation}
	n(L) = n_\infty + \frac{c_0}{L} + \frac{c_1}{L^2} + \mathcal{O}\left(\frac{1}{L^3}\right).
	\end{equation}
	Thus, we assume that energy gaps extrapolated to infinite size systems are comparisons of energy at the same density despite the fact that finite size system densities generally differ. This strategy is verified in \cref{Sec::Results/sub::hybrid} and \cref{fig:Tc_from_ratio_comp}, as two methods to obtain superconducting $T_c$, one purely numeric, the other using $\Delta$ and the field theory of \cref{Sec::Bosonization}, are shown to coincide.
	
	The upshot is that once a self-consistent Hamiltonian has been produced from MPS+MF for ground states, excited states will cost no more than one additional DMRG-run, making the computation of $\Delta$ relatively cheap.
	
	
	\section{Field theory description}\label{Sec::Bosonization}
	
	Let us now turn to a field theory analysis of the Hamiltonian~\cref{3dHam}. In  1D systems the effects of interactions are dramatically amplified. Additionally, in 1D the quantum and thermal fluctuations prevent the breaking of continuous symmetries~\cite{Mermin_Wagner_symmetry_broken}. The combination of these effects leads to a unique universality class for interacting 1D quantum systems, known as Tomonaga-Luttinger liquids (TLL)~\cite{Haldane_harmonic_approximation1D}.
    
    The low-energy physics of TLLs can be described in terms of two bosonic fields $\phi$ and $\theta$ related to collective excitations of density and currents. These field are  
    related by the canonical relation
    \begin{equation}
         [\phi(x), \nabla \theta(x')] = i \pi \delta (x-x'),
    \end{equation}
    which expresses the duality in 1D between density and phase fluctuations. 
    In this bosonized representation, the single-particle fermionic operator of fermions with spin $\nu$ reads 
    \begin{equation}
        \psi_{\nu}(x) = e^{i k_F x}  \psi_{R,\nu}(x) + 
                    e^{-i k_F x}  \psi_{L,\nu}(x),
    \end{equation}
    where $\psi_{R,L,\nu}(x)$ are slowly varying field describing excitations close to the Fermi points $\pm k_F$ (right and left movers) and $\nu = \uparrow, \downarrow$ denotes the spin. 
    These fields are expressed as \cite{Giamarchi2003}
    \begin{equation}
        \psi_{r,\nu}(x) = \frac{U_{r, \nu}}{\sqrt{2 \pi \alpha}}e^{-\frac{i}{\sqrt{2}}(r\phi_\rho(x)-\theta_\rho(x)+\nu(r\phi_\sigma(x)-\theta_\sigma(x))},
        \label{single_particle_operator_field_theory}
    \end{equation}
    where $U_{r,\nu}$ are Klein factors, $r=R,L$ and $\alpha$ is a cut-off proportional to the lattice spacing, which simulates a finite bandwidth. The fields $\phi_{\rho,\sigma}$ are given by 
    \begin{equation}
        \begin{aligned}
            & \phi_\rho(x) = \frac{1}{\sqrt{2}}[\phi_\uparrow(x)+ \phi_\downarrow(x)],\\
            & \phi_\sigma(x) = \frac{1}{\sqrt{2}}[\phi_\uparrow(x)- \phi_\downarrow(x)],
        \end{aligned}
    \end{equation}
    and similarly for the field $\theta$.
    In the basis $\{ \theta_{\rho,\sigma}, \phi_{\rho,\sigma}\}$, the 1D Hubbard model has the peculiarity of decoupling charge and spin sectors:
    	\begin{equation}
	    \mathcal{H} = \mathcal{H}_\rho + \mathcal{H}_\sigma.
	\end{equation}
    Away from half-filling (one particle per site) and for negative interactions $U<0$, the spin sector is gapped while the charge sector is gapless. In the bosonic language, the gapless Hamiltonian is the universal TLL Hamiltonian
	\begin{equation}
	     \mathcal{H_\rho} = \frac{u_\rho}{2\pi} \int dx \, \left(K_\rho (\nabla \theta_\rho(x))^2 + \frac{1}{K_\rho} (\nabla \phi_\rho(x))^2 \right),
	     \label{quadratic_bosonization_H}
	\end{equation}
	and the spectrum is linear $\omega = u_\rho |k|$. Such a Hamiltonian is fully described by two non-universal parameters which depend on the microscopic model: $u_\rho$ the charge velocity of the collective excitation and $K_\rho$ the Luttinger parameter that controls the algebraic decay of the correlations~\cite{Giamarchi2003}.

	On the contrary, the gapped sector is described by a sine-Gordon Hamiltonian which with respect to \cref{quadratic_bosonization_H} has an additional term proportional to  $\cos{(2\sqrt{2}\phi_\sigma)}$ that wants to lock the field $\phi_\sigma$ to one of its minima:
	\begin{equation}
	     \begin{aligned}
	         \mathcal{H_\sigma} &= \frac{u_\sigma}{2\pi} \int dx \, \left(K_\sigma (\nabla \theta_\sigma(x))^2 + \frac{1}{K_\sigma} (\nabla \phi_\sigma(x))^2\right) \\
	         & \qquad \qquad + \frac{2 U}{(2 \pi \alpha)^2}\int dx \, \cos{(2 \sqrt{2}\phi_\sigma(x))}.
	     \label{gapped_bosonization_H}
	     \end{aligned}
	\end{equation}
	Physically, it means that the system tends to form Cooper pairs with opposite spins and as a direct consequence suppresses spin excitations. The energy of the bound state is the gap $\Delta_\sigma$ in the spin sector. Notably, $\Delta_\sigma$ is kept distinct in notation from $\Delta E_s$ in \cref{spingap} as their definitions differ. The field theory spin gap can be computed from the sine-Gordon model by various methods~\cite{Giamarchi2003} and is exactly known for the microscopic attractive Hubbard model with $U<0$ by Bethe ansatz~\cite{Lieb_Wu_Bethe_ansatz}.
	
    As described in the previous sections we assume here that the spin gap $\Delta_\sigma$ is larger than the interchain coupling $t_\perp$. Our system has thus a low and high-energy sector. The former relative to hopping in the transverse direction ($t_\perp \ll t$) and the latter to break the pairs. 
    
    Let us first consider the case when $\Delta_\sigma \gg t$ (or $|U| \gg t$). In this case it is necessary to eliminate the spin sector before bosonizing the Hamiltonian. This can be achieved by a Schrieffer–Wolff transformation~\cite{Schrieffer_Wolff_transformation}. The resulting Hamiltonian is rewritten as
    \begin{equation}\label{analytical_wolff_sch_trasformation}
        \begin{aligned}
            \mathcal{H} &\sim  \mathcal{H}_0 + t_{\perp} \mathcal{H}_{\perp}\\ 
            &= \mathcal{H}_0  -  \frac{t_\perp^2}{\Delta_\sigma} \sum_{ \langle\textbf{R}, \textbf{R'} \rangle} \sum_{n, n', s} \\
            &  \qquad  \Big [ c_{n,   \textbf{R'} ,s}^{\dagger} c_{n,  \textbf{R},s}^{ } c_{n',  \textbf{R'}, -s}^{\dagger} c_{n',  \textbf{R}, -s}^{ } +  \text{h.c.} \Big]
        \end{aligned}
    \end{equation}
    where $\mathcal{H}_0$ is the 1D quadratic Hamiltonian  \cref{quadratic_bosonization_H}.
    The effective coupling is now proportional to the gained energy over the cost of breaking the pair $t_\perp^2/\Delta_\sigma$ and indeed it expresses local pairs hopping in the transverse direction. The spin excitations are exponentially suppressed, while the charge sector is massless. In \cref{analytical_wolff_sch_trasformation}, we neglect the terms corresponding to the formation of charge-density-wave (CDW), because the superconductive correlation (SS) decays slower. Indeed, the corresponding correlations are~\cite{Giamarchi2003}
    \begin{equation}
         \begin{aligned}
            \langle O_{\text{SS}}^{\dagger}(r)O_{\text{SS}}(0)\rangle &\sim \left ( \frac{1}{r} \right )^{\frac{1}{K_\rho}} \\
            \langle O_{\text{CDW}}^{\dagger}(r)O_{\text{CDW}}(0)\rangle &\sim \left ( \frac{1}{r} \right )^{K_\rho}
         \end{aligned}
    \end{equation}
    where the corresponding operators are defined as $O_{\text{CDW}}=\psi^{\dagger}_{R, \uparrow}\psi^{ }_{L, \uparrow} + \psi^{\dagger}_{R, \downarrow}\psi^{ }_{L, \downarrow}$ and $O_{\text{SS}}=\psi^{\dagger}_{R, \uparrow}\psi^{\dagger}_{L, \downarrow} + \psi^{\dagger}_{L, \uparrow}\psi^{\dagger}_{R, \downarrow}$. For negative $U$, the TLL charge parameter is larger than one $K_\rho >1$ , meaning that CDW formation is a sub-dominant instability. 
    By using \cref{analytical_wolff_sch_trasformation} and considering the leading terms $n=n'$, 
    since the spin gap is larger than the bandwidth (approximation of local pairs) 
    we now use the mean-field approximation whose order parameter reads
    \begin{equation} \label{order_parameter_analytical}
    \langle \psi ^\dagger_{\textbf{R}, \uparrow} (x) \psi ^\dagger_{\textbf{R}, \downarrow} (x) \rangle = C e^{-i\sqrt{2}\theta_\rho(x)}
    \end{equation}
    with $C$ a constant that depends on the spin gap $\Delta_\sigma$ but is of order one in this regime (see Appendix~\ref{App::Renormalization_group}). In this limit, since  only the charge sector survives, the problem maps onto a system of hard-core bosons (Cooper pairs) with a transverse hopping described by the field $\sqrt{2}\theta(x)$ instead of $\theta(x)$. Indeed, the physics is similar: the Pauli principle forbids to have two pairs in the same site and two hard-core bosons never occupy the same site.
    
    All chains are now identical and, in the bosonized version, the 1D effective Hamiltonian reads
    \begin{equation} \label{bosonized_mean_field_hamiltonian}
        \mathcal{H}= \mathcal{H}_{\rho} - \rho_0 ^2 C^2 \frac{ t_{\perp}^2}{\Delta_\sigma} z_c \langle  e^{ -i\sqrt{2} \theta_{\rho}} \rangle \int_0^L dx \cos{(\sqrt{2} \theta_{\rho}(x))}
    \end{equation}
	where $z_c$ is the number of nearest neighbours in the transverse direction, $L$ is the system size and $\rho_0$ is the unperturbed density. We find a sine-Gordon-like Hamiltonian and therefore, at $T<T_c$, $t_\perp$ opens a gap in the spectrum because the cosine wants to lock the field $\theta_\rho$ to one of the minima. In the thermodynamic limit, the zero temperature gap $\Delta_\rho$ is known analytically~\cite{Lukyanov_gap_sine_gordon}. This gap in the charge sector should equal the gap to the first excited state in \cref{first_exc_gap} yet notation is kept distinct due to the differing definitions of the gaps.
    
    The dimensional crossover~\cite{Bollmark2020} that occurs in such systems is represented by the mean-field critical temperature $T_c$ above which $\langle \psi ^\dagger_{\uparrow} (x) \psi ^\dagger_{\downarrow} (x) \rangle = 0$ and the system is made of incoherent and decoupled 1D chains \cite{Cazalilla2006a}.
    This means that for $T>T_c$, the thermal fluctuations wash out the transverse coherence due to the presence of $t_\perp$. The system behaves essentially as if it was an isolated chain. Notably, the critical temperature scales like the charge gap at zero temperature, see appendix \ref{App:critical_temperature_and_charge_gap_analitycal} for more details. Even though the prefactors in both $T_c$ and the charge gap are partially unknown, because of the constant $C$, the ratio is completely controlled by the Luttinger parameter $K_\rho$ only which can be computed from numerical calculations (DMRG, Bethe Ansatz, \dots)
    \begin{widetext}
	\begin{equation} \label{analytic_fermionic_ratio}
    R(K_\rho) = \frac{\Delta_\rho(T=0)}{T_c}= 2 \pi \left [ \frac{K_\rho \tan{\left ( \frac{\pi}{2} \frac{1}{4K_\rho-1}\right )} }{ 2 \kappa^2(K_\rho/2)(4K_\rho - 1) \sin \left ( \frac{\pi}{2K_\rho} \right )  B^2 \left (\frac{1}{4K_\rho}, 1 - \frac{1}{2K_\rho} \right ) }  \right ]^{\frac{K_\rho}{2K_\rho-1}} \sin \left ( \frac{\pi}{4K_\rho-1}\right ) 
    \end{equation}
	\end{widetext}
    with $B(x,y)$ the Beta function and $\kappa (K)$ a combination of gamma functions $\Gamma(K)$, defined in \cref{combination_gamma_functions_ratio}.
    
    A more challenging case arises when $|U| \ll t$. Indeed in that case since $\Delta_\sigma \ll t$ the pairs are non local and the full bosonized form of the Hamiltonian with both charge and spin sectors (\cref{quadratic_bosonization_H} and \cref{gapped_bosonization_H}) must be used. To deal with this situation we use a renormalization group (RG) procedure (see Appendix \ref{App::Renormalization_group}), in which we eliminate all degrees of freedom from the initial bandwidth of the system, down to the spin gap. At that scale, since the running ultra-violet cutoff is now identical to the spin gap we are back to the situation where ``local'' pairs (hard-core bosons) hop in the transverse direction. The single particle hopping thus disappears at that scale and simply leads to an effective Josephson coupling between the various 1D units. This coupling can be computed from the RG as detailed in Appendix~\ref{App::Renormalization_group}. The limit $\Delta_\sigma \ll W$, where $W \sim 2t$ is the bandwidth of the 1D system, corresponds naturally to the case of weak interactions $|U| \ll W$. In this regime, the spin gap either obtained from Bethe Ansatz for the Hubbard model, or more generally from the RG~\cite{Larkin_small_U_gap_from_RG}, is naturally exponentially small in the interactions $\Delta_\sigma \sim \sqrt{|U|} e^{-1/|U|}$. Thus, in order to be in the relevant regime for the whole procedure to work a very weak interchain hopping $t_\perp$ is needed. Specifically, the RG flow needs to be cut by the spin gap and it may not be cut by the interchain hopping. If this condition is satisfied we recover a model analogous to the one of the strong coupling albeit with a different Josephson parameter. As discussed above the ratio is insensitive to the precise value of the coupling and thus the ratio \cref{analytic_fermionic_ratio} is expected to hold for all values of $U$.

	\section{\label{Sec::Results}Results}
	The primary focus of this section is to make use of and test the developed MPS+MF framework on the study case of negative-$U$ Hubbard chain arrays. The methods described in \cref{Sec::Model} are tested thoroughly to produce a robust routine for determining critical temperature for the onset of superconductivity in Q1D systems of fermions with a gapped spin sector. These results are subsequently leveraged in \cref{Sec::HubbardLadder} to obtain $T_c$ for USC in weakly coupled doped Hubbard ladders with repulsive interactions. The negative-$U$ Hubbard chain array results are split into 4 sub-sections which are ordered as follows: (\textit{A}) The Hamiltonian \cref{mf_ham} depends on a range parameter of the mean-fields which is studied in this sub-section, (\textit{B}) a numerical study of the ground state superconducting energy gap and critical temperature from thermal states, as well as (\textit{C}) comparing those numerical results with a more efficient mixture of analytics and numerics, and finally (\textit{D}) benchmarking MPS+MF against AFQMC in a 2D system where the latter approach yields quasi-exact results.
	
	As there are enough parameters to consider already, in the following results for chain arrays we are targeting a fixed density 
	\begin{equation}
		n = \frac{1}{L}\sum_{i}^L\sum_\sigma\braket{c^\dagger_{i\sigma}c_{i\sigma}} = 0.5,
	\end{equation}
	i.e., a quarter-filled system. We expect other close-lying densities will not yield markedly different results due to the nature of the isotropic case having weak dependence on density around this filling in the 2D case~\cite{Paiva2004}.
	
	\subsection{Hamiltonian range dependence\label{Sec::Results/sub::range_dep}}
\begin{figure*}
        \ifthenelse{\boolean{buildtikzpics}}%
	    {%
		\tikzsetnextfilename{RangeDepGroupFigure}
		\tikzset{external/export next=true}
		\begin{tikzpicture}
			\begin{groupplot}
			[
				group style = {group size=3 by 2}
			]
			\nextgroupplot
			[
				width=0.3\textwidth,
				height=0.2\textheight,
				title = {$U=-2t$},
				ylabel = {$\langle cc\rangle_0$},
				legend pos = south east,
				legend style={nodes={scale=0.7, transform shape}},
				xticklabel style = {/pgf/number format/precision=3,
					/pgf/number format/fixed},
			]
			\addplot
			[
				red,
				thick,
				only marks,
				mark=o,
				error bars/.cd,
				y dir = both, y explicit
			]
			table
			[
				y expr = \thisrowno{1},
				y error expr = \thisrowno{2},
				x expr = \thisrowno{0}
			]
			{Data/range_comparison_U_-2.0_range_1_orp_fig_1_1.dat};
			\addlegendentry{Range 1}
			\addplot
			[
				green,
				thick,
				only marks,
				mark=o,
				error bars/.cd,
				y dir = both, y explicit
			]
			table
			[
			y expr = \thisrowno{1},
			y error expr = \thisrowno{2},
			x expr = \thisrowno{0}
			]
			{Data/range_comparison_U_-2.0_range_4_orp_fig_1_1.dat};
			\addlegendentry{Range 4}
			\addplot
			[
				blue,
				thick,
				only marks,
				mark=o,
				error bars/.cd,
				y dir = both, y explicit
			]
			table
			[
			y expr = \thisrowno{1},
			y error expr = \thisrowno{2},
			x expr = \thisrowno{0}
			]
			{Data/range_comparison_U_-2.0_range_6_orp_fig_1_1.dat};
			\addlegendentry{Range 6}
			\addplot
			[
				orange,
				thick,
				only marks,
				mark=o,
				error bars/.cd,
				y dir = both, y explicit
			]
			table
			[
			y expr = \thisrowno{1},
			y error expr = \thisrowno{2},
			x expr = \thisrowno{0}
			]
			{Data/range_comparison_U_-2.0_range_8_orp_fig_1_1.dat};
			\addlegendentry{Range 8}
		\nextgroupplot
		[
			width=0.3\textwidth,
			height=0.2\textheight,
			title = {$U=-4t$},
			legend pos = south east,
			legend style={nodes={scale=0.7, transform shape}},
			xticklabel style = {/pgf/number format/precision=3,
				/pgf/number format/fixed},
		]
		\addplot
		[
			red,
			thick,
			only marks,
			mark=o,
			error bars/.cd,
			y dir = both, y explicit
		]
		table
		[
			y expr = \thisrowno{1},
			y error expr = \thisrowno{2},
			x expr = \thisrowno{0}
		]
		{Data/range_comparison_U_-4.0_range_1_orp_fig_2_1.dat};
		\addlegendentry{Range 1}
		\addplot
			[
			green,
			thick,
			only marks,
			mark=o,
			error bars/.cd,
			y dir = both, y explicit
			]
			table
			[
			y expr = \thisrowno{1},
			y error expr = \thisrowno{2},
			x expr = \thisrowno{0}
			]
			{Data/range_comparison_U_-4.0_range_4_orp_fig_2_1.dat};
			\addlegendentry{Range 4}
			\addplot
			[
			blue,
			thick,
			only marks,
			mark=o,
			error bars/.cd,
			y dir = both, y explicit
			]
			table
			[
			y expr = \thisrowno{1},
			y error expr = \thisrowno{2},
			x expr = \thisrowno{0}
			]
			{Data/range_comparison_U_-4.0_range_6_orp_fig_2_1.dat};
			\addlegendentry{Range 6}
		\nextgroupplot
		[
		width=0.3\textwidth,
		height=0.2\textheight,
		title = {$U=-10t$},
		legend pos = south east,
		legend style = {nodes={scale=0.7, transform shape}},
		]
		\addplot
		[
		red,
		thick,
		only marks,
		mark=o,
		error bars/.cd,
		y dir = both, y explicit
		]
		table
		[
		y expr = \thisrowno{1},
		y error expr = \thisrowno{2},
		x expr = \thisrowno{0}
		]
		{Data/range_comparison_U_-10.0_range_1_orp_fig_3_1.dat};
		\addlegendentry{Range 1}
		\addplot
		[
		green,
		thick,
		only marks,
		mark=o,
		error bars/.cd,
		y dir = both, y explicit
		]
		table
		[
		y expr = \thisrowno{1},
		y error expr = \thisrowno{2},
		x expr = \thisrowno{0}
		]
		{Data/range_comparison_U_-10.0_range_4_orp_fig_3_1.dat};
		\addlegendentry{Range 4}
		\nextgroupplot
		[
			width=0.3\textwidth,
			height=0.2\textheight,
			ylabel = {$\Delta/t$},
			xlabel = {$t_\perp/t$},
			legend pos = north west,
			legend style={nodes={scale=0.7, transform shape}},
			xticklabel style = {/pgf/number format/precision=3,
				/pgf/number format/fixed},
		]
		\addplot
		[
			red,
			thick,
			only marks,
			mark=o,
			error bars/.cd,
			y dir = both, y explicit
		]
		table
		[
			y expr = \thisrowno{1},
			y error expr = \thisrowno{2},
			x expr = \thisrowno{0}
		]
		{Data/range_comparison_U_-2.0_range_1_gap_fig_1_2.dat};
		\addlegendentry{Range 1}
		\addplot
		[
			green,
			thick,
			only marks,
			mark=o,
			error bars/.cd,
			y dir = both, y explicit
		]
		table
		[
			y expr = \thisrowno{1},
			y error expr = \thisrowno{2},
			x expr = \thisrowno{0}
		]
		{Data/range_comparison_U_-2.0_range_4_gap_fig_1_2.dat};
		\addlegendentry{Range 4}
		\addplot
		[
			blue,
			thick,
			only marks,
			mark=o,
			error bars/.cd,
			y dir = both, y explicit
		]
		table
		[
			y expr = \thisrowno{1},
			y error expr = \thisrowno{2},
			x expr = \thisrowno{0}
		]
		{Data/range_comparison_U_-2.0_range_6_gap_fig_1_2.dat};
		\addlegendentry{Range 6}
		\addplot
		[
			orange,
			thick,
			only marks,
			mark=o,
			error bars/.cd,
			y dir = both, y explicit
		]
		table
		[
			y expr = \thisrowno{1},
			y error expr = \thisrowno{2},
			x expr = \thisrowno{0}
		]
		{Data/range_comparison_U_-2.0_range_8_gap_fig_1_2.dat};
		\addlegendentry{Range 8}
		\nextgroupplot
		[
		width=0.3\textwidth,
		height=0.2\textheight,
		xlabel = {$t_\perp/t$},
		legend pos = north west,
		legend style = {nodes={scale=0.7, transform shape}},
		xticklabel style = {/pgf/number format/precision=3,
			/pgf/number format/fixed},
		]
		\addplot
		[
		red,
		thick,
		only marks,
		mark=o,
		error bars/.cd,
		y dir = both, y explicit
		]
		table
		[
		y expr = \thisrowno{1},
		y error expr = \thisrowno{2},
		x expr = \thisrowno{0}
		]
		{Data/range_comparison_U_-4.0_range_1_gap_fig_2_2.dat};
		\addlegendentry{Range 1}
		\addplot
		[
		green,
		thick,
		only marks,
		mark=o,
		error bars/.cd,
		y dir = both, y explicit
		]
		table
		[
		y expr = \thisrowno{1},
		y error expr = \thisrowno{2},
		x expr = \thisrowno{0}
		]
		{Data/range_comparison_U_-4.0_range_4_gap_fig_2_2.dat};
		\addlegendentry{Range 4}
		\addplot
		[
		blue,
		thick,
		only marks,
		mark=o,
		error bars/.cd,
		y dir = both, y explicit
		]
		table
		[
		y expr = \thisrowno{1},
		y error expr = \thisrowno{2},
		x expr = \thisrowno{0}
		]
		{Data/range_comparison_U_-4.0_range_6_gap_fig_2_2.dat};
		\addlegendentry{Range 6}
		\nextgroupplot
		[
		width=0.3\textwidth,
		height=0.2\textheight,
		xlabel = {$t_\perp/t$},
		legend pos = north west,
		legend style = {nodes={scale=0.7, transform shape}},
		]
		\addplot
		[
		red,
		thick,
		only marks,
		mark=o,
		error bars/.cd,
		y dir = both, y explicit
		]
		table
		[
		y expr = \thisrowno{1},
		y error expr = \thisrowno{2},
		x expr = \thisrowno{0}
		]
		{Data/range_comparison_U_-10.0_range_1_gap_fig_3_2.dat};
		\addlegendentry{Range 1}
		\addplot
		[
		green,
		thick,
		only marks,
		mark=o,
		error bars/.cd,
		y dir = both, y explicit
		]
		table
		[
		y expr = \thisrowno{1},
		y error expr = \thisrowno{2},
		x expr = \thisrowno{0}
		]
		{Data/range_comparison_U_-10.0_range_4_gap_fig_3_2.dat};
		\addlegendentry{Range 4}
		\end{groupplot}
		\end{tikzpicture}
    	}%
    	{%
    		\includegraphics{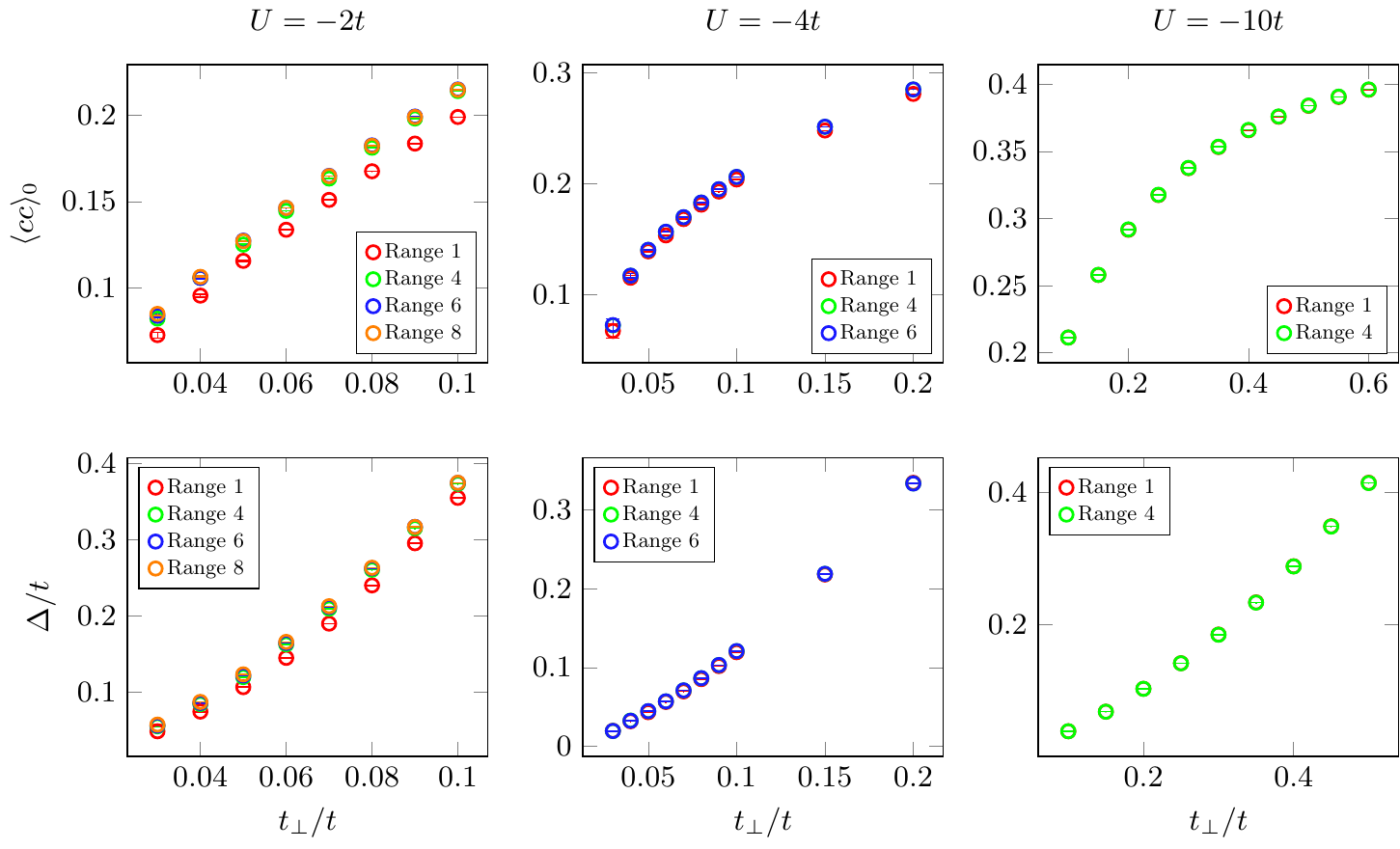}%
    	}%
		\caption{Range dependence of the order parameter defined in \cref{orderpar} and first excitation energy defined in \cref{first_exc_gap} for different values of interaction $U/t=-2,-4,-10$ and density $n=0.5$. Notably the difference between the two largest ranges is not visible which indicates sufficient range in the Hamiltonian.}
		\label{fig:range_depend}
	\end{figure*}
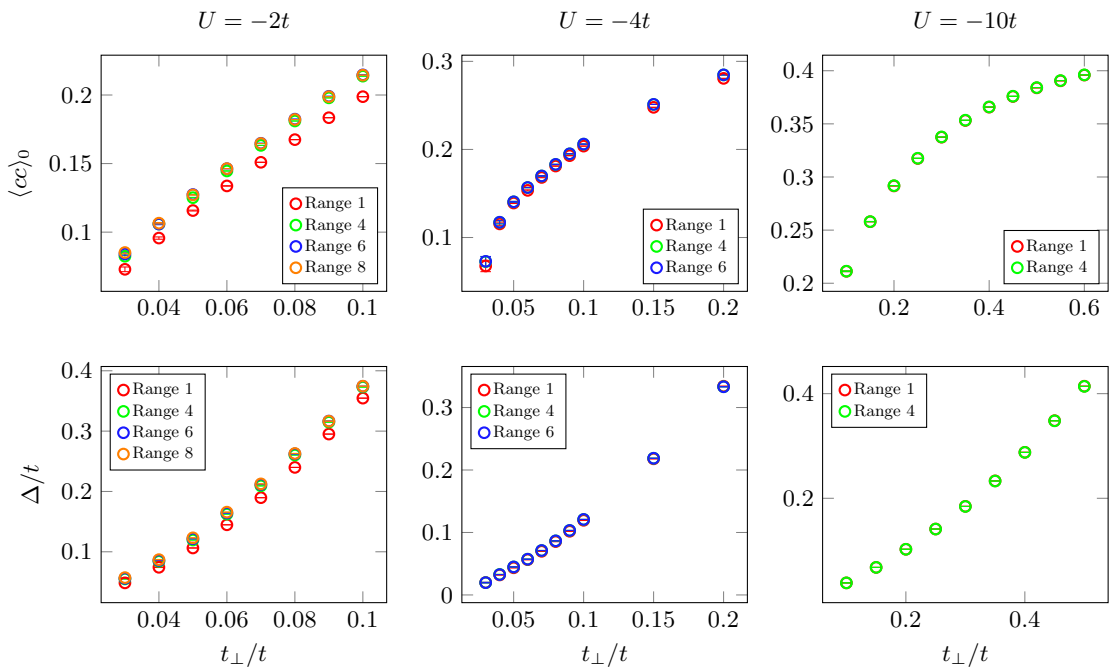

	Considering the effective 1D Hamiltonian of \cref{mf_ham} the first question to be answered is how long-range the mean-field terms can be. As discussed at the end of \cref{Sec::Model}, the MF-amplitudes will always decay with an exponential envelope as a function of distance between the operators appearing in them. Since longer range terms are more difficult to simulate, we want to find a minimal range for each parameter set with which longer-ranged Hamiltonians agree. As a metric for determining agreement we use the order parameter and energy gap $\Delta$: The former defined by
	\begin{equation}\label{orderpar}
	\braket{cc}_r=\frac{1}{i_{l}-i_{f}}\sum_{i=i_{f}}^{i_{l}}\braket{c_{i\uparrow}c_{i+r\downarrow}},
	\end{equation}
	with $r=0$ and where $i_f$ and $i_l$ chosen to avoid boundary effects of the open boundary conditions typical in DMRG. As can be seen from \cref{fig:range_depend} the minimal range required varies with the strength of interaction. This is natural since weaker attraction makes electron pairs more dispersive and thus less localized.
	
	We note that longer-ranged terms can still be finite and ignoring them should yield at the very least a difference in ground state energies. However, the primary question is rather the degree at which these terms affect the ground state wave function and the critical temperature. In \cref{fig:range_depend} we show that the order parameter and excited state gap $\Delta$ do not change appreciably beyond the ranges that are displayed. While we find  $\alpha$-terms at longer ranges to be finite, $\Delta$ and the onsite order parameter are largely unaffected. 
	
	\subsection{\label{Sec::Results/sub::numerical}Numerical results}
\begin{figure*}
		\centering
        \ifthenelse{\boolean{buildtikzpics}}%
	    {%
		\tikzsetnextfilename{DMRGAnalyticCompTc}
		\tikzset{external/export next=true}
		\begin{tikzpicture}
			\pgfplotstableread[header=false]{Data/Pairing_energy_U_-2.0_n_0.5.dat}{\epu2tab}
			\pgfplotstablegetelem{0}{[index]0}\of\epu2tab
			\edef\EpUTwo{\pgfplotsretval}
			\pgfplotstableread[header=false]{Data/Pairing_energy_U_-4.0_n_0.5.dat}{\epu4tab}
			\pgfplotstablegetelem{0}{[index]0}\of\epu4tab
			\edef\EpUFour{\pgfplotsretval}
			\pgfplotstableread[header=false]{Data/Pairing_energy_U_-10.0_n_0.5.dat}{\epu10tab}
			\pgfplotstablegetelem{0}{[index]0}\of\epu10tab
			\edef\EpUTen{\pgfplotsretval}
            
			\begin{axis}
			[
			    width	=	0.485\textwidth,
			    height	=	0.35\textheight,
				ylabel={$k_BT_c/t$},
				xlabel={$t_\perp^2/\Delta E_p$},
				legend pos = north west,
				legend columns = 3,
				xmode=log,
				title = (a),
				every axis title/.style={above right, at={(0,1)}},
    			legend cell align={left},
    			transpose legend,
			]
			\addlegendimage{color=colorA, thick};
			\addlegendentry{$U=-2t$};
			\addlegendimage{color=colorB, thick};
			\addlegendentry{$U=-4t$};
			\addlegendimage{color=colorC, thick};
			\addlegendentry{$U=-10t$};
			
			\addlegendimage{color=black, mark=x, only marks, error bars/.cd};
			\addlegendentry{$T_c=\Delta/R(K_\rho)$};
			\addlegendimage{color=black, mark=o, only marks, error bars/.cd};
			\addlegendentry{$T_c$: direct calculation};
			\addlegendimage{color=black, dashed};
			\addlegendentry{$R(K_\rho)$};
			\addplot
			[
				mark=x,
				only marks,
				color=colorA,
				forget plot,
			]
				table
				[
					x expr = \thisrow{t_perp}*\thisrow{t_perp}/\EpUTwo,
					y expr = \thisrow{DeltaE}/\thisrow{Ratio},
				]
					{Data/fits_U_-2.0.fix.single.dat};
			
						\addplot
			[
				mark=o,
				only marks,
				color=colorA,
				forget plot,
			]
				table
				[
					x expr	=	\thisrow{t_perp}*\thisrow{t_perp}/\EpUTwo,
					y expr	=	\thisrow{T_c_numeric},
				]
					{Data/fits_U_-2.0.fix.single.dat}; 

			\addplot
			[
				mark=o,
				only marks,
				color = colorA,
				forget plot,
				error bars/.cd, 
				y dir=minus,
				y explicit,
			]
				table
				[
					x expr	=	\thisrow{t_perp}*\thisrow{t_perp}/\EpUTwo,
					y expr	=	\thisrow{T_c_numeric},
					y error expr = abs(\thisrow{T_c_numeric}-\thisrow{Tcnum_lower_limit}), 
				]
					{Data/fits_U_-2.0.fix.single.dat}; 
			\addplot
			[
				mark=o,
				only marks,
				color = colorA,
				forget plot,
				error bars/.cd, 
				y dir=plus,
				y explicit,
			]
				table
				[
					x expr	=	\thisrow{t_perp}*\thisrow{t_perp}/\EpUTwo,
					y expr	=	\thisrow{T_c_numeric},
					y error expr = abs(\thisrow{T_c_numeric}-\thisrow{Tcnum_upper_limit}), 
				]
					{Data/fits_U_-2.0.fix.single.dat}; 

			\addplot
			[
				mark=x,
				only marks,
				color=colorB,
				forget plot,
			]
				table
				[
					x expr = \thisrow{t_perp}*\thisrow{t_perp}/\EpUFour,
					y expr = \thisrow{DeltaE}/\thisrow{Ratio},
				]
					{Data/fits_U_-4.0.fix.single.dat};
			
			\addplot
			[
				mark=o,
				only marks,
				color = colorB,
				forget plot,
			]
				table
				[
					x expr	=	\thisrow{t_perp}*\thisrow{t_perp}/\EpUFour,
					y expr	=	\thisrow{T_c_numeric},
				]
					{Data/fits_U_-4.0.fix.single.dat}; 
					
			\addplot
			[
				mark=o,
				color = colorB,
				forget plot,
				only marks,
				error bars/.cd, 
				y dir=minus,
				y explicit,
			]
				table
				[
					x expr	=	\thisrow{t_perp}*\thisrow{t_perp}/\EpUFour,
					y expr	=	\thisrow{T_c_numeric},
					y error expr = abs(\thisrow{T_c_numeric}-\thisrow{Tcnum_lower_limit}), 
				]
					{Data/fits_U_-4.0.fix.single.dat}; 
			\addplot
			[
				mark=o,
				color = colorB,
				forget plot,
				only marks,
				error bars/.cd, 
				y dir=plus,
				y explicit,
			]
				table
				[
					x expr	=	\thisrow{t_perp}*\thisrow{t_perp}/\EpUFour,
					y expr	=	\thisrow{T_c_numeric},
					y error expr = abs(\thisrow{T_c_numeric}-\thisrow{Tcnum_upper_limit}),
				]
				{Data/fits_U_-4.0.fix.single.dat}; 
			\addplot+
			[
				color = colorC,
				thick,
				mark=x,
				forget plot,
				only marks,
				error bars/.cd,
				y dir = both, y explicit
			]
			table
			[
				y expr = \thisrowno{1},
				y error expr = \thisrowno{2},
				x expr = \thisrowno{0}*\thisrowno{0}/\EpUTen
			]
			{Data/analytic_and_numeric_Tc_comparison_gap.dat};
			\addplot+
			[
			    color = colorC,
				mark=o,
				only marks,
				forget plot,
				thick,
				error bars/.cd,
				y dir = both, y explicit
			]
			table
			[
				y expr = \thisrowno{1},
				y error expr = \thisrowno{2},
				x expr = \thisrowno{0}*\thisrowno{0}/\EpUTen
			]
			{Data/analytic_and_numeric_Tc_comparison_Tc.dat};
			\end{axis}
		\end{tikzpicture}
    	}%
    	{%
    		\includegraphics{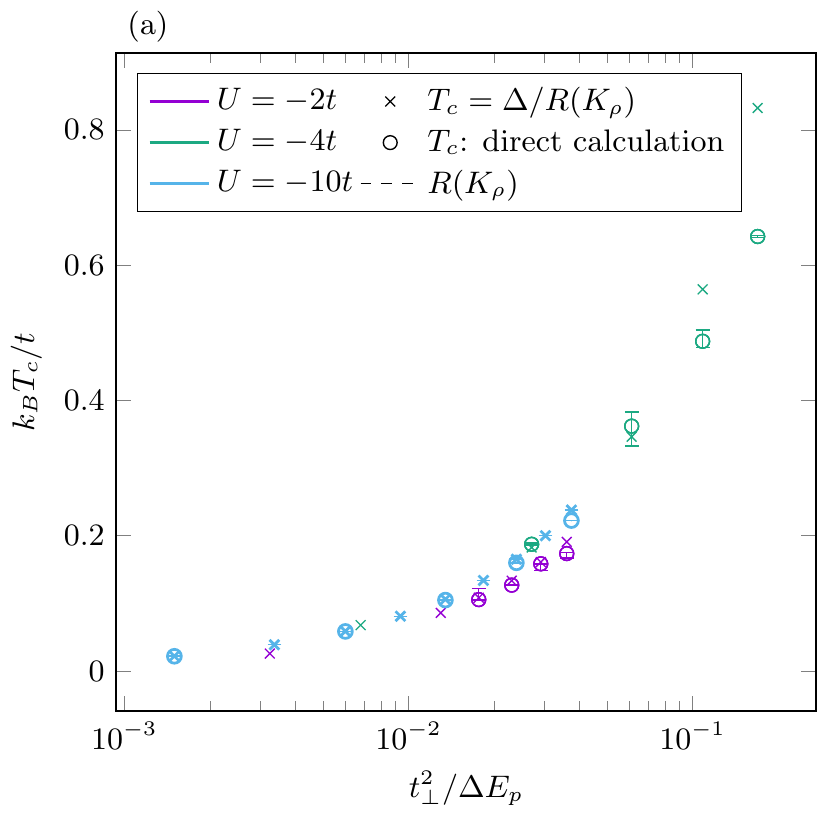}%
    	}%
    	\ifthenelse{\boolean{buildtikzpics}}%
	    {%
		\tikzsetnextfilename{DMRGAnalyticCompRatio}
		\tikzset{external/export next=true}
		\begin{tikzpicture}
    		\begin{axis}
    		[
    				width = 0.485\textwidth,
    				height = 0.35\textheight,
    				xlabel = {$t_\perp^2/\Delta E_p$},
    				ylabel = {R},
    				legend pos = north west,
    				xmode=log,
    				title = (b),
    				every axis title/.style={above right, at={(0,1)}},
    		]
    		\pgfplotstableread[header=false]{Data/Pairing_energy_U_-2.0_n_0.5.dat}{\epu2tab}
    		\pgfplotstablegetelem{0}{[index]0}\of\epu2tab
    		\edef\EpUTwo{\pgfplotsretval}
    		\pgfplotstableread[header=false]{Data/Pairing_energy_U_-4.0_n_0.5.dat}{\epu4tab}
    		\pgfplotstablegetelem{0}{[index]0}\of\epu4tab
    		\edef\EpUFour{\pgfplotsretval}
    		\pgfplotstableread[header=false]{Data/Pairing_energy_U_-10.0_n_0.5.dat}{\epu10tab}
    		\pgfplotstablegetelem{0}{[index]0}\of\epu10tab
    		\edef\EpUTen{\pgfplotsretval}
    		\pgfplotstableread[header=false]{Data/analytic_and_numeric_ratio_comparison_analytic_ratio.dat}{\rattable}
    		\pgfplotstablegetelem{0}{[index]0}\of\rattable
    		\edef\arat{\pgfplotsretval}
    		\pgfplotstableread[header=true]{Data/analytic_and_numeric_ratio_comparison.dat}{\dmrgtable}
    		\pgfplotstablegetrowsof{\dmrgtable}
    		\pgfmathsetmacro{\Utenlastrow}{\pgfplotsretval-1}
    		\pgfplotstablegetelem{\Utenlastrow}{[index]0}\of\dmrgtable
    		\edef\Utentperpmax{\pgfplotsretval}
    		
    		\def\EpUTwo{0.27860280706331136}
    		\addplot
    		[
    			colorA,
    			only marks,
    			forget plot,
    			mark=o,
    			thick,
    		]
    			table
    			[
    				y expr = \thisrow{DeltaE}/\thisrow{T_c_numeric},
    				x expr = \thisrow{t_perp}*\thisrow{t_perp}/\EpUTwo,
    			]
    				{Data/fits_U_-2.0.fix.single.dat};
    		\addplot
    		[
    			colorA,
    			only marks,
    			mark=o,
    			thick,
    			forget plot,
    			error bars/.cd, 
    			y dir=plus,
    			y explicit,
    		]
    			table
    			[
    				y expr = \thisrow{DeltaE}/\thisrow{T_c_numeric},
    				x expr = \thisrow{t_perp}*\thisrow{t_perp}/\EpUTwo,
    				y error expr = abs(\thisrow{DeltaE}/\thisrow{T_c_numeric}-\thisrow{DeltaE}/\thisrow{Tcnum_upper_limit}),
    			]
    				{Data/fits_U_-2.0.fix.single.dat};
    		\addplot
    		[
    			colorA,
    			only marks,
    			mark=o,
    			thick,
    			forget plot,
    			error bars/.cd, 
    			y dir=minus,
    			y explicit,
    		]
    			table
    			[
    				y expr = \thisrow{DeltaE}/\thisrow{T_c_numeric},
    				x expr = \thisrow{t_perp}*\thisrow{t_perp}/\EpUTwo,
    				y error expr = abs(\thisrow{DeltaE}/\thisrow{T_c_numeric}-\thisrow{DeltaE}/\thisrow{Tcnum_lower_limit}),
    			]
    				{Data/fits_U_-2.0.fix.single.dat};
    		\addplot
    		[
    			colorA,
    			dashed,
    			no marks,
    			forget plot,
    			domain=8e-4:0.2,
    			thick
    		]
    			{1.94941855652991};
    		
    		\def\EpUFour{1.4737628688241569}
    		\addplot
    		[
    			colorB,
    			only marks,
    			mark=o,
    			forget plot,
    			thick,
    		]
    			table
    			[
    				y expr = \thisrow{DeltaE}/\thisrow{T_c_numeric},
    				x expr = \thisrow{t_perp}*\thisrow{t_perp}/\EpUFour,
    			]
    				{Data/fits_U_-4.0.fix.single.dat};
    		\addplot
    		[
    			colorB,
    			only marks,
    			mark=o,
    			thick,
    			forget plot,
    			error bars/.cd, 
    			y dir=plus,
    			y explicit,
    		]
    			table
    			[
    				y expr = \thisrow{DeltaE}/\thisrow{T_c_numeric},
    				x expr = \thisrow{t_perp}*\thisrow{t_perp}/\EpUFour,
    				y error expr = abs(\thisrow{DeltaE}/\thisrow{T_c_numeric}-\thisrow{DeltaE}/\thisrow{Tcnum_upper_limit}),
    			]
    				{Data/fits_U_-4.0.fix.single.dat};	
    		\addplot
    		[
    			colorB,
    			only marks,
    			mark=o,
    			thick,
    			forget plot,
    			error bars/.cd, 
    			y dir=minus,
    			y explicit,
    		]
    			table
    			[
    				y expr = \thisrow{DeltaE}/\thisrow{T_c_numeric},
    				x expr = \thisrow{t_perp}*\thisrow{t_perp}/\EpUFour,
    				y error expr = abs(\thisrow{DeltaE}/\thisrow{T_c_numeric}-\thisrow{DeltaE}/\thisrow{Tcnum_lower_limit}),
    			]
    				{Data/fits_U_-4.0.fix.single.dat};		
    		
    		\addplot
    		[
    			colorB,
    			dashed,
    			forget plot,
    			no marks,
    			domain=8e-4:0.2,
    			thick
    		]
    			{1.82191753705144};
    		\addplot
    		[
        		blue,
        		only marks,
        		mark=o,
        		thick,
        		forget plot,
        		error bars/.cd,
        		y dir = both, 
        		y explicit,
    		]
        		table
        		[
            		y expr = \thisrowno{1},
            		x expr = \thisrowno{0}*\thisrowno{0}/\EpUTen,
            		y error expr = \thisrowno{2}
        		]
        		{\dmrgtable};
    		\addplot
    		[
        		blue,
        		dashed,
    			forget plot,
        		no marks,
                domain=8e-4:0.2,
        		thick,
    		]
    		    {linearFct(0,\arat)};
    		\end{axis}
		\end{tikzpicture}
    	}%
    	{%
    		\includegraphics{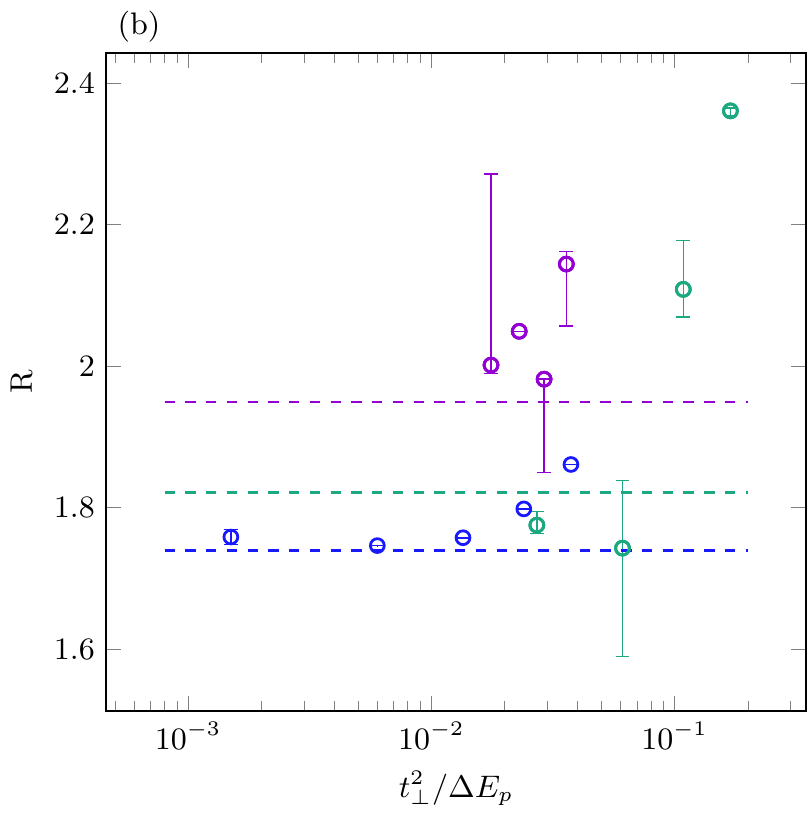}%
    	}%
		\caption{A comparison of analytical methods and combination with numerical methods for several values of interaction. Sub-figure (a) shows critical temperature $T_c$ vs transverse tunneling $t_\perp^2/\Delta E_p$. Circles show data obtained via direct calculation of thermal states, while crosses show $T_c$ computed from zero-temperature calculations, using the excitation gap $\Delta$ and \cref{analytic_fermionic_ratio}. Sub-figure (b) shows the ratio $R=\Delta/T_c$ of ground state excitation gap to critical temperature, both computed separately, compared with the analytical ratio \cref{analytic_fermionic_ratio} (dashed lines) vs $t_\perp^2/\Delta E_p$. }
		\label{fig:Tc_from_ratio_comp}
	\end{figure*}
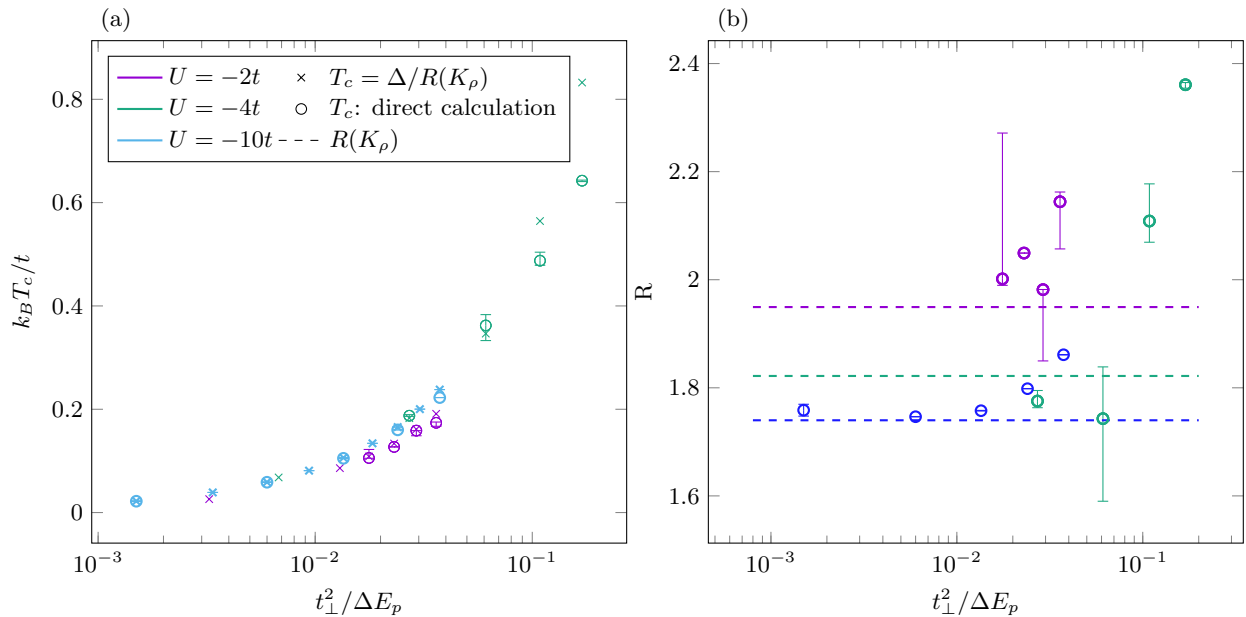

	Using the minimal ranges we may compute ground and finite temperature states for the Hamiltonian \cref{mf_ham} using the MPS+MF framework. As shown in \cref{fig:Tc_from_ratio_comp}a we find the critical temperature decreasing with transverse tunneling $t_\perp$, vanishing as expected for $t_\perp\to0$.
	
	The excited state gap $\Delta$ disappears in the same manner. Notably, the zero-temperature gap in field theory $\Delta_\rho$ should have the same meaning as the excited state gap $\Delta$, which has been verified. Thus, from field theory it is expected that both $T_c$ and $\Delta$ scale with $t_\perp$ with the same exponent, as can be seen from \cref{scaling_eq_Tc} and \cref{scaling_eq_charge_gap}.
	
 	In determining $T_c$ numerically via state-purification within the MPS-approach, we have to contend with the increase in inverse temperature $\beta=1/T$ required as $t_\perp$ is decreased. This results not just in longer imaginary-time evolutions, but also in increased finite size effects (see \cref{App::Finite_size_extrapolation}). 
 	
 	Reaching large system sizes for these thermal state calculations can be challenging. We used two different approaches. In the first, used for $U=-10t$, the infinite-temperature purified starting state of the imaginary time evolution is constructed to be in the $S=0$ subspace, which allows to keep exploiting the conserved spin quantum number during the evolution. The drawback is that this $\beta=0$ initial state has entanglement growing strongly with system size, limiting the practically attainable system lengths. 
 	
 	The second approach, employed for $U=-2t,-4t$, is to sacrifice the spin-conservation. This makes the purified initial state into a trivial-to-construct product state, and thus arbitrary system lengths are accessible. However, this makes the imaginary time evolution more costly, as there are no conserved quantum numbers anymore. In order to alleviate this issue we use the PP-DMRG framework~\cite{PP} expounded upon in~\cref{App::ppDMRG_and_purification}. We moreover rely on the ability of $H_\rm{MF}$ to filter out the $S=0$ subspace at temperatures at and below $T_c$. We ascertain this to be correct, with violations of $\langle S_z \rangle = 0$ reaching at most $10^{-2}$, and typically much less, across all calculations.
	
	
	\subsection{\label{Sec::Results/sub::hybrid}Numerical-analytical hybrid results}
	For the field theory of \cref{Sec::Bosonization} it is difficult to quantify the superconducting $T_c$ or the excited state gap $\Delta$, due to the unknown pre-factors arising from the massive spin-sector. However, forming the ratio $\Delta/T_c=R(K_\rho)$ is free from these unknown pre-factors, depending just on the TLL parameter $K_\rho$ and is, strikingly, constant in $t_\perp$.
    
    \begin{table}[h]
    \begin{tabular}{ l @{\qquad} c @{\qquad} c }
    \toprule
    \textrm{$U/t$}&
    \textrm{$K_\rho$}&
    \textrm{$\Delta_\sigma / t$}\\
    \colrule
    -2.0 & 1.26  & 0.289 \\
    -4.0 & 1.41 & 1.476 \\
    -10.0 & 1.51 & 6.671 \\
    \botrule
    \end{tabular}
    \caption{\label{parameters_Bethe_ansatz} Results from Bethe Ansatz solution of 1D Hubbard model with attractive interaction at quarter-filling. Here, we compute the TLL parameter of charge sector $K_\rho$ and the field theory spin gap $\Delta_\sigma$ as a function of interactions.}
    \label{tab:1D_hub_parms}
    \end{table}
	
	Generating $R$ from the numerically determined data of \cref{fig:Tc_from_ratio_comp}a and the analytical expression \cref{analytic_fermionic_ratio} yields \cref{fig:Tc_from_ratio_comp}b. Notably, while the analytical ratio is not agreeing with numerical estimates exactly, the constant nature of the analytics is likely approximative. Achieving a ratio which lies close enough to the data is sufficient in order to obtain critical temperatures. With this knowledge the new $T_c$ estimate becomes
	\begin{equation}\label{ratio_estimate}
		T_c = \frac{\Delta(T=0)}{R(K_\rho)}.
	\end{equation}
	Since the primary issues of $T_c$ computation came from the imaginary time evolution we may now obtain $T_c$ estimates from ground-state DMRG by computing $\Delta(T=0)$. As shown in \cref{fig:Tc_from_ratio_comp}a the estimation scheme \cref{ratio_estimate} is agreeing very well with the numerical $T_c$ values obtained from thermal state calculations.
	
	Since the more efficient ground-state DMRG is the main tool of this alternative way to obtain $T_c$, it can be brought to even lower values of $t_\perp$ and greater system sizes, allowing for greater precision at larger parameter ranges. 
	
	\subsection{\label{Sec::Results/sub::afqmc}Comparison with AFQMC}
	In \cref{mf_ham} the coordination number $z_c$ tracks the dimensionality of the underlying Q1D array. For the calculations on 3D Q1D systems, as performed so far, we have $z_c=4$. Lattices of other dimension can be simulated just as well, just by changing $z_c$ to the appropriate value.
	
	We exploit this for benchmarking the MPS+MF approach against quasi-exact results, which AFQMC is able to obtain in the absence of a sign problem. These benchmarks are necessarily done for 2D models, as the finite temperature algorithm used scales cubically in the number of lattice sites~\cite{Assaad2021}. We stress that the MPS+MF method describes ordering in the Ginzburg-Landau sense, via the pairing mean-fields $\alpha_{ik}$, while AFQMC detects the actual 2D BKT transition. We interpret the mean-field $T_c$ of MPS+MF as an approximation of the $T_{BKT}$ which occurs for the 2D model.
	
\begin{figure}
	    \centering
	    \ifthenelse{\boolean{buildtikzpics}}%
	    {%
		\tikzsetnextfilename{DMRGAFQMCComp}
		\tikzset{external/export next=true}
		\begin{tikzpicture}
			\begin{groupplot}
			[
				group style = 
				{
				    group size=1 by 2,
    				vertical sep		=	4.5em,
    				horizontal sep		=	4mm,
    			},
				width=0.41\textwidth,
			    legend cell align={left},
			    xlabel=$t_\perp/t$,
			]
    			\nextgroupplot
    			[
    				ylabel=$\braket{cc}_0$,
    				legend pos = north west,
    				legend style = {nodes={scale=0.7, transform shape}},
    				xticklabel style = {/pgf/number format/precision=3,
    					/pgf/number format/fixed},
    				clip mode=individual
    			]
    			\node[anchor=south west] at (rel axis cs:0,1) {(a)};
    			\addplot
    			[
    				mark=o,
    				thick,
    				black
    			]
    			table
    			[
    				y expr = \thisrowno{1},
    				x expr = sqrt(2)*\thisrowno{0}
    			]
    			{Data/afqmc_gs_comp_orp.dat};
    			\addlegendentry{MPS+MF}
    			\addplot+
    			[
    				mark=o,
    				thick,
    				red,
    				error bars/.cd,
    				y dir = both, y explicit
    			]
    			table
    			[
    				y expr = \thisrowno{1},
    				x expr = sqrt(2)*\thisrowno{0},
    				y error expr = \thisrowno{2}
    			]
    			{Data/afqmc_gs_orp_20x4.dat};
    			\addlegendentry{AFQMC 20x4}
    			\addplot+
    			[
    				mark=o,
    				thick,
    				blue,
    				error bars/.cd,
    				y dir = both, y explicit
    			]
    			table
    			[
    				y expr = \thisrowno{1},
    				x expr = sqrt(2)*\thisrowno{0},
    				y error expr = \thisrowno{2}
    			]
    			{Data/afqmc_gs_orp_30x4.dat};
    			\addlegendentry{AFQMC 30x4}
    			\addplot
    			[
    				mark=o,
    				thick,
    				green,
    				error bars/.cd,
    				y dir = both, y explicit
    			]
    			table
    			[
    				y expr = \thisrowno{1},
    				x expr = sqrt(2)*\thisrowno{0},
    				y error expr = \thisrowno{2}
    			]
    			{Data/afqmc_gs_orp_30x8.dat};
    			\addlegendentry{AFQMC 30x8}
    			\addplot
    			[
    				mark=o,
    				thick,
    				orange,
    				error bars/.cd,
    				y dir = both, y explicit
    			]
    			table
    			[
    				y expr = \thisrowno{1},
    				x expr = sqrt(2)*\thisrowno{0},
    				y error expr = \thisrowno{2}
    			]
    			{Data/afqmc_gs_orp_40x4.dat};
    			\addlegendentry{AFQMC 40x4}
    			\nextgroupplot
    			[
    				xlabel = {$t_\perp/t$},
    				ylabel = {$T_{BKT}$},
    				legend pos = north west,
    				legend style = {nodes={scale=0.7, transform shape, inner xsep = 5pt}},
    				legend columns = 3,
    				ytick style={draw=none},
    				clip mode=individual
    			]
    			\node[anchor=south west] at (rel axis cs:0,1) {(b)};
    			\addplot+
    			[
    				thick,
    				mark=o,
    				blue,
    				error bars/.cd,
    				y dir = both, y explicit
    			]
    			table
    			[
    				y expr = \thisrowno{1},
    				x expr = \thisrowno{0},
    				y error expr = \thisrowno{2}
    			]
    			{Data/afqmc_mpsmf_comp_Tc_vs_tp_U_-4.0_n_0.5_ax1_mpsmf.dat};
    			\addlegendentry{MPS+MF}
    			\addplot+
    			[
    				thick,
    				mark=o,
    				red,
    				error bars/.cd,
    				y dir = both, y explicit
    			]
    			table
    			[
    				y expr = \thisrowno{1},
    				x expr = \thisrowno{0},
    				y error expr = \thisrowno{2}
    			]
    			{Data/afqmc_mpsmf_comp_Tc_vs_tp_U_-4.0_n_0.5_ax1_afqmc.dat};
    			\addlegendentry{AFQMC}
    			\addlegendimage{color=black, only marks, mark=x, thick};
    			\addlegendentry{Ratio}
			\end{groupplot}
			\begin{groupplot}
			[
				group style = 
				{
				    group size=1 by 2,
    				vertical sep		=	4.5em,
    				horizontal sep		=	4mm,
    			},
				width=0.41\textwidth,
			    legend cell align={left},
			]
    			\nextgroupplot
    			[
    				axis x line=none,
    				axis y line=none
    			]
    			\nextgroupplot
    			[
    				axis y line*=right,
    				axis x line=none,
    				ylabel = {Ratio},
    				ymin = 0,
    				ymax = 0.3,
    				legend pos = south east,
    				legend style = {nodes={scale=0.7, transform shape}},
    				ytick style={draw=none}
    			]
    			\addplot+
    			[
    				thick,
    				mark=x,
    				only marks,
    				black,
    				forget plot,
    				error bars/.cd,
    				y dir = both, y explicit
    			]
    			table
    			[
    				y expr = \thisrowno{1},
    				x expr = \thisrowno{0},
    				y error expr = \thisrowno{2}
    			]
    			{Data/afqmc_mpsmf_comp_Tc_vs_tp_U_-4.0_n_0.5_ax2_ratio.dat};
			\end{groupplot}
		\end{tikzpicture}
    	}%
    	{%
    		\includegraphics{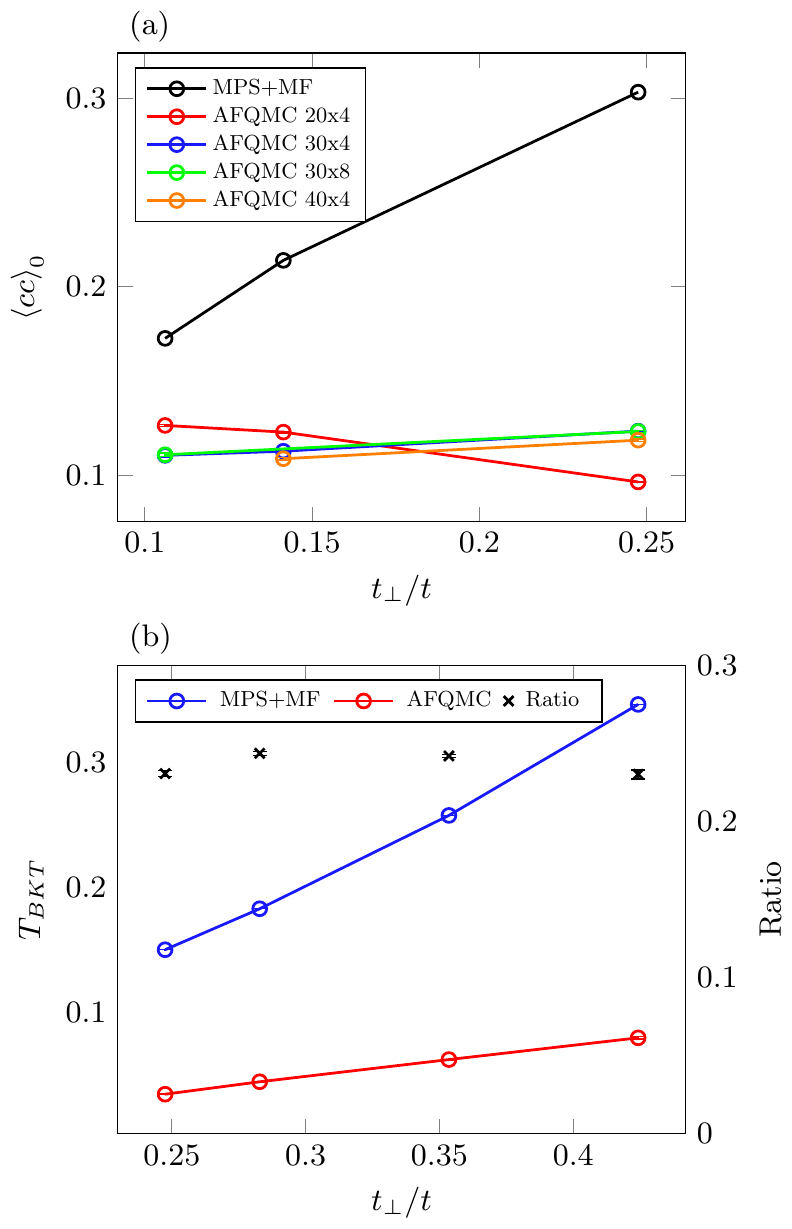}%
    	}%
		\caption{Comparison of AFQMC and MPS+MF for (a) ground state order parameter at $U=-2t$, $n=0.5$ and (b) $T_c$ from MPS+MF and $T_{BKT}$ from AFQMC at $U=-4t$ and $n=0.5$. Black crosses denote the ratio $T_{BKT}/T_c$.}
		\label{fig:afqmc_mpsmf_comp}
	\end{figure}
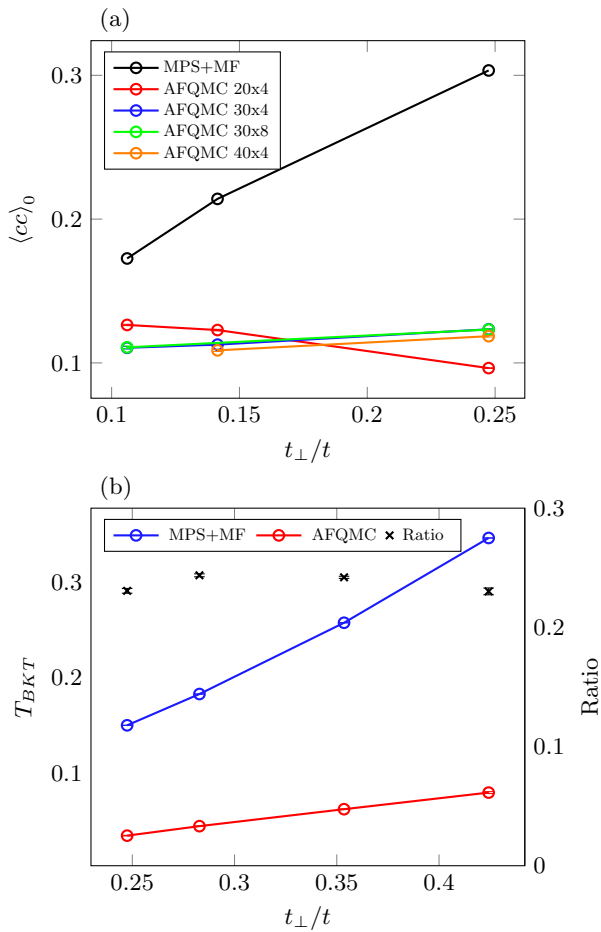

	We compare the two algorithms both for ground state and finite temperature calculations. For ground states, which do achieve superconducting LRO even in 2D, the onsite order parameter is compared between the two algorithms as shown in \cref{fig:afqmc_mpsmf_comp}a. Since 2D is the lower critical dimension, quantum fluctuations around any mean will be especially strong. Considering this fact we note that the smallest simulated $t_\perp$ values have a modest overestimation as MPS+MF neglects transverse quantum fluctuations by design.
	
	For finite temperature states we compute the BKT transition termperature, $T_{BKT}$, using AFQMC and compare it to $T_c$ from MPS+MF. The strategy for obtaining the $T_{BKT}$ temperature from AFQMC is standard and elaborated in~\cref{App::AFQMC_procedure}. Both $T_{BKT}$ and $T_c$, as well as their ratio, are shown in~\cref{fig:afqmc_mpsmf_comp}b.
	Over the range of $t_\perp$ for which we simulate, the near-constant ratio between $T_c$ and $T_{BKT}$ is striking, being approximately
	\begin{equation}
		\frac{T_{BKT}}{T_c}\approx 0.25.
	\end{equation}
	This is in line with previous work where we also found such a ratio to be robust to changes in parameters~\cite{Bollmark2020}. Similarly to the comparison to zero-temperature AFQMC, MPS+MF will overestimate the transition due to the neglect of both quantum and thermal fluctuations, and these will again be especially pronounced, given that these are 2D systems. Previous work on bosonic systems, as well as the generally known dependence of phase transitions on spacial dimensionality of a system, indicates that these will be strongly reduced for a Q1D 3D system~\cite{Bollmark2020}. Specifically, there we found that $T_c^{\rm{QMC}}/T_c^{\rm{MPS+MF}}\approx 0.7$. For 3D fermionic Q1D systems we thus expect that a correction factor for $T_c$ computed from the MPS+MF framework to obtain the true $T_c$ will lie somewhere between these two extremes, and probably closer to $0.7$ than to $0.25$.
	
	\section{3D array of weakly doped repulsive-U Hubbard ladder} \label{Sec::HubbardLadder}
	With the MPS+MF framework for fermions developed on 3D Q1D systems of negative-$U$ Hubbard chains this section applies it to a much more demanding system: 3D arrays of weakly coupled, doped, repulsive-$U$ Hubbard-ladders. These systems have been investigated via field theory as an alternative to Q2D systems in the study of USC and high-$T_c$ superconductivity. The microscopic mechanism of repulsively mediated pairing is understood from field theory, and the strength of this pairing can be quantified reliably via MPS-based methods. Despite these critical advantages over the Q2D models, there was no quantitative method to study these 3D arrays that incorporates the pairing physics at the microscopic level, as the TLL field theory in practice yields largely qualitative results.

	The MPS+MF framework supplies that ability. Just like the negative-$U$ Hubbard chains, the isolated, doped repulsive-$U$ Hubbard ladders have finite $\Delta E_s$ and $\Delta E_p$, manifesting the repulsively mediated pairing. Further, when analyzed via renormalization group theory within the TLL approach, the low-energy physics of both these 1D sub-units is structurally analogous. Both exhibit an ungapped charge sector, characterized by a TLL coefficient $K_\rho$ for the chain, and $K_{\rho,+}$ for the ladder. Furthermore, both can be computed from the microscopic Hamiltonians via MPS-methods. For the Hubbard ladders, there are three additional gapped modes, a charge one and two spin ones, the smaller having minimal energy $\Delta E_s$, as for the single spin mode in the negative-$U$ chains.
	
	The MPS+MF framework thus can be applied to the Hubbard-ladder arrays. For a proof-of-principle treatment, we will focus on plain Hubbard ladders depicted in \cref{fig:geometry:ladders}, with $U=8t$ and average density fixed at $n=0.9375$. The characteristic energy scales and TLL parameters can be computed with DMRG which results in ${\Delta E_s\approx0.078t}$, ${\Delta E_p\approx0.134t}$ and ${K_{\rho,+}=0.77}$ (the latter extracted from~\cite{Dolfi2015b}). The study of optimized Hubbard-ladder arrays, engineered for high $T_c$'s via deliberate optimization of $\Delta E_s$, $\Delta E_p$ and $K_\rho$, as well as examining the possibility of charge-density order competing with USC within two-channel MPS+MF, is the subject of forthcoming future work.
	
	As ladder geometries require much larger MPS resources than chains, direct calculation of $T_c$, while feasible, is challenging. We thus use MPS+MF combined with the analytics developed and tested in \cref{Sec::Model}-\cref{Sec::Results}. In this manner we obtain $T_c$ for USC in these arrays by computing excited state energy gaps $\Delta$ using DMRG for ground states, then applying \cref{ratio_estimate}. Compared to \cref{Sec::Bosonization} and \cref{Sec::Results}, this procedure requires only marginal adjustments for the ladder as negative-$U$ chain and Hubbard ladder look largely identical in the low-energy parts of their respective field theories. For the former, the order-parameter scales as $e^{-i\sqrt{2} \theta_\rho(x)}$ in the phase-operator of the ungapped charge mode, while for the latter it is $e^{-i\theta_{\rho,+}(x)}$. From that, it follows that the ratio-function for the ladder is
	\begin{equation}\label{analytic_ratio_ladder}
	    R_\rm{ladder}(K_{\rho,+}) = R(2K_{\rho,+}),
	\end{equation}
	where $R$ is given by \cref{analytic_fermionic_ratio}. Thus, $R_\rm{ladder}$ will retain its dependence on a single TLL parameter: $K_{\rho,+}$. This is due to the gapped spin-sectors entering the order-parameter in the same manner both for $T_c$ and $\Delta$. These non-universal contributions thus cancel when forming $R$, analogous to the derivation in \cref{Sec::Bosonization} and \cref{App:critical_temperature_and_charge_gap_analitycal}. The same analysis yields that for the ladder we have 
	\begin{equation}\label{ladder_Tc_scaling}
	    T_c \propto t_\perp^{\frac{4K_{\rho,+}}{4K_{\rho,+}-1}}.
	\end{equation}
	The effective MF-Hamiltonian for the Q1D array of the ladders is given by
	\begin{flalign}
		\nonumber H_{\mathrm{HL}} =& -t\sum_{i=1}^{L-1}\sum_{j=0}^1\sum_\sigma\Bigl(c^\dagger_{i+1,j\sigma} c_{ij\sigma} + c^\dagger_{ij\sigma} c_{i+1,j\sigma}\Bigr) \\\nonumber& - t\sum_{i=1}^L\sum_\sigma \left(c^\dagger_{i1\sigma} c_{i0\sigma} + c^\dagger_{i0\sigma} c_{i1\sigma}\right) \\\nonumber&-\mu \sum_{ij}n_{ij} +U\sum_{ij} n_{ij\uparrow} n_{ij\downarrow} \\&- H_{pair,MF}-H_{exc,MF},
	\end{flalign}
	\begin{sloppypar}
	where $i,j$ are leg and rung indices respectively and $H_{Pair}$ and $H_{PH}$ are derived in \cref{App::Ladder_SMF_model} and defined by
	\end{sloppypar}
	\begin{align}
	H_{\mathrm{pair,MF}} = \sum_{ii',jj'}\alpha_{ii',jj'}\left(c^\dagger_{ij\downarrow} c^\dagger_{i'j'\uparrow} + c_{i'j'\uparrow} c_{ij\downarrow} \right), \\
	H_{exc,MF} = -\sum_{ii',jj',\sigma} \beta_{ii',jj',\sigma}c^\dagger_{ij\sigma}c_{i'j'\sigma},
	\end{align}
	and the pairing amplitudes are given by
	\begin{gather}
	\alpha_{ii',00} = \frac{2t_\perp^2}{\Delta E_p}\left(\braket{c_{i'1\uparrow}c_{i1\downarrow}} + 2\braket{c_{i'0\uparrow}c_{i0\downarrow}}\right) \\
	\alpha_{ii',11} = \frac{2t_\perp^2}{\Delta E_p}\left(\braket{c_{i'0\uparrow}c_{i0\downarrow}} + 2\braket{c_{i'1\uparrow}c_{i1\downarrow}}\right) \\
	\alpha_{ii',10} = \frac{4t_\perp^2}{\Delta E_p}\braket{c_{i'0\uparrow}c_{i1\downarrow}} \\
	\alpha_{ii',01} = \frac{4t_\perp^2}{\Delta E_p}\braket{c_{i'1\uparrow}c_{i0\downarrow}},
	\end{gather}
	whereas the exchange terms are given by
	\begin{gather}
	\beta_{ii',00,\sigma} = \frac{2t_\perp^2}{\Delta E_p}\left(\braket{c^\dagger_{i1\sigma}c_{i'1\sigma}} + 2\braket{c^\dagger_{i0\sigma}c_{i'0\sigma}}\right), \\
	\beta_{ii',11,\sigma} = \frac{2t_\perp^2}{\Delta E_p}\left(\braket{c^\dagger_{i0\sigma}c_{i'0\sigma}} + 2\braket{c^\dagger_{i1\sigma}c_{i'1\sigma}}\right), \\
	\beta_{ii',10,\sigma} = \frac{4t_\perp^2}{\Delta E_p}\braket{c^\dagger_{i'0\sigma}c_{i1\sigma}}, \\
	\beta_{ii',01,\sigma} =\frac{4t_\perp^2}{\Delta E_p}\braket{c^\dagger_{i'1\sigma}c_{i0\sigma}}.
	\end{gather}
	In this work we exclude the possibility for a CDW phase such that density is independent of which rung we measure and is constant throughout the system. This is obtained by the restriction $\beta_{i_1i_1,j_1j_1,\sigma}=\beta_{i_2i_2,j_1j_1,\sigma}$. We note that inclusion of the exchange terms $\beta_{ii',ll',\sigma}$ has previously not been possible by analytical methods.
	
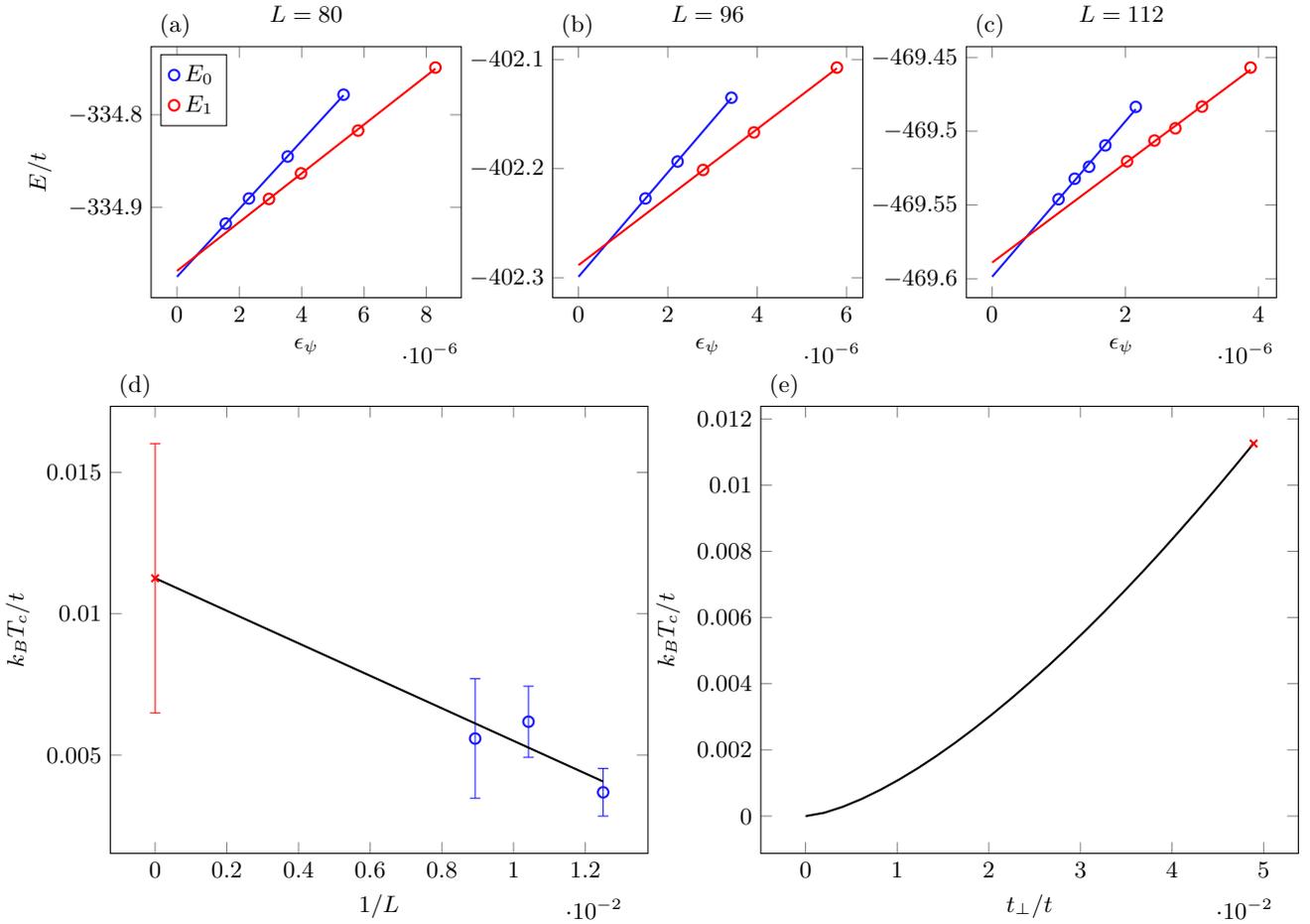
\begin{figure*}
        \ifthenelse{\boolean{buildtikzpics}}%
	    {%
	    \tikzsetnextfilename{HubLadderTruncFitL_80}
		\tikzset{external/export next=true}
		\begin{tikzpicture}
			\begin{axis}
			[
				xlabel = $\epsilon_\psi$,
				ylabel = $E/t$,
				width = 0.32\textwidth,
				clip mode=individual,
				title = {$L=80$},
				legend pos = north west
			]
			\node[anchor=south west] at (rel axis cs:0,1) {(a)};
			\pgfplotstableread[header=false]{Data/hub_ladder_gs_ener_fit_params_L_80.dat}{\fitgs}
			\pgfplotstablegetelem{0}{[index]0}\of\fitgs
			\edef\kgs{\pgfplotsretval}
			\pgfplotstablegetelem{0}{[index]1}\of\fitgs
			\edef\mgs{\pgfplotsretval}
			\pgfplotstableread[header=false]{Data/hub_ladder_exc_ener_fit_params_L_80.dat}{\fitexc}
			\pgfplotstablegetelem{0}{[index]0}\of\fitexc
			\edef\kexc{\pgfplotsretval}
			\pgfplotstablegetelem{0}{[index]1}\of\fitexc
			\edef\mexc{\pgfplotsretval}
			\pgfplotstableread[header=true]{Data/hub_ladder_gs_ener_L_80.dat}{\gsdat}
			\pgfplotstablegetelem{0}{[index]0}\of\gsdat
			\edef\maxgstrunc{\pgfplotsretval}
			\pgfplotstableread[header=true]{Data/hub_ladder_exc_ener_L_80.dat}{\excdat}
			\pgfplotstablegetelem{0}{[index]0}\of\excdat
			\edef\maxexctrunc{\pgfplotsretval}
			\addplot
			[
				only marks,
				mark=o,
				thick,
				blue
			]
			table
			[
				y expr = \thisrowno{1},
				x expr = \thisrowno{0}
			]
			{\gsdat};
			\addlegendentry{$E_0$}
			\addplot
			[
				only marks,
				mark=o,
				thick,
				red
			]
			table
			[
				y expr = \thisrowno{1},
				x expr = \thisrowno{0}
			]
			{\excdat};
			\addlegendentry{$E_1$}
			\addplot
			[
			thick,
			no marks,
			domain = 0:\maxgstrunc,
			blue
			]
			{linearFct(\kgs,\mgs)};
			\addplot
			[
				thick,
				no marks,
				domain = 0:\maxexctrunc,
				red
			]
			{linearFct(\kexc,\mexc)};
			
			\end{axis}
		\end{tikzpicture}
    	}%
    	{%
    		\includegraphics{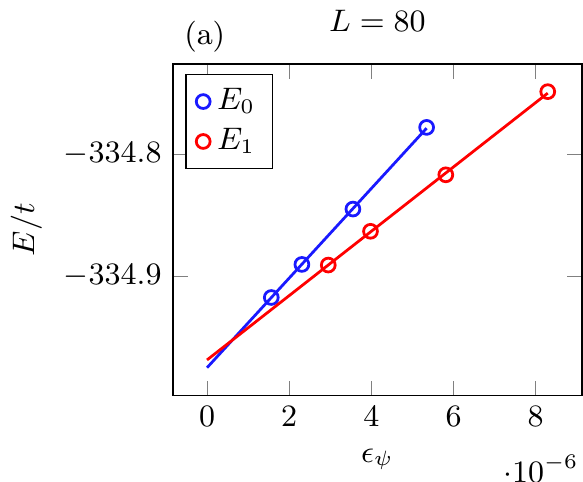}%
    	}%
    	\ifthenelse{\boolean{buildtikzpics}}%
	    {%
		\tikzsetnextfilename{HubLadderTruncFitL_96}
		\tikzset{external/export next=true}
		\begin{tikzpicture}
			\begin{axis}
			[
				xlabel = $\epsilon_\psi$,
				width = 0.32\textwidth,
				clip mode=individual,
				title = {$L=96$},
			]
			\node[anchor=south west] at (rel axis cs:0,1) {(b)};
			\pgfplotstableread[header=false]{Data/hub_ladder_gs_ener_fit_params_L_96.dat}{\fitgs}
			\pgfplotstablegetelem{0}{[index]0}\of\fitgs
			\edef\kgs{\pgfplotsretval}
			\pgfplotstablegetelem{0}{[index]1}\of\fitgs
			\edef\mgs{\pgfplotsretval}
			\pgfplotstableread[header=false]{Data/hub_ladder_exc_ener_fit_params_L_96.dat}{\fitexc}
			\pgfplotstablegetelem{0}{[index]0}\of\fitexc
			\edef\kexc{\pgfplotsretval}
			\pgfplotstablegetelem{0}{[index]1}\of\fitexc
			\edef\mexc{\pgfplotsretval}
			\pgfplotstableread[header=true]{Data/hub_ladder_gs_ener_L_96.dat}{\gsdat}
			\pgfplotstablegetelem{0}{[index]0}\of\gsdat
			\edef\maxgstrunc{\pgfplotsretval}
			\pgfplotstableread[header=true]{Data/hub_ladder_exc_ener_L_96.dat}{\excdat}
			\pgfplotstablegetelem{0}{[index]0}\of\excdat
			\edef\maxexctrunc{\pgfplotsretval}
			\addplot
			[
				only marks,
				mark=o,
				thick,
				blue
			]
			table
			[
				y expr = \thisrowno{1},
				x expr = \thisrowno{0}
			]
			{\gsdat};
			\addplot
			[
				thick,
				no marks,
				domain = 0:\maxgstrunc,
				blue
			]
			{linearFct(\kgs,\mgs)};
			\addplot
			[
				only marks,
				mark=o,
				thick,
				red
			]
			table
			[
				y expr = \thisrowno{1},
				x expr = \thisrowno{0}
			]
			{\excdat};
			\addplot
			[
				thick,
				no marks,
				domain = 0:\maxexctrunc,
				red
			]
			{linearFct(\kexc,\mexc)};
			\end{axis}
		\end{tikzpicture}
    	}%
    	{%
    		\includegraphics{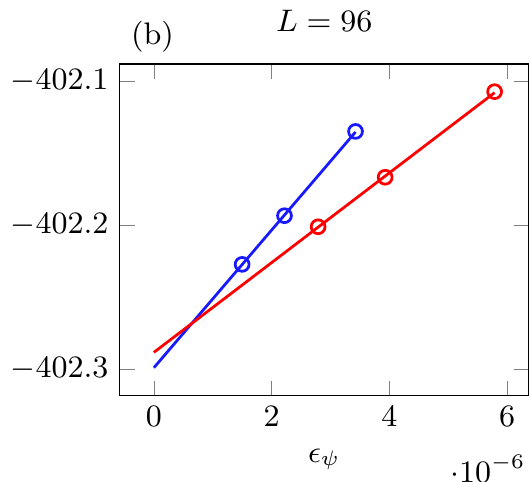}%
    	}%
    	\ifthenelse{\boolean{buildtikzpics}}%
    	{%
		\tikzsetnextfilename{HubLadderTruncFitL_112}
		\tikzset{external/export next=true}
		\begin{tikzpicture}
			\begin{axis}
			[
				xlabel = $\epsilon_\psi$,
				width = 0.32\textwidth,
				clip mode=individual,
				title = {$L=112$},
			]
			\node[anchor=south west] at (rel axis cs:0,1) {(c)};
			\pgfplotstableread[header=false]{Data/hub_ladder_gs_ener_fit_params_L_112.dat}{\fitgs}
			\pgfplotstablegetelem{0}{[index]0}\of\fitgs
			\edef\kgs{\pgfplotsretval}
			\pgfplotstablegetelem{0}{[index]1}\of\fitgs
			\edef\mgs{\pgfplotsretval}
			\pgfplotstableread[header=false]{Data/hub_ladder_exc_ener_fit_params_L_112.dat}{\fitexc}
			\pgfplotstablegetelem{0}{[index]0}\of\fitexc
			\edef\kexc{\pgfplotsretval}
			\pgfplotstablegetelem{0}{[index]1}\of\fitexc
			\edef\mexc{\pgfplotsretval}
			\pgfplotstableread[header=true]{Data/hub_ladder_gs_ener_L_112.dat}{\gsdat}
			\pgfplotstablegetelem{0}{[index]0}\of\gsdat
			\edef\maxgstrunc{\pgfplotsretval}
			\pgfplotstableread[header=true]{Data/hub_ladder_exc_ener_L_112.dat}{\excdat}
			\pgfplotstablegetelem{0}{[index]0}\of\excdat
			\edef\maxexctrunc{\pgfplotsretval}
			\addplot
			[
				only marks,
				mark=o,
				thick,
				blue
			]
			table
			[
				y expr = \thisrowno{1},
				x expr = \thisrowno{0}
			]
			{\gsdat};
			\addplot
			[
				only marks,
				mark=o,
				thick,
				red
			]
			table
			[
				y expr = \thisrowno{1},
				x expr = \thisrowno{0}
			]
			{\excdat};
			\addplot
			[
				thick,
				no marks,
				domain = 0:\maxgstrunc,
				blue
			]
			{linearFct(\kgs,\mgs)};
			\addplot
			[
				thick,
				no marks,
				domain = 0:\maxexctrunc,
				red
			]
			{linearFct(\kexc,\mexc)};
			\end{axis}
		\end{tikzpicture}
    	}%
    	{%
    		\includegraphics{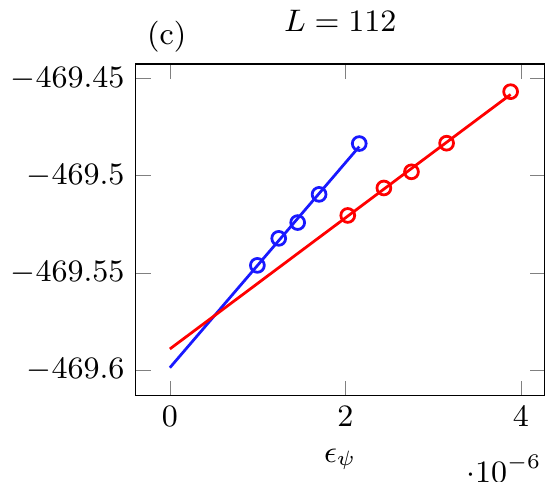}%
    	}%
    	
    	\ifthenelse{\boolean{buildtikzpics}}%
    	{%
		\tikzsetnextfilename{HubLadderFinSizFit}
		\tikzset{external/export next=true}
		\begin{tikzpicture}
			\begin{axis}
			[
				xlabel = $1/L$,
				ylabel = $k_BT_c/t$,
				width = 0.49\textwidth,
				yticklabel style = {/pgf/number format/precision=3,
					/pgf/number format/fixed},
				scaled y ticks = false,
				clip mode=individual
			]
			\node[anchor=south west] at (rel axis cs:0,1) {(d)};
			\pgfplotstableread[header=false]{Data/hub_ladder_fsize_gap_fit.dat}{\fsizefit}
			\pgfplotstablegetelem{0}{[index]0}\of\fsizefit
			\edef\b{\pgfplotsretval}
			\pgfplotstablegetelem{0}{[index]1}\of\fsizefit
			\edef\c{\pgfplotsretval}
			\pgfplotstablegetelem{1}{[index]1}\of\fsizefit
			\edef\fsizerr{\pgfplotsretval}
			\pgfplotstableread[header=false]{Data/hub_ladder_analyt_ratio.dat}{\tab}
			\pgfplotstablegetelem{0}{[index]0}\of\tab
			\edef\ladrat{\pgfplotsretval}
			\addplot+
			[
				only marks,
				mark=o,
				thick,
				error bars/.cd,
				y dir = both, y explicit
			]
			table
			[
				y expr = \thisrowno{1}/\ladrat,
				y error expr = \thisrowno{2}/\ladrat,
				x expr = \thisrowno{0}
			]
			{Data/hub_ladder_fsize_gap.dat};
			\addplot
			[
				no marks,
				thick,
				domain = 0:1/80
			]
			{linearFct(\b,\c)/\ladrat};
			\addplot+
			[
				only marks,
				mark=x,
				red,
				thick,
				error bars/.cd,
				y dir = both, y explicit
			]
			coordinates {(0,\c/\ladrat) +- (0,\fsizerr/\ladrat) };
			\end{axis}
		\end{tikzpicture}
    	}%
    	{%
    		\includegraphics{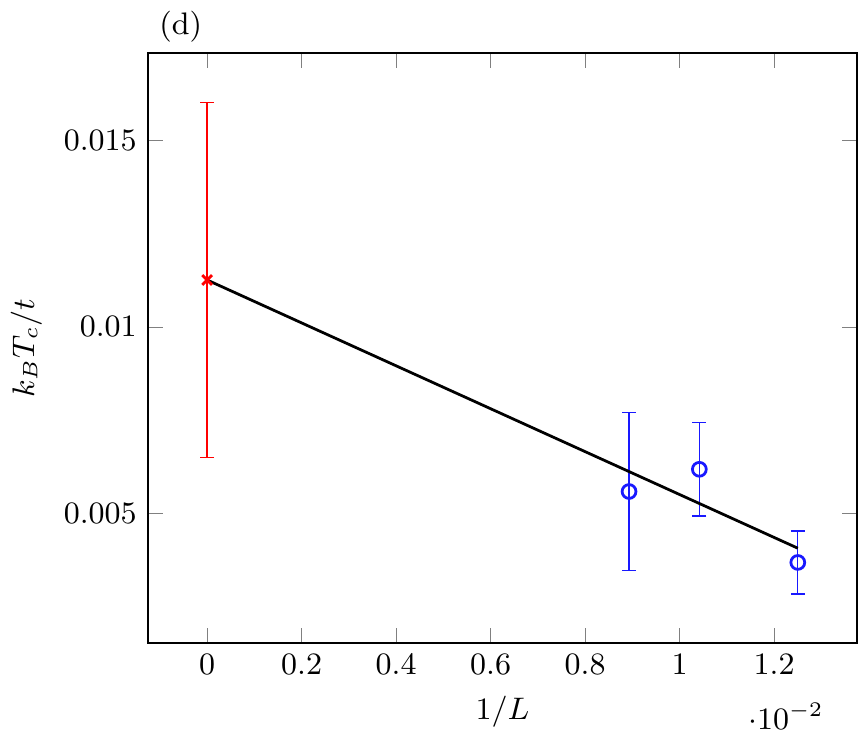}%
    	}%
    	\ifthenelse{\boolean{buildtikzpics}}%
    	{%
		\tikzsetnextfilename{HubLaddertperpscaling}
		\tikzset{external/export next=true}
		\begin{tikzpicture}
		\begin{axis}
		[
		xlabel = $t_\perp/t$,
		ylabel = $k_BT_c/t$,
		width = 0.49\textwidth,
		yticklabel style = {/pgf/number format/precision=3,
			/pgf/number format/fixed},
		scaled y ticks = false,
		clip mode=individual
		]
		\node[anchor=south west] at (rel axis cs:0,1) {(e)};
		\pgfplotstableread[header=false]{Data/hub_ladder_fsize_gap_fit.dat}{\fsizefit}
		\pgfplotstablegetelem{0}{[index]0}\of\fsizefit
		\edef\b{\pgfplotsretval}
		\pgfplotstablegetelem{0}{[index]1}\of\fsizefit
		\edef\c{\pgfplotsretval}
		\pgfplotstablegetelem{1}{[index]1}\of\fsizefit
		\edef\fsizerr{\pgfplotsretval}
		\pgfplotstableread[header=false]{Data/hub_ladder_analyt_ratio.dat}{\tab}
		\pgfplotstablegetelem{0}{[index]0}\of\tab
		\edef\ladrat{\pgfplotsretval}
		\edef\Krho{0.77}
		\addplot
		[
		no marks,
		thick,
		domain = 0:0.0489
		]
		{(\c/\ladrat)*(x/0.0489)^(4*\Krho/(4*\Krho-1))};
		\addplot
		[
			only marks,
			thick,
			mark=x,
			red
		]
		coordinates {(0.0489,\c/\ladrat)};

		\end{axis}
		\end{tikzpicture}
    	}%
    	{%
    		\includegraphics{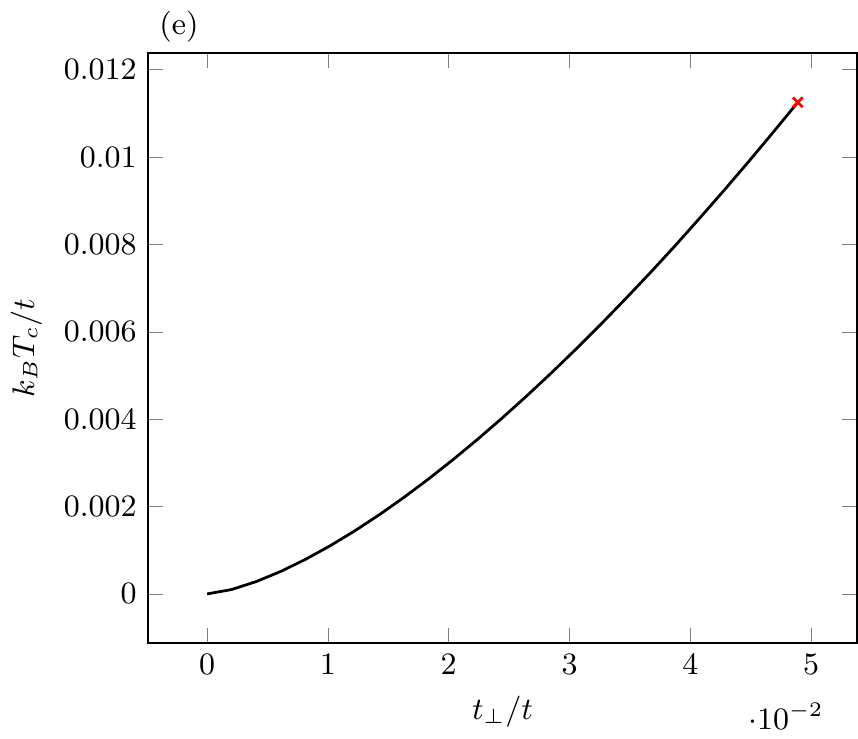}%
    	}%
		\caption{Energy gap to first excited state and critical temperature of the Hubbard ladder for $U=8t$, $n=0.9375$ and $t_\perp=0.0489t$. The sub-figures show ground state energy $E_0$ (blue circles) and excited state energy $E_1$ (red circles) extrapolated to zero truncation error (blue/red lines) for a ladder with length (a) $L=80$, (b) $L=96$, (c) $L=112$. Sub-figure (d) shows the critical temperature of the 3D ladder array $T_c=\Delta/R_\rm{ladder}(K_{\rho,+})=\frac{E_1-E_0}{R_\rm{ladder}(K_{\rho,+})}$ extrapolated to infinite size where $R_\rm{ladder}(K_{\rho,+})$ is given by \cref{analytic_ratio_ladder} and $K_{\rho,+}=0.77$~\cite{Dolfi2015b}, while (e) shows the critical temperature vs. $t_\perp$ exploiting the scaling known from \cref{scaling_eq_Tc}, using the $T_c$ value obtained in sub-figure (d) (red cross).}
		\label{fig:hub_lad_gap}
	\end{figure*}

	In isolation, Hubbard ladders typically require large bond dimensions for accurate simulations (see, e.g.,~\cite{Dolfi2015b}). With the included superconducting MF ordering channel we have found this requirement to be relaxed. This is expected, as any long-range order, be it in real space or momentum space, requires fewer retained Schmidt components than that same system without such order. However, at these lowered bond dimensions, converged MF-amplitudes exhibit non-negligible dependence on the bond dimension despite modest truncation errors. Regardless of this, the linear scaling of energy with truncation error, typically found in DMRG, remains intact even here, as shown in~\cref{fig:hub_lad_gap}a-c. Thus, it is possible to compute the mean-field amplitudes at modest bond dimension and obtain energy measurements, and thus $\Delta$, extrapolated to zero truncation error at different system sizes. 
	
	Finally, we obtain $\Delta$, and thus $T_c$ after rescaling with~\cref{analytic_ratio_ladder}, in the thermodynamic regime via infinite-size extrapolation as shown in~\cref{fig:hub_lad_gap}d. We note that $t_\perp$ is chosen quite close to $\Delta E_p$ and $\Delta E_s$. The primary reason are the small energy gaps we need to resolve. At smaller values of $t_\perp$ the discrete energy gaps of the finite-length systems mask $\Delta$ even for ladders with ${L=80}$ or even larger. However, using \cref{ladder_Tc_scaling} it is possible to extrapolate thermodynamic $T_c$ to smaller, physically more reasonable values of $t_\perp$ as shown in \cref{fig:hub_lad_gap}e.

	\section{\label{Sec::Resource_requirements}Resource requirements}
	In the present section we summarize time and resource consumption for the different algorithms used in this work. The MPS+MF routines feature repeated DMRG solutions. Generally, the total number of CPU-core hours $\tau$ scales with the same parameters as DMRG does, i.e.,
	\begin{equation}
		\tau_{MPS}\sim Ld^2\chi^3
	\end{equation}
	where $L$ is the number of lattice sites (and thus MPS-tensors), $d$ the size of the local Hilbert space, and $\chi$ is bond dimension. The number of loops required to reach self-consistency, $N_{tot}$ is shown for representative examples in~\cref{fig:loop_number} and subject to the optimizations mentioned in~\cref{subsec::init_guess}. The total resources consumed by the algorithm are thus given by
	\begin{equation}
		\tau_{MPS+MF} \sim N_{\mathrm{tot}}\tau_{MPS}.
	\end{equation}
	For the ground states of the negative-$U$ Hubbard chain with MF-amplitudes we have used MPS-based DMRG from the Matrix Product Toolkit package~\cite{MPTK_webpage}. The data has been computed using Intel Xeon E5 2630 v4 at 2.20 GHz CPU-cores. Most results are obtained at a bond dimension of $\chi=300$. A certain speed-up was obtained using 2 CPU-cores and threads used in the LAPACK and BLAS routines on which the algorithm rests. For a typical run  a single MF-loop takes about $8000s$ of wallclock time for $\chi=300$ and $L=100$. Notably $d=4$ for the negative-$U$ Hubbard chain. With 2 CPU-cores in use for this calculation, the total required resources for solution are $\tau_{MPS+MF}\sim 5N_\mathrm{tot}$ CPU core-hours.
	
	Additionally, in order to obtain critical temperatures directly we have calculated thermal states using imaginary time evolution of purified states. For simpler Hamiltonians where a shorter range (maximum of 1) for both $\alpha_{ik}$ and $\beta_{ir\sigma}$ is possible, trotterized time evolution suffices~\cite{Schollwock2011, Paeckel2019}. For this case we have used a time step of $\delta\tau=0.1$ and a fourth order Trotter discretization. With additional linear scaling in the length of the imaginary time simulated (equal to half of inverse temperature $\beta$) a typical solution requires about $12000s$ of wallclock time for $L=60$, $\chi=200$ and $\beta=9.5$. Having used  2 CPU-cores, the required resources in total are  $\tau_{MPS+MF}\sim 7N_{tot}$ CPU-core hours. Notably, the range of inverse temperature $\beta$ to  be simulated to determine $T_c$ changes markedly with $t_\perp$, leading to commensurate changes in $\tau$.
	
	In the case of Hamiltonians with long\hyp range terms it is necessary to apply more advanced time evolution schemes.
	With that the resource requirements increase significantly, i.e., a typical solution requires about $32000s$ on $6$ CPU cores for $L=60$, $\beta=9.5$, and $\chi=100$, leading to an overall $\tau_{MPS+MF}=53N_{tot}$ CPU-core hours.
	In part, this increase is due to the longer-range couplings. Another cause is the increased effective system size. This increase in turn is down to the specific  MPS-implementation~\cite{symmps} used,  which always requires full quantum number conservation. In our use-case, where both charge- and spin-conservation are discarded (c.f. subsec.~\ref{Sec::Results/sub::numerical}), quantum-numbers are restored artificially via the use of \gls{PP-DMRG}\cite{PP}, at the price of a larger effective system.
	Further details on obtaining finite temperature results are provided in the \cref{App::Finite_size_extrapolation,App::ppDMRG_and_purification}.

	Exact ground-state order parameters in finite lattices were computed with an AFQMC method using generalized Metropolis with force bias [\onlinecite{HaoPRA2015}]. This algorithm scales quadratically with the number of electrons, $N_e$, and linearly with the number of lattice sites $N_l$:
	\begin{equation}
	    \tau_{AFQMC,GS}\sim N_l N_e^2.
	\end{equation}
	For a $N_l=40\cdot 4=160$ system with 80 electrons, we performed calculations of 2000 sweeps for measurements after 10 sweeps of thermalization, with 100 independent repeats. Such a calculation, with an imaginary propagation time $\beta$ set to 64, took 49 hours of wallclock time on 100 Intel Xeon E5-2640 v4 2.4 GHz cores yielding a requirement of $\tau_{AFQMC,GS}\simeq4900$ CPU-core hours.
	
	In order to obtain exact values for $T_{BKT}$ in two-dimensional Q1D for benching MPS+MF against (c.f. subsec.~\ref{Sec::Results/sub::afqmc}), we have used finite temperature AFQMC. The package utilized for this purpose is called Algorithms for Lattice Fermions (ALF)~\cite{Assaad2021}. Data was obtained with an algorithm which scales cubically in the number of lattice sites, $N$, and linearly with inverse temperature $\beta$:
	\begin{equation}
		\tau_{AFQMC}\sim \beta N^3.
	\end{equation}
	An improvement of this scaling is available~\cite{He2019} but was not implemented for our calculations. We find that a single instance of sampling  requires ${\simeq 200s}$ for ${N=48\cdot8=384}$ and ${\beta=20}$ using a
	Intel Xeon E5-2698 v3 2.30GHz CPU-core. Statistical error bars are sufficiently small for a sample size of $\sim100000$ for the parameter set we study, requiring $\tau_{AFQMC}\simeq 5000$ CPU-core hours. The ALF-package, like most QMC-implementations, can of course parallelize this workload near-perfectly.
	
	When performing MPS+MF on the weakly coupled repulsive-U Hubbard ladders, MPS+MF becomes more resource-intensive: as earlier, one conserved quantum number is lost and the Matrix Product Operator representing the Hamiltonian is significantly larger than any for the negative-$U$ Hubbard chains. We use a DMRG-implementation offering distributed-memory parallelism (pDMRG) to obtain faster solutions~\cite{Kantian2019a}. Different from QMC-type algorithms, any such parallelisation will inevitably show non-trivial communication overheads, and thus not scale linearly in the number of MPI-processes. For a single converged ground state from pDMRG we thus require ${\simeq 27000s}$ using $32\cdot8=256$
	Intel Xeon E5-2698 v3 2.30GHz CPU-cores, at $\chi=1000$ and $L=96$. The total cost thus becomes $\tau\sim 2000N_{tot}$ CPU core-hours. The optimizations of \cref{subsec::init_guess} particularly apply to the Hubbard ladder systems. Without these optimizations, $N_\mathrm{tot}$ can be as high as $N_{\mathrm{tot}}=25$, but by employing them, where possible, can drop as low as $N_{\mathrm{tot}}=6$.
	\section{\label{Sec::Conclusion}Conclusion}
	In this work we have developed a numerical framework combining MPS-based numerics with MF and perturbation theory to solve correlated quasi-one-dimenional fermionic systems, constructed out of weakly coupled 1D sub-units, in two and three spacial dimensions. This method relies on MF-approximating tunneling processes occurring transverse to the 1D sub-units with amplitude $t_\perp$. The requirement for this approximation being reasonable is that $t_\perp$ be weaker than any gap on the 1D sub-units that suppresses first-order tunneling between 1D sub-units. Using the example of superconductivity in such Q1D arrays, we show how this framework allows to map otherwise difficult or even intractable correlated-fermion models in 2D and 3D onto a self-consistent 1D problem. We then demonstrate how these can be effectively solved both for ground states and thermal states.
	
	We test the framework on a model of attractive fermions on a 1D chain extensively comparing to both analytical methods and AFQMC. We obtain analytical expressions for superconducting $T_c$ of the model and the gap, $\Delta$, to the first excited state. Utilizing that a ratio $R=\Delta/T_c$ of these two quantities remains constant over $t_\perp$, we obtain $R$ analytically and greatly speed up calculation of $T_c$ via the use of $\Delta$ and $R$. Comparing this value with $T_c$ obtained directly from thermal state calculations shows that obtaining $T_c$ from $\Delta$ and $R$ yields excellent agreement, especially at small $t_\perp$. This allows MPS+MF to obtain $T_c$ without using imaginary time evolution, which is numerically more costly than obtaining $\Delta$ via ground state calculations.
	
	Subsequently, we use the gap and ratio method to obtain $T_c$ from MPS+MF and compare with $T_{BKT}$ from AFQMC. We find a semi-constant ratio of these over a range of $t_\perp$, in line with previous comparisons to QMC~\cite{Bollmark2020}. As expected, the MF-approximation yields greater over-estimation of the ordering temperature for lower-dimensional systems. With this in mind the method seems able to provide reliable estimates of $T_c$ in fermionic systems for an appropriate choice of parameters.
	
	Utilizing the developed MPS+MF framework we treat the case of a 3D array of weakly coupled, doped Hubbard ladders with strong repulsive interaction. With the tools developed in this work we are able to calculate $T_c$ quantitatively for the first time for these systems. The MPS+MF framework thus allows the simulation of a subgroup of 3D systems of strongly correlated fermions, namely the Q1D models, which were previously out of reach for any quantitative method.
	
	Notably, the 3D arrays of weakly coupled Hubbard ladders studied in this work have not been optimized to yield large critical temperatures. Previous work has indicated that by modifying the ladder-parameters larger $T_c$'s may be achieved~\cite{Karakonstantakis2011}. With the MPS+MF framework it is possible  to systematically search for improved critical temperatures starting from the microscopic models. This allows not just to deliberately search for optimal high-$T_c$ prototype-materials in the Q1D space, something that remains elusive for Q2D materials. It likewise permits to design ultra cold gas experiments capable of observing analogue states of high-$T_c$ superconductivity within current or near-future experimental constraints. We are pursuing both possibilities in current follow-up work.
	
	In this work we have focused on the physics of superconductivity using the MPS+MF routine. However, MPS+MF can be used for any Q1D system in which tunneling in-between 1D sub-units is suppressed at first order by a gap . This gap can be of any physical nature, such as, e.g., the charge gap present in the insulating phases of the Bechgaard and Fabre salts. The MPS+MF framework can thus also be deployed to understand, e.g., the antiferromagnet to spin density wave transition in these materials. This potential application of the framework highlights its power to incorporate multiple ordering channels simultaneously at the mean-field level, and thus its ability to resolve the competition between competing orders. 
	
	Finally, the capacity of MPS-numerics to address real-time dynamics of many-body systems both near and far from equilibrium opens the possibility to use the MPS+MF algorithm to study real-time dynamics of correlated fermions in high-dimensional Q1D systems. Such forthcoming work is currently in preparation on dynamically induced superconductivity in such systems~\cite{Martens2022}.
	\begin{acknowledgments}
    This work was funded by an ERC Starting Grant from the European Union's Horizon 2020 research and innovation programme under grant agreement No. 758935, and by the Swiss NSF under Division II. The computations were enabled by resources provided by the Swedish National Infrastructure for Computing (SNIC), partially funded by the Swedish Research Council through grant agreement no. 2018-05973, as well as by compute-time awarded by the United Kingdom's EPSRC-UKRI on the ARCHER2, Peta4-Skylake and Cirrus compute-clusters through the ''Access to HPC´´ and ''Scottish Academic Access´´calls. A.K. would like to thank Shintaro Takayoshi and Pierre Bouillot for helpful discussions. Y.Y. acknowledges support from US ONR (Contract No. N00014-17-1-2237). The
Flatiron Institute is a division of the Simons Foundation.	
    J.H. was supported by the European Research Council (ERC) under grant HQMAT (Grant Agreement No. 817799), the Israel-US Binational Science Foundation (BSF), and by a Research grant from Irving and Cherna Moskowitz.
    \end{acknowledgments}
	
	\newpage
	\appendix
	
	\section{\label{App::Finite_size_extrapolation}Finite size extrapolation}
	The Q1D models we consider in this paper suffer from finite size effects like all numerics on finite systems. In particular, as the connection between 1D systems weaken so does the strength of the resulting superconductor resulting in increased healing lengths. Additionally, we have found that finite size effects persist even at large $t_\perp$ albeit reduced in size. In order to accurately simulate these systems extrapolation has to be performed on the finite size observables to the limit of infinite size.
	
	Using MPS-numerics we compute the observables to be measured for several system sizes. We then utilize various strategies to obtain infinte size values depending on the type of finite size effect and observable. The strategies are outlined in this appendix.
	\subsection{Local observables}
\begin{figure}
    	\ifthenelse{\boolean{buildtikzpics}}%
    	{%
		\tikzsetnextfilename{GapExtrapGroupFigure}
		\tikzset{external/export next=true}
		\begin{tikzpicture}
			\begin{groupplot}
			[
				group style={group size=2 by 2}
			]
			\nextgroupplot
			[
				xlabel = $1/L$,
				ylabel = $\Delta/t$,
				width = 0.24\textwidth,
				yticklabel style = {/pgf/number format/precision=3,
					/pgf/number format/fixed},
				scaled y ticks = false,
				clip mode=individual
			]
			\node[anchor=south west] at (rel axis cs:0,1) {(a)};
			\pgfplotstableread[header=true]{Data/gap_over_invL_fit_tperp_0.05_U_-4.0_Prange_4_PHrange_4_n_0.5.dat}{\fittable}
			\pgfplotstablegetelem{0}{[index]0}\of\fittable
			\edef\a{\pgfplotsretval}
			\pgfplotstablegetelem{0}{[index]1}\of\fittable
			\edef\b{\pgfplotsretval}
			\pgfplotstablegetelem{0}{[index]2}\of\fittable
			\edef\c{\pgfplotsretval}
			\addplot+
			[
				mark=o,
				only marks,
				thick,
				error bars/.cd,
				y dir = both, y explicit
			]
			table
			[
				y expr = \thisrowno{1},
				x expr = \thisrowno{0},
				y error expr = \thisrowno{2}
			]
			{Data/gap_over_invL_tperp_0.05_U_-4.0_Prange_4_PHrange_4_n_0.5.dat};
			\addplot
			[
				no marks,
				thick,
				domain = 0:0.05
			]
			{quadFct(\a,\b,\c)};
			\nextgroupplot
			[
				xlabel = $1/L$,
				width = 0.24\textwidth,
				clip mode=individual
			]
			\node[anchor=south west] at (rel axis cs:0,1) {(b)};
			\pgfplotstableread[header=true]{Data/gap_over_invL_fit_tperp_0.175_U_-4.0_Prange_4_PHrange_4_n_0.5.dat}{\fittable}
			\pgfplotstablegetelem{0}{[index]0}\of\fittable
			\edef\a{\pgfplotsretval}
			\pgfplotstablegetelem{0}{[index]1}\of\fittable
			\edef\b{\pgfplotsretval}
			\pgfplotstablegetelem{0}{[index]2}\of\fittable
			\edef\c{\pgfplotsretval}
			\addplot+
			[
			mark=o,
			only marks,
			thick,
			error bars/.cd,
			y dir = both, y explicit
			]
			table
			[
			y expr = \thisrowno{1},
			x expr = \thisrowno{0},
			y error expr = \thisrowno{2}
			]
			{Data/gap_over_invL_tperp_0.175_U_-4.0_Prange_4_PHrange_4_n_0.5.dat};
			\addplot
			[
			no marks,
			thick,
			domain = 0:0.05
			]
			{quadFct(\a,\b,\c)};
			\nextgroupplot
			[
				xlabel = $1/L$,
				ylabel = $\Delta/t$,
				yticklabel style = {/pgf/number format/precision=4,
									/pgf/number format/fixed},
				width = 0.24\textwidth,
				clip mode=individual
			]
			\node[anchor=south west] at (rel axis cs:0,1) {(c)};
			\pgfplotstableread[header=true]{Data/gap_over_invL_fit_tperp_0.2_U_-10.0_Prange_1_PHrange_1_n_0.5.dat}{\fittable}
			\pgfplotstablegetelem{0}{[index]0}\of\fittable
			\edef\a{\pgfplotsretval}
			\pgfplotstablegetelem{0}{[index]1}\of\fittable
			\edef\b{\pgfplotsretval}
			\pgfplotstablegetelem{0}{[index]2}\of\fittable
			\edef\c{\pgfplotsretval}
			\addplot+
			[
			mark=o,
			only marks,
			thick,
			error bars/.cd,
			y dir = both, y explicit
			]
			table
			[
			y expr = \thisrowno{1},
			x expr = \thisrowno{0},
			y error expr = \thisrowno{2}
			]
			{Data/gap_over_invL_tperp_0.2_U_-10.0_Prange_1_PHrange_1_n_0.5.dat};
			\addplot
			[
			no marks,
			thick,
			domain = 0:0.05
			]
			{quadFct(\a,\b,\c)};
			\nextgroupplot
			[
				xlabel = $1/L$,
				width = 0.24\textwidth,
				clip mode=individual
			]
			\node[anchor=south west] at (rel axis cs:0,1) {(d)};
			\pgfplotstableread[header=true]{Data/gap_over_invL_fit_tperp_0.05_U_-2.0_Prange_8_PHrange_8_n_0.5.dat}{\fittable}
			\pgfplotstablegetelem{0}{[index]0}\of\fittable
			\edef\a{\pgfplotsretval}
			\pgfplotstablegetelem{0}{[index]1}\of\fittable
			\edef\b{\pgfplotsretval}
			\pgfplotstablegetelem{0}{[index]2}\of\fittable
			\edef\c{\pgfplotsretval}
			\addplot+
			[
			mark=o,
			only marks,
			thick,
			error bars/.cd,
			y dir = both, y explicit
			]
			table
			[
			y expr = \thisrowno{1},
			x expr = \thisrowno{0},
			y error expr = \thisrowno{2}
			]
			{Data/gap_over_invL_tperp_0.05_U_-2.0_Prange_8_PHrange_8_n_0.5.dat};
			\addplot
			[
			no marks,
			thick,
			domain = 0:0.05
			]
			{quadFct(\a,\b,\c)};
		\end{groupplot}
		\end{tikzpicture}
    	}%
    	{%
    		\includegraphics{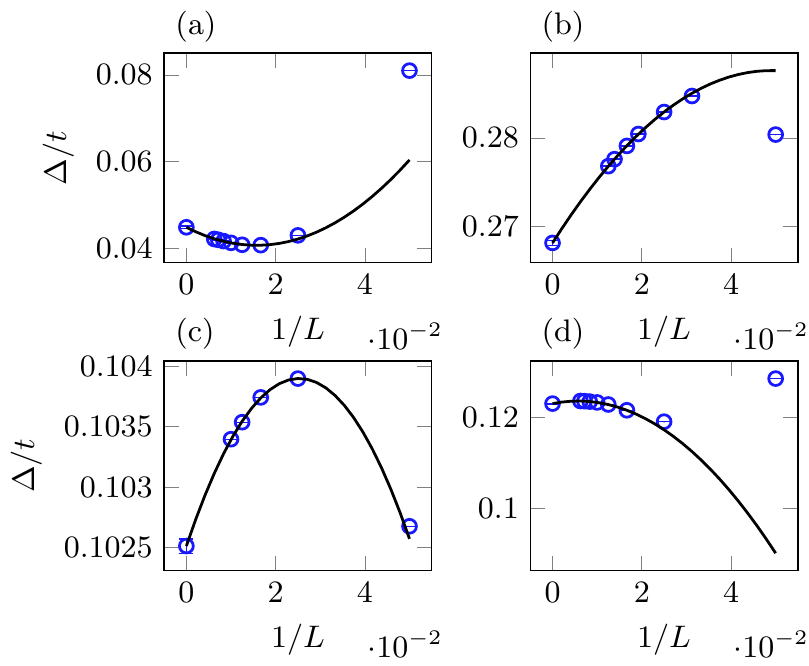}%
    	}%
		\caption{Examples of finite size extrapolations of the gap from the ground state to the first excited state ($\Delta$) for (a) $t_\perp=0.05t$, $U=-4t$, (b) $t_\perp=0.175t$, $U=-4t$, (c) $t_\perp=0.2t$, $U=-10t$ and (d) $t_\perp=0.05t$, $U=-2t$. All data at $n=0.5$.}
		\label{fig_app:locobs_fit}
	\end{figure}
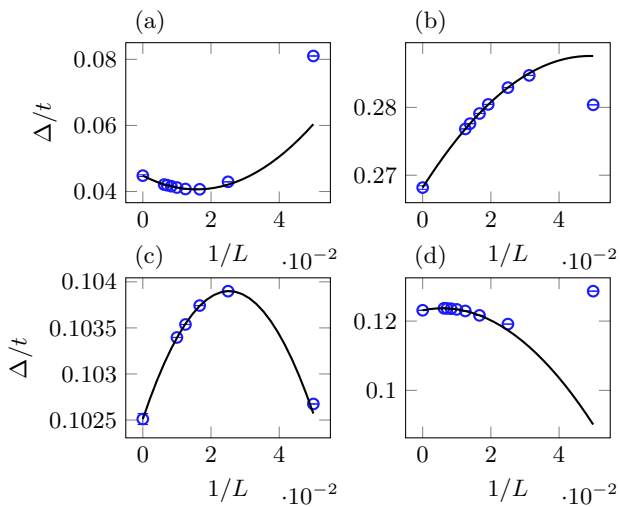

	For local quantities such as the order parameter and energy we use a common heuristic
	\begin{equation}
		O(L) = O_\infty+c_0\frac{1}{L}+c_1\frac{1}{L^2}+\mathcal{O}(L^{-3}),
	\end{equation}
	where $O(L)$ is a measurement of any local observable for a system size $L$ and $O_\infty$ the thermodynamic limit of that observable. Thus, we fit measurements at finite size to a quadratic polynomial in inverse size. We find data fits the heuristic pretty well as shown in \cref{fig_app:locobs_fit}.
	
	\subsection{Finite temperature at criticality}
	Several results in this work obtain the critical temperature of a system by evolving in imaginary time. The phase transition point is dependent on system size. For the case of significant finite size effects we consider the critical behaviour of the superconducting order parameter in particular as that determines when the system enters superconductivity.
	\subsubsection*{Significant finite size effects}
\begin{figure}
    	\ifthenelse{\boolean{buildtikzpics}}%
    	{%
		\tikzsetnextfilename{FinSizDataCollapse}
		\tikzset{external/export next=true}
		\begin{tikzpicture}
			\begin{axis}
			[
				ylabel = $L^{\beta/\nu}\psi$,
				xlabel = $L^{1/\nu}t$,
				legend pos = north east
			]
			\addplot+
			[
				thin,
				only marks,
				mark=x,
				mark options={scale=0.7},
				red
			]
			table
			[
				y expr = \thisrowno{1},
				x expr = \thisrowno{0}
			]
			{Data/resc_orp_vs_resc_T_collapse_L_60_tp_0.3_chi_200.dat};
			\addlegendentry{$L=60$}
			\addplot+
			[
			thin,
			only marks,
			mark=x,
			mark options={scale=0.7},
			blue
			]
			table
			[
			y expr = \thisrowno{1},
			x expr = \thisrowno{0}
			]
			{Data/resc_orp_vs_resc_T_collapse_L_50_tp_0.3_chi_200.dat};
			\addlegendentry{$L=50$}
			\addplot+
			[
			thin,
			only marks,
			mark=x,
			mark options={scale=0.7},
			green
			]
			table
			[
			y expr = \thisrowno{1},
			x expr = \thisrowno{0}
			]
			{Data/resc_orp_vs_resc_T_collapse_L_40_tp_0.3_chi_200.dat};
			\addlegendentry{$L=40$}
			\addplot+
			[
			thin,
			only marks,
			mark=x,
			mark options={scale=0.7},
			yellow
			]
			table
			[
			y expr = \thisrowno{1},
			x expr = \thisrowno{0}
			]
			{Data/resc_orp_vs_resc_T_collapse_L_30_tp_0.3_chi_200.dat};
			\addlegendentry{$L=30$}
			\addplot+
			[
			thin,
			only marks,
			mark=x,
			mark options={scale=0.7},
			purple
			]
			table
			[
			y expr = \thisrowno{1},
			x expr = \thisrowno{0}
			]
			{Data/resc_orp_vs_resc_T_collapse_L_20_tp_0.3_chi_200.dat};
			\addlegendentry{$L=20$}
			\end{axis}
		\end{tikzpicture}
    	}%
    	{%
    		\includegraphics{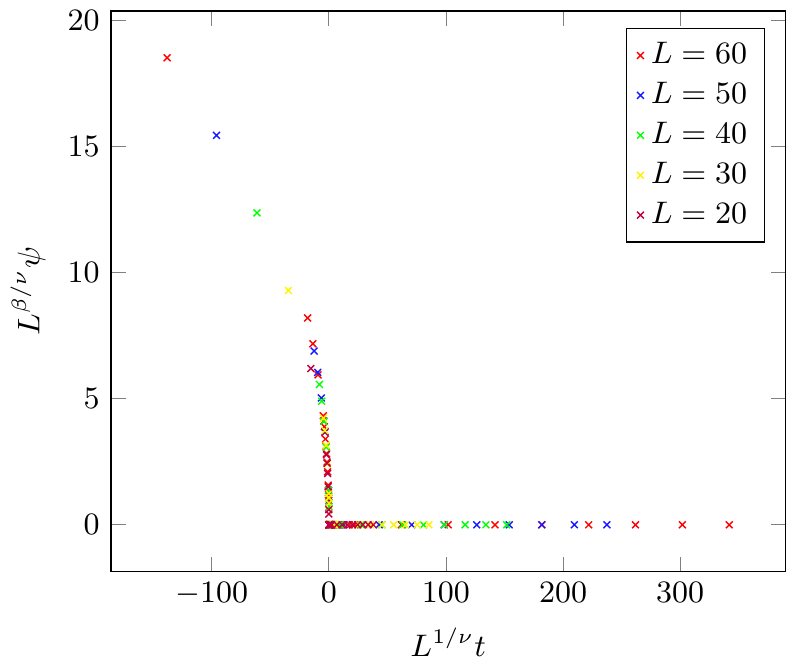}%
    	}%
		\caption{The rescaled order parameter plotted against reduced temperature $t=\frac{T-T_c}{T_c}$ for $U=-10t$, $t_\perp=0.3t$ and $n=0.5$. The critical temperature has been obtained by minimizing \cref{variance_for_scaling_func}.}
		\label{fig_app:curvecollapse}
	\end{figure}
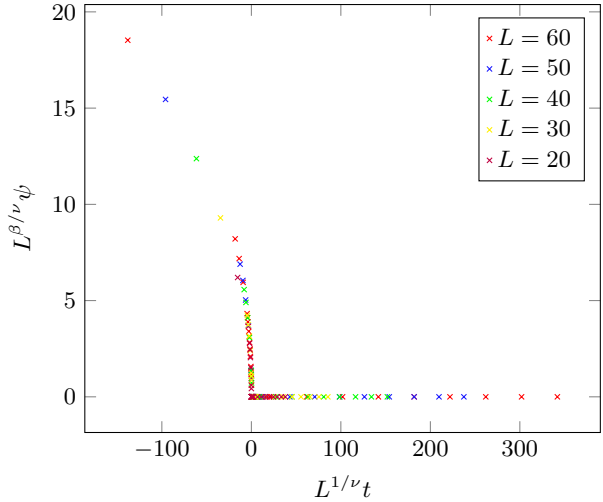

	When finite size effects must be considered we follow a common strategy used, e.g., on QMC results~\cite{Newman1999}. We study a second-order phase transition, where the critical behaviour of the order parameter (here named $\psi$ for simpler notation) is given by
	\begin{equation}
		\psi_L(t) = L^{-\beta/\nu}\tilde{\psi}(L^{-1/\nu}t).
	\end{equation}
	The reduced temperature, $t$, is given by
	\begin{equation}
		t=\frac{T-T_c}{T_c}
	\end{equation}
	and critical exponents $\beta,\nu$ are given by
	\begin{align}
		\label{orpscale}\psi\sim |t|^\beta, \\
		\label{corrlenscale}\xi \sim |t|^\nu,
	\end{align}
	where $\xi$ is the correlation length of the ordering field. Notably, on the unordered side the order parameter is zero. In order to extract $T_c$ we assume that our systems critical behaviour belongs to the mean-field universality class which is consistent with fits to \cref{orpscale} close to transition such that $\beta=\nu=0.5$.
	Using these exponents the finite size order parameter is rescaled by $L^{\beta/\nu}$ and plotted over $L^{1/\nu}t$
	\begin{equation}
		\tilde{\psi}(x) = L^{\beta/\nu}\psi_L(L^{-1/\nu}x).
	\end{equation}
	The function $\tilde{\psi}(x)$ is the scaling function of the order parameter and is system size independent. Thus, for a correct choice of $T_c$ all curves overlap close to transition as shown in \cref{fig_app:curvecollapse}.
	The quality of the collapse has been determined using
	\begin{multline}\label{variance_for_scaling_func}
		s_\psi^2=\frac{1}{x_{max}-x_{min}}\times\\\int_{x_{min}}^{x_{max}}\frac{\sum_L\tilde{\psi}_L(x)^2}{N}-\left[\frac{\sum_L\tilde{\psi}_L(x)}{N}\right]^2dx,
	\end{multline}
	where $N$ is the number of system sizes, $s_\psi^2$ is the variance integrated over a range $[x_{min},x_{max}]$ and $\tilde{\psi}_L(x)$ is the scaling function for size $L$ given a $T_c$. The range is chosen to lie around the proposed value of $T_c$. The critical temperature is then obtained for different widths of the integration interval and extrapolated to zero. The final error bar shown in values for $T_c$ is taken to be the fitting error for decreasing integration range. This ultimately gives a small error bar. Added data for larger sizes might move the result outside this error bar.
	
	For parameter sets where $t_\perp$ is particularly small generating data close to transition is tricky due to slow convergence of the mean-field amplitudes. One consequence of this is that the data used for collapse can end up too far from transition for finite size scaling to apply. On such occasion the analysis fails to produce reliable collapse of data and another strategy is needed.
	
	\subsubsection*{Critical temperature interval}
	\begin{figure}
		\centering
		\def\tperp{0.4}
		\def\U{-4.0}
		\def\Prange{4}%
		\def\PHrange{4}%
		\def\Nfac{8}%
		\def\S{0.0}%
		\def\V{0.0}%
		\def\Ls{{20,50,60}}%
		\edef\betadatadir{Data/SymMPS-FH-RANGE/Prange_\noexpand\Prange/PHrange_\noexpand\PHrange/tperp_\noexpand\tperp/L_\noexpand\L/N_\noexpand\N/S_\noexpand\S/U_\noexpand\U/V_\noexpand\V/}%
		\ifthenelse{\boolean{buildtikzpics}}%
	    {%
		\tikzsetnextfilename{TcFit_U_\U_tperp_\tperp}
		\begin{tikzpicture}
			\begin{axis}
			[
				width	=	0.51\textwidth-1.9pt,
				height	=	0.3\textheight,
				samples	=	100,
				xlabel	=	{temperature $T$},
				ylabel	=	{Order parameter},
				legend pos = north east,
				legend columns = 2,
				legend style = %
				{%
					font	=	\tiny,%
				}
			]
				\pgfmathtruncatemacro{\LsDim}{dim(\Ls)-1}%
				\pgfplotsforeachungrouped \i in {0,...,\LsDim}%
				{
					\pgfmathsetmacro{\L}{\Ls[\i]}%
					\pgfmathsetmacro\N{int(\Nfac*\L/16)}%
					\pgfmathsetmacro{\marker}{\gpmarkers[mod(\i,dim(\gpmarkers))]}%
					\pgfmathsetmacro{\cl}{\gpcolors[mod(\i,dim(\gpcolors))]}%
					\edef\tmp
					{
						\noexpand\addplot
						[
							mark	=	\marker,%
							color	=	\cl,
							error bars/.cd, 
							y dir=both,
							y explicit,	
						]
							table
							[
								x expr	=	1.0/\noexpand\thisrowno{0},
								y expr	=	\noexpand\thisrowno{7},
								y error expr = \noexpand\thisrowno{8}, 
							]
								{\betadatadir/fits.dat};
						\noexpand\addlegendentry{$L=\L$};
					}\tmp
					\edef\tmp
					{		
						\noexpand\pgfplotstableread[header=true]{\betadatadir/fits.dat.fix.fit}{\noexpand\fit}
						\noexpand\pgfplotstablegetelem{0}{amp}\noexpand\of\noexpand\fit 
						\noexpand\edef\noexpand\fittedAmp{\noexpand\pgfplotsretval}
						\noexpand\pgfplotstablegetelem{0}{amperror}\noexpand\of\noexpand\fit 
						\noexpand\edef\noexpand\fittedAmpError{\noexpand\pgfplotsretval}
						\noexpand\pgfplotstablegetelem{0}{tc}\noexpand\of\noexpand\fit 
						\noexpand\edef\noexpand\fittedTc{\noexpand\pgfplotsretval}
						\noexpand\pgfplotstablegetelem{0}{tcerror}\noexpand\of\noexpand\fit 
						\noexpand\edef\noexpand\fittedTcError{\noexpand\pgfplotsretval}
						\noexpand\pgfplotstablegetelem{0}{exponent}\noexpand\of\noexpand\fit 
						\noexpand\edef\noexpand\fittedExponent{\noexpand\pgfplotsretval}
						\noexpand\pgfplotstablegetelem{0}{exponenterror}\noexpand\of\noexpand\fit 
						\noexpand\edef\noexpand\fittedExponentError{\noexpand\pgfplotsretval}
						\noexpand\addplot
						[
							domain = 0.25:0.55,
							color=blue,
							dashed,
							color	=	\cl,
							thick,
						]
						{critExponentFct(\noexpand\fittedAmp,\noexpand\fittedTc,\noexpand\fittedExponent)};
					}\tmp
					\pgfplotstableread[header=true]{\betadatadir/fits.dat.fix.fit}{\fit}
					\pgfplotstablegetelem{0}{amp}\of\fit 
					\edef\fittedAmp{\pgfplotsretval}
					\pgfplotstablegetelem{0}{amperror}\of\fit 
					\edef\fittedAmpError{\pgfplotsretval}
					\pgfplotstablegetelem{0}{tc}\of\fit 
					\edef\fittedTc{\pgfplotsretval}
					\pgfplotstablegetelem{0}{tcerror}\of\fit 
					\edef\fittedTcError{\pgfplotsretval}
					\edef\tmp
					{
						\noexpand\addlegendentry{$\noexpand\printpgfnumberwithouterrorInMath{\fittedAmp}{\fittedAmpError}\cdot(T\noexpand\printpgfnumberwithouterrorInMath[2]{\fittedTc}{\fittedTcError})^{\noexpand\nicefrac{1}{2}}$};							
					}\tmp
					\node[] at (axis cs: 0.275,0.025) (insetPos) {};
				}%
			\end{axis}%
			\pgfmathsetmacro{\L}{50}%
			\pgfmathsetmacro\N{int(\Nfac*\L/16)}%
			\def\betas{2.0,2.1,2.2,2.25,2.5,2.75,3.0,3.25,3.5,4.0}%
			\def\chis{100}%
			\edef\datadir{Data/SymMPS-FH-RANGE/Prange_\noexpand\Prange/PHrange_\noexpand\PHrange/tperp_\noexpand\tperp/L_\noexpand\L/N_\noexpand\N/S_\noexpand\S/U_\noexpand\U/V_\noexpand\V/beta_\noexpand\pbeta/chi_\noexpand\chimax/}%
			\edef\datafile{\noexpand\datadir/checkpoints}%
			
			\pgfplotsset
			{
				/pgfplots/colormap={temp}{
					rgb255=(255,0,0) 		
					rgb255=(0,0,255) 		
				}
			}
			
			\begin{axis}
			[
				at		=	(insetPos),
				width	=	0.21\textwidth,
				height	=	0.15\textheight,
				xmin	=	-0,
				ymin	=	-0.00001,
				xticklabel style = {font=\tiny},
				yticklabel style = {font=\tiny},
				xlabel	=	{\tiny inv. self-consistency it},
				ylabel	=	{\tiny Order parameter},
				title	=	{\footnotesize $L=50$},
				every axis title/.style={anchor=north east,at={(1,1)}},
				colorbar right,
				colormap name	= temp,
				every colorbar/.append style =
				{
					width				=	1.5mm,
 					xshift				=	-1.25em,
					scaled y ticks		= 	false,
					ylabel style		= 	{font=\tiny,yshift=0pt},
					ylabel				=	{inv. temperature $\beta$},
					ytick				= 	{2,3,4},
					yticklabels			=	{\tiny2,\tiny3,\tiny4},
					ylabel shift 		=	-4pt,
				},
				point meta min=2,
				point meta max=4,
			]
				\foreach \pbeta [count=\i] in \betas%
				{
					\foreach \chimax [count=\j] in \chis%
					{								
						\pgfmathsetmacro{\marker}{\gpmarkers[mod(\j,dim(\gpmarkers))]}%
						\pgfmathsetmacro{\cl}{(1.0/(\pbeta)-0.25)/0.25*100}%
						\edef\tmp
						{
							\noexpand\addplot
							[
								color	=	red!\cl!blue,
								thin,
							] 
								table
								[
									header	= 	false,
									x expr	=	1.0/(\noexpand\coordindex+1),
									y expr	=	\noexpand\thisrowno{1},
								]
								{\noexpand\datafile};

							\noexpand\pgfplotstableread[header=true]{\datafile.fit}{\noexpand\fit}
							\noexpand\pgfplotstablegetelem{0}{extrapolation}\noexpand\of\noexpand\fit 
							\noexpand\edef\noexpand\fittedExtrapolation{\noexpand\pgfplotsretval}
							\noexpand\pgfplotstablegetelem{0}{extrapolerror}\noexpand\of\noexpand\fit 
							\noexpand\edef\noexpand\fittedExtapolationError{\noexpand\pgfplotsretval}
							\noexpand\pgfplotstablegetelem{0}{expamp}\noexpand\of\noexpand\fit 
							\noexpand\edef\noexpand\fittedExpAmplitude{\noexpand\pgfplotsretval}
							\noexpand\pgfplotstablegetelem{0}{expamperror}\noexpand\of\noexpand\fit 
							\noexpand\edef\noexpand\fittedExpAmplitudeError{\noexpand\pgfplotsretval}
							\noexpand\pgfplotstablegetelem{0}{comp}\noexpand\of\noexpand\fit 
							\noexpand\edef\noexpand\fittedCompression{\noexpand\pgfplotsretval}
							\noexpand\pgfplotstablegetelem{0}{comperror}\noexpand\of\noexpand\fit 
							\noexpand\edef\noexpand\fittedCompressionError{\noexpand\pgfplotsretval}
											
							\noexpand\addplot
							[
								domain = 0.05:1,
								color=black,
								dashed,
								forget plot,
								samples=100,
							]
								{exponentialInFct(\noexpand\fittedExtrapolation,\noexpand\fittedExpAmplitude,\noexpand\fittedCompression)};
						}\tmp
					}
				}
			\end{axis}%
		\end{tikzpicture}
		}%
    	{%
    		\includegraphics{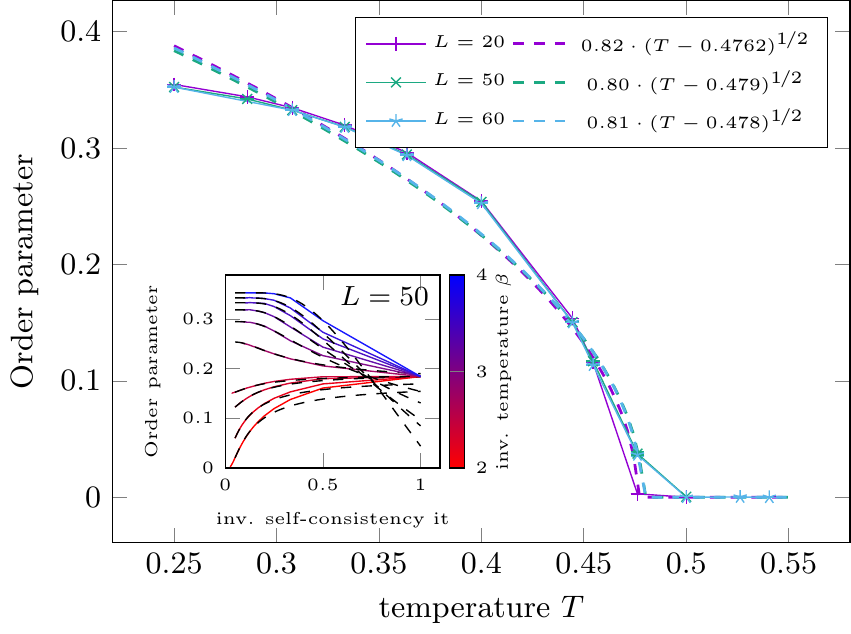}%
    	}%
	
		\caption
		{
			\label{fig_app:TcFit_U_-4.0_tperp_0.4}
			Example obtaining $T_c$ from, fits without finite size scaling for $U=-4t$, $t_\perp=0.4t$, $n=0.5$. Symbols denote the onsite order parameter extrapolated to infinite number of iterations, for $L=20$ (plus), $L=50$ (cross) and $L=60$ (star) computed via MPS+MF for thermal states. Solid lines are guide to the eye. Dashed lines are fits with ansatz \cref{mf_scaling_ansatz}. Inset shows an example of extrapolating the order parameter to infinite numbers of iterations for $L=50$
		}
	\end{figure}
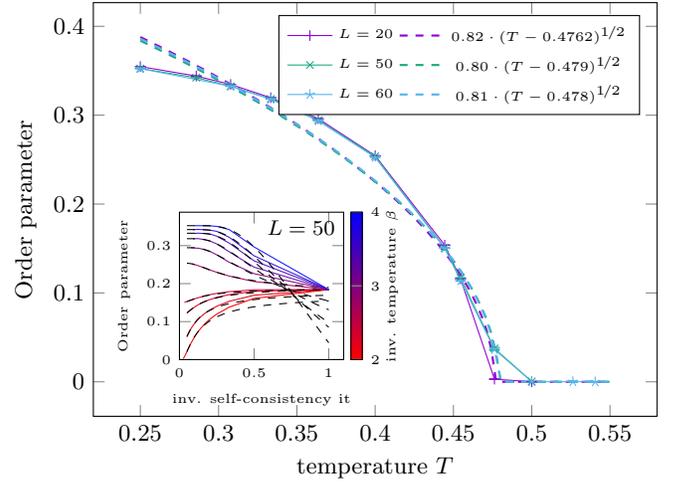
		When the previous analysis fails to produce a reasonable collapse we generate a grid around the estimated phase transition. 
	Since the convergence close to the phase transition is especially demanding, we focus on surrounding temperatures.
	In order to reduce the number of necessary self-consistency iterations we extrapolate the order parameter via $O(n)=O_{n\rightarrow\infty}+a\cdot\exp(-b n)$ in the number of self-consistency iterations $n$.
	Those extrapolations are shown as an example in the inset of \cref{fig_app:TcFit_U_-4.0_tperp_0.4} for $L=50$, but for convenience plotted over the inverse number of iterations.
	The results of these extrapolations are then used within to fit the data with
	\begin{equation}\label{mf_scaling_ansatz}
		O(T)=
		\begin{cases}
			a \lvert T - T_c \rvert^{\frac12}, \quad \mathrm{if }\, T < \lvert T_c \rvert\\
			0, \quad \mathrm{otherwise}
		\end{cases} \;,
	\end{equation}
	as can be seen in \cref{fig_app:TcFit_U_-4.0_tperp_0.4}.
	The final result interval for the critical temperature is then given by the estimated values for different system sizes.	
	This procedure causes significant errors compared to the previous method and fails to account for finite size effects that may occur. 
	Nevertheless, the errors are small enough to permit analysis as can be seen in \cref{fig:Tc_from_ratio_comp}.
	
	\section{\label{App::ppDMRG_and_purification}Finite temperature ppDMRG}
 	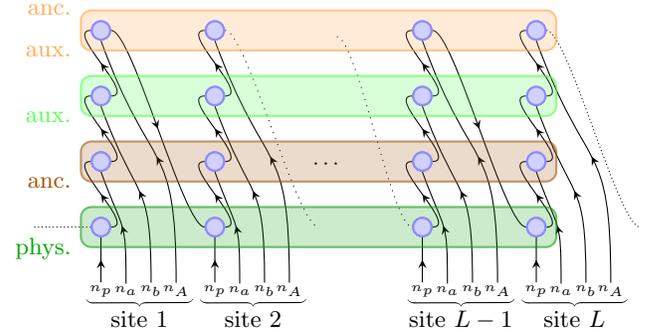
\begin{figure}[!t]
 		\centering
 		\ifthenelse{\boolean{buildtikzpics}}%
	    {%
 		\tikzsetnextfilename{mps2quadruple}
 		\begin{tikzpicture}
 			\begin{scope}[node distance = 0.6 and 1.2425]
 				\node[ghost] (site0a) {};
 				\node[site] (site1a) [right=0.75 of site0a] {};
 				\node[site] (site1b) [above=of site1a] {};
 				\node[site] (site1c) [above=of site1b] {};
 				\node[site] (site1d) [above=of site1c] {};
 				\node[site] (site2a) [right=of site1a] {};
 				\node[site] (site2b) [above=of site2a] {};
 				\node[site] (site2c) [above=of site2b] {};
 				\node[site] (site2d) [above=of site2c] {};
 				\node[unsite] (site4a) [right=of site2a] {};
 				\node[unsite] (site4b) [above=of site4a] {};
 				\node[unsite] (site4c) [above=of site4b] {};
 				\node[unsite] (site4d) [above=of site4c] {};
 				\node[ghost] at ($(site4a)!0.5!(site4c)$) {$\cdots$};
 				\node[site] (site5a) [right=1.0 of site4a] {};
 				\node[site] (site5b) [above=of site5a] {};
 				\node[site] (site5c) [above=of site5b] {};
 				\node[site] (site5d) [above=of site5c] {};
 				\node[site] (site6a) [right=of site5a] {};
 				\node[site] (site6b) [above=of site6a] {};
 				\node[site] (site6c) [above=of site6b] {};
 				\node[site] (site6d) [above=of site6c] {};
  				\node[unsite] (site7a) [right=of site6a] {};
 				
 				\node[ld] (sigma1a) [below=of site1a] {\tiny$\ingoing{n}_{p}$};
 				\node[ld] (sigma2a) [below=of site2a] {\tiny$\ingoing{n}_{p}$};
 				\node[ld] (sigma4a) at (site4a|-sigma2a) {};
 				\node[ld] (sigma5a) at (site5a|-sigma2a) {\tiny$\ingoing{n}_{p}$};
 				\node[ld] (sigma6a) at (site6a|-sigma2a) {\tiny$\ingoing{n}_{p}$};
 				\node[ld] (sigma7a) at (site7a|-sigma2a) {};
 				
 				\node[ld] (sigma1b) at ($(sigma1a)!0.22!(sigma2a)$) {\tiny$\ingoing{n}_{a}$};
 				\node[ld] (sigma2b) at ($(sigma2a)!0.22!(sigma4a)$) {\tiny$\ingoing{n}_{a}$};
 				\node[ld] (sigma5b) at ($(sigma5a)!0.22!(sigma6a)$) {\tiny$\ingoing{n}_{a}$};
 				\node[ld] (sigma6b) at ($(sigma6a)!0.22!(sigma7a)$) {\tiny$\ingoing{n}_{a}$};
 				
 				\node[ld] (sigma1c) at ($(sigma1a)!0.44!(sigma2a)$) {\tiny$\ingoing{n}_{b}$};
 				\node[ld] (sigma2c) at ($(sigma2a)!0.44!(sigma4a)$) {\tiny$\ingoing{n}_{b}$};
 				\node[ld] (sigma5c) at ($(sigma5a)!0.44!(sigma6a)$) {\tiny$\ingoing{n}_{b}$};
 				\node[ld] (sigma6c) at ($(sigma6a)!0.44!(sigma7a)$) {\tiny$\ingoing{n}_{b}$};
 				
 				\node[ld] (sigma1d) at ($(sigma1a)!0.66!(sigma2a)$) {\tiny$\ingoing{n}_{A}$};
				\node[ld] (sigma2d) at ($(sigma2a)!0.66!(sigma4a)$) {\tiny$\ingoing{n}_{A}$};
				\node[ld] (sigma5d) at ($(sigma5a)!0.66!(sigma6a)$) {\tiny$\ingoing{n}_{A}$};
				\node[ld] (sigma6d) at ($(sigma6a)!0.66!(sigma7a)$) {\tiny$\ingoing{n}_{A}$};
 				
 				\draw[densely dotted] (site0a) -- (site1a);
 				
 				\draw[->-,out=0,in=180] (site1a.east) to (site1b.west);
 				\draw[->-,out=0,in=180] (site2a.east) to (site2b.west);
 				\draw[->-,out=0,in=180] (site5a.east) to (site5b.west);
 				\draw[->-,out=0,in=180] (site6a.east) to (site6b.west);
 				
 				\draw[->-,out=0,in=180] (site1b.east) to (site1c.west);
 				\draw[->-,out=0,in=180] (site2b.east) to (site2c.west);
 				\draw[->-,out=0,in=180] (site5b.east) to (site5c.west);
 				\draw[->-,out=0,in=180] (site6b.east) to (site6c.west);
 				
 				\draw[->-,out=0,in=180] (site1c.east) to (site1d.west);
 				\draw[->-,out=0,in=180] (site2c.east) to (site2d.west);
 				\draw[->-,out=0,in=180] (site5c.east) to (site5d.west);
 				\draw[->-,out=0,in=180] (site6c.east) to (site6d.west);
 				
 				\draw[->-,out=0,in=180,looseness=0.25] (site1d.east) to (site2a.west);
 				\draw[->-,out=0,in=180,looseness=0.25] (site5d.east) to (site6a.west);
 				\draw[out=0,in=180,densely dotted,looseness=0.25] (site6d.east) to (site7a.west);
 				\draw[out=0,in=180,dotted,looseness=0.25] (site2d.east) to (site4a.west);
 				\draw[out=0,in=180,dotted,looseness=0.25] (site4d.east) to (site5a.west);
 				
 				\draw[->-] (sigma1a) to (site1a);
 				\draw[->-] (sigma2a) to (site2a);
 				\draw[->-] (sigma5a) to (site5a);
 				\draw[->-] (sigma6a) to (site6a);
 				\draw[->-,out=90,in=-70,looseness=1.5] (sigma1b.north) to (site1b.south);
 				\draw[->-,out=90,in=-70,looseness=1.5] (sigma2b.north) to (site2b.south);
 				\draw[->-,out=90,in=-70,looseness=1.5] (sigma5b.north) to (site5b.south);
 				\draw[->-,out=90,in=-70,looseness=1.5] (sigma6b.north) to (site6b.south);
 				\draw[->-,out=90,in=-70,looseness=1.5] (sigma1c.north) to (site1c.south);
 				\draw[->-,out=90,in=-70,looseness=1.5] (sigma2c.north) to (site2c.south);
 				\draw[->-,out=90,in=-70,looseness=1.5] (sigma5c.north) to (site5c.south);
 				\draw[->-,out=90,in=-70,looseness=1.5] (sigma6c.north) to (site6c.south);
 				\draw[->-,out=90,in=-70,looseness=1.5] (sigma1d.north) to (site1d.south);
 				\draw[->-,out=90,in=-70,looseness=1.5] (sigma2d.north) to (site2d.south);
 				\draw[->-,out=90,in=-70,looseness=1.5] (sigma5d.north) to (site5d.south);
 				\draw[->-,out=90,in=-70,looseness=1.5] (sigma6d.north) to (site6d.south);
 				
 				\node[draw,fill=green!60!black,draw=green!60!black,fill opacity=.2,draw opacity=0.6,thick,rounded corners,fit=(site1a) (site6a),inner sep=0.4em] (phys) {};
 							
 				\node[draw,fill=orange!60!black,draw=orange!60!black,fill opacity=0.2,draw opacity=0.6,thick,rounded corners,fit=(site1b) (site6b),inner sep=0.4em] (anc) {};
 				
 				\node[draw,,fill=green!75,draw=green!75,fill opacity=.2,draw opacity=.6,thick,rounded corners,fit=(site1c) (site6c),inner sep=0.4em] (aux) {};
 				
 				\node[draw,,fill=orange!75,draw=orange!75,fill opacity=0.2,draw opacity=.6,thick,rounded corners,fit=(site1d) (site6d),inner sep=0.4em] (ancaux) {};
 				
 				\node[site] (site1a) at (site1a) {};
 				\node[site] (site1b) at (site1b) {};
 				\node[site] (site1c) at (site1c) {};
 				\node[site] (site1d) at (site1d) {};
 				\node[site] (site2a) at (site2a) {};
 				\node[site] (site2b) at (site2b) {};
 				\node[site] (site2c) at (site2c) {};
 				\node[site] (site2d) at (site2d) {};
 				\node[site] (site5a) at (site5a) {};
 				\node[site] (site5b) at (site5b) {};
 				\node[site] (site5c) at (site5c) {};
 				\node[site] (site5d) at (site5d) {};
 				\node[site] (site6a) at (site6a) {};
 				\node[site] (site6b) at (site6b) {};
 				\node[site] (site6c) at (site6c) {};
 				\node[site] (site6d) at (site6d) {};
 				
 				\node[anchor=east, green!65!black] at (phys.south west) {phys.};
 				\node[anchor=east, orange!65!black] at (anc.south west) {anc.};
 				\node[anchor=east, green!65] at (aux.south west) {aux.};
 				\node[anchor=east, orange!65] at (ancaux.north west) {anc.};
 				\node[anchor=east, orange!65] at (ancaux.south west) {aux.};
				\draw[decorate, decoration = {calligraphic brace}] (sigma1d.south east) -- node[below,ld,inner sep=4pt] {site $1$} (sigma1a.south west);
				\draw[decorate, decoration = {calligraphic brace}] (sigma2d.south east) -- node[below,ld,inner sep=4pt] {site $2$} (sigma2a.south west);
				\draw[decorate, decoration = {calligraphic brace}] (sigma5d.south east) -- node[below,ld,inner sep=4pt] {site $L-1$} (sigma5a.south west);
				\draw[decorate, decoration = {calligraphic brace}] (sigma6d.south east) -- node[below,ld,inner sep=4pt] {site $L$} (sigma6a.south west);
 			\end{scope}
 		\end{tikzpicture}
    	}%
    	{%
    		\includegraphics{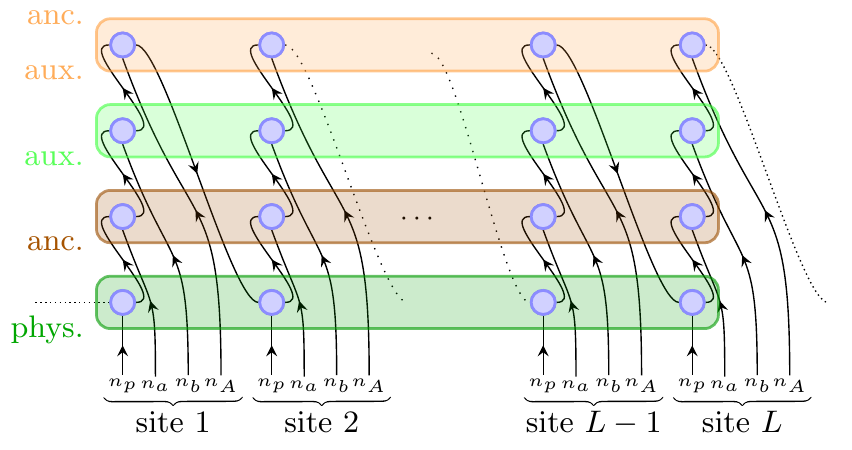}%
    	}%
 		\caption
 		{
 			\label{fig_app:mps2quadruple}
 			Stucture of MPS for thermal state calculations using state purification and imaginary time evolution via TDVP, which requires introducing ancilla sites. The use of the SymMPS-package~\cite{symmps} for TDVP also requires using  PP-DMRG~\cite{PP}, meaning adding auxiliary sites for both physical and ancilla sites to recover conserved charge and spin quantum numbers. 
 		}
 	\end{figure}
	In order to use time\hyp evolution methods that are designed for long\hyp range interactions we choose the SymMPS toolkit \cite{symmps}.
	This choice comes with the caveat that conserved $U(1)$ quantum numbers are necessary.
	Since the MPS+MF Hamiltonian does violate those symmetries, we applied \gls{PP-DMRG} \cite{PP} to restore them.
	That means we not only double the system size in order to represent density matrices instead of states, as it is usually the case in state purification \cite{PhysRevLett.93.207204,PhysRevB.72.220401}, but also double it again in order to have an auxiliary bath to restore the conservation-laws via these.
	Hence, as can be seen in \cref{fig_app:mps2quadruple}, we have the physical and the ancilla system, which represent density matrices and the auxiliary and the ancilla\hyp auxiliary system, which restore the quantum number conservation.
	Since the long\hyp range interactions are thus increased in their range by a factor of four, another obstacle needs to be circumvented: 
	this is the loss of particles from the physical (and its auxiliary) system into the ancilla (and its auxiliary) system \cite{Paeckel2019}.
	This leakage occurs due to the accummulation of numerical errors, which we prevent by having the $U(1)$ symmetries conserved on the physical and the ancilla system separately.
	To achieve that we increased the local basis from four to $16$ states, by adding a separate fermionic particle species, which is supposed to only occur on the ancilla systems.
	
	Evolving a product state, such as an infinite temperature state, with single\hyp site \gls{TDVP} lead to significant projection errors \cite{Paeckel2019}. 
	Hence one needs to increase the bond dimension first via a different time\hyp evolution method \cite{PhysRevX.11.031007,Paeckel2020}.
	We chose to use two\hyp site \gls{TDVP} for the first ten time steps, which are chosen to be very small $\delta t=10^{-7}$.
	This way the bond dimension grows, but the projection error, which is scaled within the exponential by the time step, stays small.
	Afterwards one time step is performed to go to the usual time step grid $\delta t = 0.05$.
	Then all following time steps are executed by single\hyp site \gls{TDVP}, since that is faster.
	\section{\label{App::Truncation_error_extrapolation}Truncation error extrapolation}
	For any local quantity it is possible to perform an extrapolation in truncation error. For large enough bond dimension a general measurement follows
	\begin{equation}
		\braket{O}(\epsilon_\psi) = \braket{O} + c_0\epsilon_\psi,
	\end{equation}
	i.e., a measurement of a DMRG state typically depends linearly on its truncation error~\cite{White1993b,Legeza1996,Schollwock2005}.

	\section{\label{App::AFQMC_procedure}Auxiliary Field Quantum Monte Carlo (AFQMC)}
	There are a number of different flavors of the AFQMC method, which are documented in the literature [\onlinecite{Assaad2021,Shiwei2019,HaoPRE2016}].
In this work, the ground-state order parameters from
AFQMC were obtained using a 
generalized Metropolis approach with force bias [\onlinecite{HaoPRA2015}],
while the  $T_{BKT}$ was obtained with standard finite-temperature AFQMC method 
[\onlinecite{Scalapino1993}, \onlinecite{Paiva2004}].
In this appendix we briefly describe our ground-state and finite-temperature 
calculations in two separate subsections below.
	\subsection{Ground state order parameter}
	\begin{figure}
    	\ifthenelse{\boolean{buildtikzpics}}%
    	{%
		\tikzsetnextfilename{AFQMC_GSPairCorr}
		\tikzset{external/export next=true}
		\begin{tikzpicture}
			\pgfplotstableread[header=true]{Data/afqmc_gs_orp_40x4.dat}{\orptable}
			\pgfplotstablegetrowsof{\orptable}
			\pgfplotstablegetelem{0}{[index]1}\of\orptable
			\edef\orp{\pgfplotsretval}
			\begin{axis}
			[
				xlabel = $|i-j|$,
				title = $\braket{c^\dagger_{i\uparrow}c^\dagger_{i\downarrow}c_{j\downarrow}c_{j\uparrow}}$
			]
			\addplot
			[
				thick,
				mark=x
			]
			table
			[
				y expr = \thisrowno{1},
				x expr = \thisrowno{0}
			]
			{Data/afqmc_gs_tperp_0.1_Lx_40_Ly_4.dat};
			\addplot
			[
				thick,
				dashed,
				mark=none,
				samples=2,
				domain = 5:35
			]
			{\orp*\orp};
			\end{axis}
		\end{tikzpicture}
    	}%
    	{%
    		\includegraphics{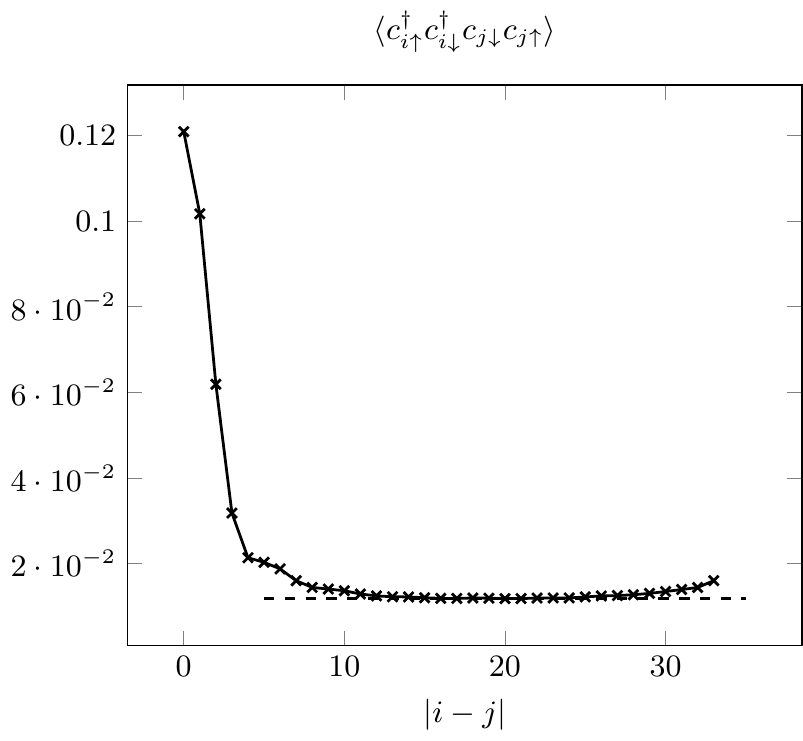}%
    	}%
		\caption{Example of on-chain pair-correlator in ground state AFQMC for a $40 \times 4$ lattice with $U=-2t$, $t_\perp\approx 0.14t$, $n=0.5$. We extract the effective order parameter for superconductivity from the square-root of the long-distance behaviour (dashed line).}
		\label{fig_app:afqmc_gs_corr}
	\end{figure}
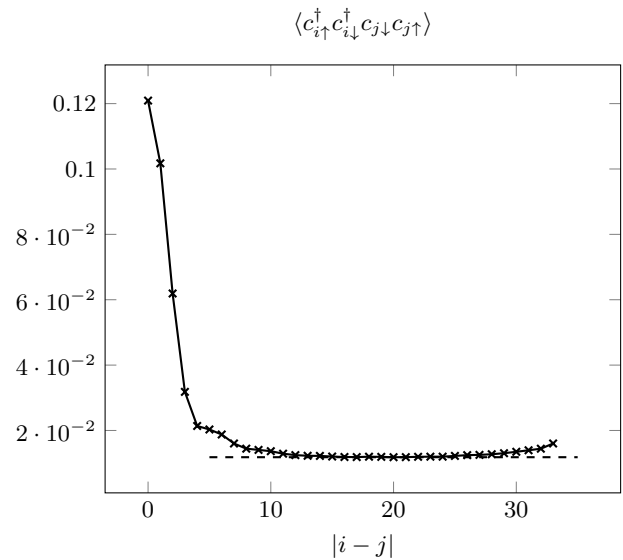
	Here we briefly describe our ground-state AFQMC calculations and how the superconducting order parameters are obtained. A comprehensive discussion of AFQMC can be found in Refs.\,[\onlinecite{Shiwei2019},\onlinecite{Assaad2002}].

The ground state AFQMC method solves the Schr\"odinger equation of the quantum many body problem by projecting out the ground state wave function $\ket{\Psi_G}$ of the system from an initial wave function $\ket{\Psi_I}$:
\begin{equation}
    e^{-\beta\hat{H}}\ket{\Psi_I}\propto\ket{\Psi_{\beta}}\rightarrow_{\beta\rightarrow\infty}\ket{\Psi_G}.
    \label{eq:GSproj}
\end{equation}
The initial wave function is generally obtained from a mean field calculation, for example, with the Hartree-Fock method.  
When the imaginary time $\beta$ is sufficiently large, the projected wave function $\ket{\Psi_{\beta}}$ approaches the 
ground state  $\ket{\Psi_G}$
of $\hat H$.

Numerically, the propagator $e^{-\beta\hat{H}}$ is rewritten in a one body form. This is achieved by first discretizing the imaginary time 
into small 
time steps $\Delta\tau$,
\begin{equation}
    e^{-\beta\hat{H}}=(e^{-\Delta\tau\hat{H}})^m.
\end{equation}
In this work, $\Delta\tau$ is set to be 0.01 for the ground state algorithm. We have verified this to give Trotter errors well within our statistical error in the final results. Then we apply Trotter-Suzuki breakup for each small imaginary step,
\begin{equation}
   e^{-\Delta\tau\hat{H}}=e^{-\Delta\tau\hat{K}/2}e^{-\Delta\tau\hat{V}}e^{-\Delta\tau\hat{K}/2}+\mathcal{O}\left(\Delta\tau^3\right)
\end{equation}
where $\hat{K}$ is the kinetic part of the Hamiltonian, containing only one-body terms, while $\hat{V}$ is the potential part, which consists of two-body terms.

To rewrite all two-body terms into one-body terms, we apply
the Hubbard-Stratonovich~(HS) transformation in a charge decomposition form:
\begin{equation}\label{eq:a4}
\begin{split}
    e^{\Delta Un_{i\uparrow}n_{i\downarrow}} &= \frac{1}{2}\sum_{x_i=\pm 1}e^{(\gamma x_i-\Delta\tau U/2)(n_{i\uparrow}+n_{i\downarrow}-1)}\\
    &\equiv\frac{1}{2}\sum_{x_i=\pm 1}\hat{b}_i(x_i),
\end{split}
\end{equation}
where $\cosh(\gamma)=\exp(-\Delta\tau U/2)$.
With the above transformation, the short-time  propagator can be written as
\begin{equation}
    e^{-\Delta\tau\hat{H}}= \int d\textbf{x}\,\,p(\textbf{x})\hat{B}(\textbf{x}),
\end{equation}
where $\hat{B}(\textbf{x})=e^{-\Delta\tau\hat{K}/2}\prod_i\hat{b}_i(x_i)\,e^{-\Delta\tau\hat{K}/2}$ is now a one-body propagator, 
and $p(\textbf{x})$ is a probability density function, which in the form above
is uniform in the auxiliary-field (AF) configurations 
$\textbf{x}=\{x_1,x_2,...,x_{N_l}\}$.

Ground state observables are given by
\begin{equation}\label{eq:a6}
 \langle\hat{O}\rangle=\frac{\bra{\Psi_L}\,\hat{O}\,\ket{\Psi_R}}{\langle \Psi_L|\Psi_R\rangle}\,,
\end{equation}
where 
$\bra{\Psi_L}=\bra{\Psi_I}\,\text{exp}(-\beta_L\hat{H})$ and 
$\ket{\Psi_R} = \text{exp}(-\beta_R\hat{H})\,\ket{\Psi_I}$. 
In our calculations with the generalized Metropolis algorithm, we choose a total projection time $\beta$,
which 
defines a fixed length of the imaginary-time path integral. The location along the path where $\hat{O}$ is measured moves with our sampling; for example, as we sweep from left to right, we start measuring when $\beta_L>\beta_{\rm eq}$, where $\beta_{\rm eq}< \beta/2$ is a parameter which ensures that
the asymptotic limit in Eq.~(\ref{eq:GSproj}) is reached (in a numerical sense), and $\beta_R\equiv \beta-\beta_L$. Conversely, when we sweep from right to left, the measurement starts when $\beta_R>\beta_{\rm eq}$ and stops when $\beta_L\equiv \beta-\beta_R$ reaches $\beta_{\rm eq}$.
The expectation $\langle O\rangle$ is expressed as path integrals in AF space:
\begin{equation}\label{eq:PI-AF}
 \langle\hat{O}\rangle=\frac{\int  \frac{\bra{\phi_L}\,\hat{O}\,\ket{\phi_R}}
 {\langle \phi_L|\phi_R\rangle}\,P(X)\,\langle \phi_L|\phi_R\rangle\,dX}
 {\int P(X)\,\langle \phi_L|\phi_R\rangle\,dX}\,.
\end{equation}
In Eq.~(\ref{eq:PI-AF}), $X\equiv \{ \textbf{x}^{(M)}, \textbf{x}^{(M-1)}, \cdots, \textbf{x}^{(2)},\textbf{x}^{(1)} \}$, which is an $MN_l$-dimensional vector, denotes the AF configuration of the  entire path, with $M\equiv \beta/\Delta\tau$ being the number of time slices in the path,
and the probability function $P(X)=\prod_{m=1}^M p(\textbf{x}^{(m)})$.
The wave functions $\ket{\phi_R}$ and  $\ket{\phi_L}$ are single Slater determinants (if we choose $\ket{\Psi_I}$ to be a single Slater determinant), and have the form:
$\ket{\phi_R} = \prod_{m=1}^{M_R}B(\textbf{x}^{(m)})\,\ket{\Psi_I}$ and,
correspondingly, 
$\bra{\phi_L} = \bra{\Psi_I} \prod_{m=1}^{M_L}B(\textbf{x}^{(M-m+1)})$,
with $M_R\equiv \beta_R/\Delta\tau$ and $M_L\equiv \beta_L/\Delta\tau$.
A heat-bath like algorithm and a cluster update scheme are incorporated in our generalized Metropolis algorithm, which is  
described in detail in Appendix A of Ref. [\onlinecite{HaoPRA2015}].

The pair-correlator (the pair-pair correlation) can be computed by the path-integral above in Eq.~(\ref{eq:PI-AF}). For each path, if we denote the expectation value
$\bra{\phi_L}\,\hat{O}\,\ket{\phi_R}/
 \langle \phi_L|\phi_R\rangle$ by 
 $\langle \hat{O}\rangle_{L,R} $, then 
$\langle c_{i\uparrow}^{\dagger}c_{i\downarrow}^{\dagger}c_{j\downarrow}c_{j\uparrow}\rangle_{L,R}$ can be decomposed by Wick's theorem into pair products of 
one-body Green's functions:
\begin{flalign}
    \nonumber\langle c_{j\sigma}^{\dagger}c_{i\sigma}\rangle_{L,R}&\equiv\frac{\bra{\phi_{L}}c_{j\sigma}^{\dagger}c_{i\sigma}\ket{\phi_R}}{\langle \phi_L|\phi_R\rangle}\\&=[\Phi_R^{\sigma}[(\Phi_L^{\sigma})^{\dagger}\Phi_R^{\sigma}]^{-1}(\Phi_L^{\sigma})^{\dagger}]_{ij}\,,
\end{flalign}
where $\Phi_L$ and $\Phi_R$ are the matrix representation of the kets $\ket{\phi_L}$ and $\ket{\phi_R}$, respectively.

We choose a reference site and then compute the pair-correlator between the reference site and different other lattice sites, as is shown in Fig. \ref{fig_app:afqmc_gs_corr}.
We then average the pair-correlation of the sites that have the longest distance. To further reduce statistical error, the reference point is averaged over the whole lattice, since each lattice site is equivalent under periodic boundary condition.
	\subsection{Kosterlitz-Thouless transition temperature}
	For finite temperature results we use ALF. The structure of the algorithm is similar to that of ground state AFQMC previously outlined and we refer the reader to the ALF documentation~\cite{Assaad2021}.
	
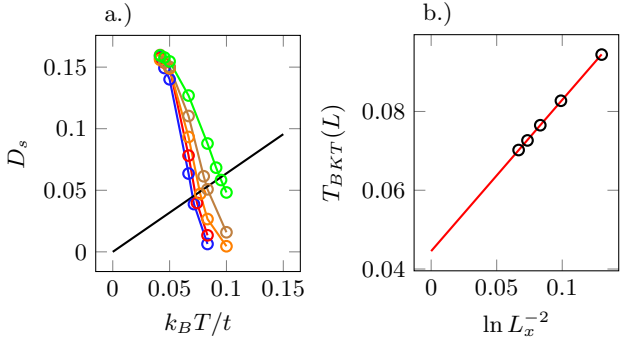
\begin{figure}
    	\ifthenelse{\boolean{buildtikzpics}}%
    	{%
		\tikzsetnextfilename{AFQMCSuperWeight}
		\tikzset{external/export next=true}
		\begin{tikzpicture}
			\begin{axis}
			[
				xlabel = {$k_BT/t$},
				ylabel = {$D_s$},
				legend pos = north east,
				xticklabel style = {/pgf/number format/precision=3,
									/pgf/number format/fixed},
				yticklabel style = {/pgf/number format/precision=3,
									/pgf/number format/fixed},
				width = 0.24\textwidth,
				height=0.2\textheight,
				title = a.),
				every axis title/.style={above right, at={(0,1)}},
			]
			\addplot
			[
				mark=o,
				thick,
				blue,
			]
			table
			[
				y expr = \thisrowno{1},
				x expr = \thisrowno{0}
			]
			{Data/SC_weight_U_-4.0_tp_0.2_mu_-0.6533_Lx_48.dat};
			\addplot
			[
			mark=o,
			thick,
			red,
			]
			table
			[
			y expr = \thisrowno{1},
			x expr = \thisrowno{0}
			]
			{Data/SC_weight_U_-4.0_tp_0.2_mu_-0.6533_Lx_40.dat};
			\addplot
			[
			mark=o,
			thick,
			orange
			]
			table
			[
			y expr = \thisrowno{1},
			x expr = \thisrowno{0}
			]
			{Data/SC_weight_U_-4.0_tp_0.2_mu_-0.6533_Lx_32.dat};
			\addplot
			[
			mark=o,
			thick,
			brown
			]
			table
			[
			y expr = \thisrowno{1},
			x expr = \thisrowno{0}
			]
			{Data/SC_weight_U_-4.0_tp_0.2_mu_-0.6533_Lx_24.dat};
			\addplot
			[
			mark=o,
			thick,
			green
			]
			table
			[
			y expr = \thisrowno{1},
			x expr = \thisrowno{0}
			]
			{Data/SC_weight_U_-4.0_tp_0.2_mu_-0.6533_Lx_16.dat};
			\addplot
			[
				thick,
				black,
				domain=0:0.15
			]
			{linearFct(2/3.14,0)};
			\end{axis}
		\end{tikzpicture}
    	}%
    	{%
    		\includegraphics{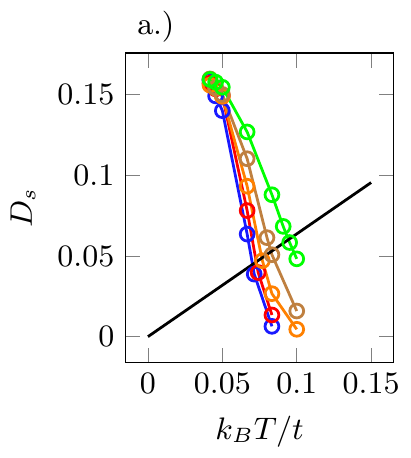}%
    	}%
    	\ifthenelse{\boolean{buildtikzpics}}%
    	{%
		\tikzsetnextfilename{AFQMCLogLFit}
		\tikzset{external/export next=true}
		\begin{tikzpicture}
		\begin{axis}
		[
		xlabel = {$\ln L_x^{-2}$},
		ylabel = {$T_{BKT}(L)$},
		scaled x ticks = false,
		scaled y ticks = false,
		xticklabel style = {/pgf/number format/precision=3,
			/pgf/number format/fixed},
		yticklabel style = {/pgf/number format/precision=3,
			/pgf/number format/fixed},
		width = 0.24\textwidth,
		height= 0.2\textheight,
		title = b.),
		every axis title/.style={above right, at={(0,1)}},
		]
		\pgfplotstableread[header=false]{Data/logL_sq_fit_SC_weight_fit1_U_-4.0_tp_0.2_mu_-0.6533.dat}{\fittable}
		\pgfplotstablegetelem{1}{[index]1}\of\fittable
		\edef\insec{\pgfplotsretval}
		\pgfplotstablegetelem{1}{[index]0}\of\fittable
		\edef\slope{\pgfplotsretval}
		\addplot
		[
		thick,
		only marks,
		mark = o
		]
		table
		[
		y expr = \thisrowno{1},
		x expr = \thisrowno{0}
		]
		{Data/logL_sq_fit_SC_weight_U_-4.0_tp_0.2_mu_-0.6533.dat};
		\addplot
		[
		thick,
		red,
		domain=0:\eval{1/(ln(16)*ln(16))},
		]
		{linearFct(\slope,\insec)};
		\end{axis}
		\end{tikzpicture}
    	}%
    	{%
    		\includegraphics{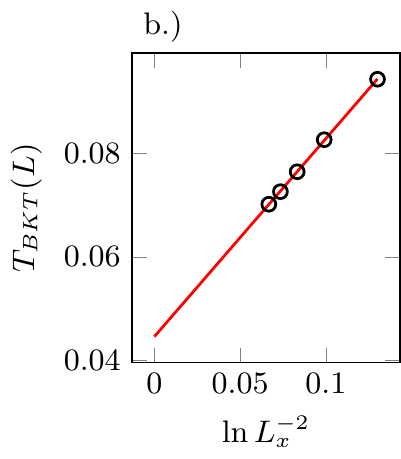}%
    	}%
		\caption{Strategy for obtaining $T_{BKT}$ for the parameters $U=-4t$, $t_\perp\approx0.28t$ and $\mu=-0.6533t$. In sub-figure \textit{a.} an intersection of $D_s$ and $2T/\pi$ indicating $T_{BKT}(L)$ and sub-figure \textit{b.} the thermodynamic limit of $T_{BKT}$ obtained through \cref{fsize_TKT}. The aspect ratio is chosen such that $L_y=L_x/4$.}
		\label{fig_app:afqmc_tkt_strat}
	\end{figure}

	Using ALF we obtain results for any value of imaginary time $\beta$. From linear response theory it is possible to relate the superfluid weight to current-current correlators and kinetic energy~\cite{Scalapino1993}:
	\begin{equation}
		\frac{D_s}{4e^2} = \braket{-k_x}-\Lambda_{xx}\left(q_x=0,q_y\to0,i\omega_m=0\right),
	\end{equation}
	\begin{equation}
		0 = \braket{-k_x}-\Lambda_{xx}\left(q_x\to0,q_y=0,i\omega_m=0\right).
	\end{equation}
	For the KT-transition we expect that~\cite{Nelson1977}
	\begin{equation}
		\lim_{T\to T_{BKT}^-}\frac{T}{D_s} = \frac{\pi}{2}.
	\end{equation}
	Thus, the straight line in $D_s-T$ space
	\begin{equation}
		D_s = \frac{2T}{\pi}
	\end{equation}
	intersects at $T=T_{BKT}$. An example of this is shown in \cref{fig_app:afqmc_tkt_strat}a.
	This calculation is performed for each size of the system, yielding a trend of the finite size $T_{BKT}$ which is then extrapolated to the thermodynamic limit using the form~\cite{Bramwell1993}
	\begin{equation}\label{fsize_TKT}
		T_{BKT}(L) = T_{BKT}+\frac{A}{\ln L^2}.
	\end{equation}
	An example of this fit is shown in \cref{fig_app:afqmc_tkt_strat}b.
	\subsection{Density}
	The AFQMC finite temperature algorithm from the ALF collaboration uses the Blancenbecler-Scalapino-Sugar (BSS) algorithm~\cite{Assaad2021, Blankenbecler1981}. This algorithm works in the grand canonical ensemble as is necessary and will not have fixed density. In general, we are interested in specifying the density to work at as $T_{BKT}$ will have some dependence on this quantity.
	
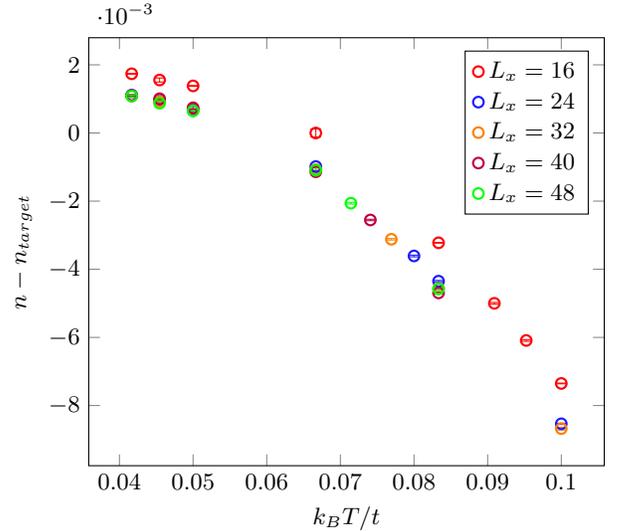
\begin{figure}
    	\ifthenelse{\boolean{buildtikzpics}}%
    	{%
		\tikzsetnextfilename{AFQMCDensOverT}
		\tikzset{external/export next=true}
		\begin{tikzpicture}
			\begin{axis}
			[
				xlabel = {$k_BT/t$},
				ylabel = {$n-n_{target}$},
				legend pos = north east,
				xticklabel style = {/pgf/number format/precision=3,
									/pgf/number format/fixed},
			]
			\addplot+
			[
				thick,
				only marks,
				mark=o,
				red,
				error bars/.cd,
				y dir = both, y explicit
			]
			table
			[
				y expr = \thisrowno{1}-0.5,
				x expr = \thisrowno{0},
				y error expr = \thisrowno{2}
			]
			{Data/afqmc_dens_vs_T_tperp_0.2_L_16_mu_-0.6533.dat};
			\addlegendentry{$L_x=16$}
			\addplot+
			[
			thick,
			only marks,
			mark=o,
			blue,
			error bars/.cd,
			y dir = both, y explicit
			]
			table
			[
			y expr = \thisrowno{1}-0.5,
			x expr = \thisrowno{0},
			y error expr = \thisrowno{2}
			]
			{Data/afqmc_dens_vs_T_tperp_0.2_L_24_mu_-0.6533.dat};
			\addlegendentry{$L_x=24$}
			\addplot+
			[
			thick,
			only marks,
			mark=o,
			orange,
			error bars/.cd,
			y dir = both, y explicit
			]
			table
			[
			y expr = \thisrowno{1}-0.5,
			x expr = \thisrowno{0},
			y error expr = \thisrowno{2}
			]
			{Data/afqmc_dens_vs_T_tperp_0.2_L_32_mu_-0.6533.dat};
			\addlegendentry{$L_x=32$}
			\addplot+
			[
			thick,
			only marks,
			mark=o,
			purple,
			error bars/.cd,
			y dir = both, y explicit
			]
			table
			[
			y expr = \thisrowno{1}-0.5,
			x expr = \thisrowno{0},
			y error expr = \thisrowno{2}
			]
			{Data/afqmc_dens_vs_T_tperp_0.2_L_40_mu_-0.6533.dat};
			\addlegendentry{$L_x=40$}
			\addplot+
			[
			thick,
			only marks,
			mark=o,
			green,
			error bars/.cd,
			y dir = both, y explicit
			]
			table
			[
			y expr = \thisrowno{1}-0.5,
			x expr = \thisrowno{0},
			y error expr = \thisrowno{2}
			]
			{Data/afqmc_dens_vs_T_tperp_0.2_L_48_mu_-0.6533.dat};
			\addlegendentry{$L_x=48$}
			\end{axis}
		\end{tikzpicture}
    	}%
    	{%
    		\includegraphics{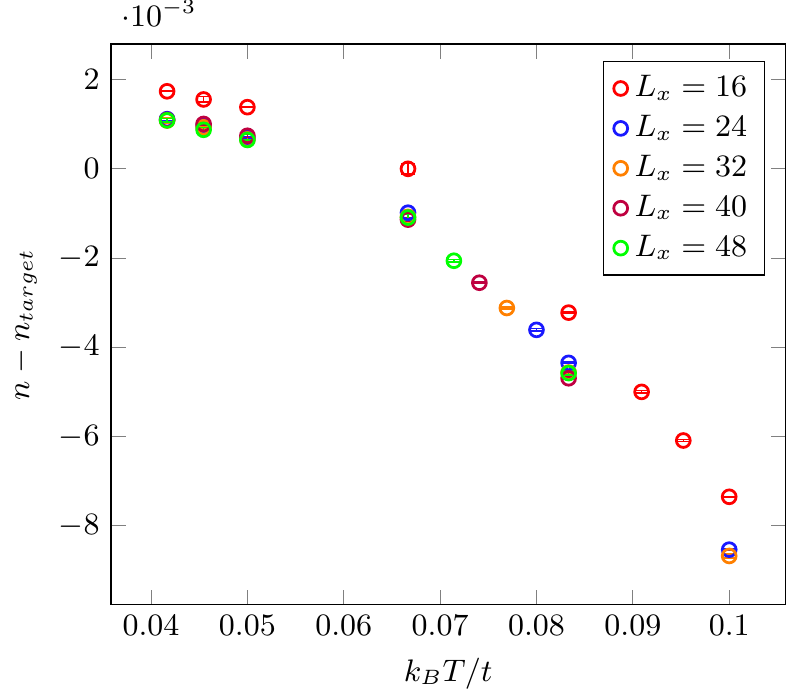}%
    	}%
		\caption{The density over temperature for a given chemical potential $\mu=-0.6533t$ at parameters $U=-4t$ and $t_\perp\approx0.28t$. The aspect ratio is chosen such that $L_y=L_x/4$.}
		\label{fig_app:afqmc_density_over_T}
	\end{figure}

	Precisely fixing the density requires simulation of a large number of chemical potential values. In order to alleviate this problem we run simulations for a small number of lattice points and determine the correct chemical potential for a given temperature. We then use this value of chemical potential for all lattice sizes and temperatures of that parameter set. This will yield a notable error in density as shown in \cref{fig_app:afqmc_density_over_T}. At the same time we find that $T_{BKT}$ is only modestly affected by density.
	\section{\label{App::Ladder_SMF_model}Array of Hubbard ladders}
	We derive the effective 1D model for the weakly coupled Hubbard ladders starting from the 3D array given by
	\begin{equation}
	H_{\mathrm{3D}} = \sum_{kl}H_{\mathrm{HL}}(\textbf{R}_{kl}) + t_\perp H_\perp,
	\end{equation}
	where
	\begin{flalign}
	\nonumber H_{\mathrm{HL}}(\textbf{R}_{kl}) =\\\nonumber -t\sum_{i=1}^{L-1}\sum_{j=0}^1\sum_\sigma&\Bigl(c^\dagger_{i+1,j\sigma}(\textbf{R}_{kl}) c_{ij\sigma}(\textbf{R}_{kl}) \\\nonumber&+ c^\dagger_{ij\sigma}(\textbf{R}_{kl}) c_{i+1,j\sigma}(\textbf{R}_{kl})\Bigr) \\\nonumber - t'\sum_{i=1}^L\sum_\sigma &\Bigl(c^\dagger_{i1\sigma}(\textbf{R}_{kl}) c_{i0\sigma}(\textbf{R}_{kl}) \\\nonumber&+ c^\dagger_{i0\sigma}(\textbf{R}_{kl}) c_{i1\sigma}(\textbf{R}_{kl})\Bigr) \\\nonumber&-\mu \sum_{ij}n_{ij}(\textbf{R}_{kl}) \\& + U\sum_{ij} n_{ij\uparrow}(\textbf{R}_{kl}) n_{ij\downarrow}(\textbf{R}_{kl}) 
	\end{flalign}
	The vector, $\textbf{R}_{kl}$, denotes the position of the ladder in a 2D grid. We have used the definition ($\textbf{R}_{kl}$ suppressed)
	\begin{equation}
	n_{ij} = \sum_\sigma c^\dagger_{ij\sigma}c_{ij\sigma} = \sum_\sigma n_{ij\sigma},
	\end{equation}
	and $c_{ij\sigma}$ follow the usual anti-commutation relations.
	
	So far, we have only described a set of Fermi-Hubbard ladders. The added Hamiltonian $H_\perp$ is defined by
	\begin{flalign}
	\nonumber H_\perp = -\sum_{il}^L\sum_\sigma\sum_{k=1}^{L-1}&\Bigl[c^\dagger_{i1\sigma}(\textbf{R}_{kl}) c_{i0\sigma}(\textbf{R}_{k+1,l}) \\\nonumber&+ c^\dagger_{i0\sigma}(\textbf{R}_{k+1,l}) c_{i1\sigma}(\textbf{R}_{kl})\Bigr] \\\nonumber - \sum_{ik}^L\sum_j\sum_\sigma\sum_{l=1}^{L-1}&\Bigl[c^\dagger_{ij\sigma}(\textbf{R}_{kl}) c_{ij\sigma}(\textbf{R}_{k,l+1}) \\\nonumber&+ c^\dagger_{ij\sigma}(\textbf{R}_{k,l+1}) c_{ij\sigma}(\textbf{R}_{kl})\Bigr].
	\end{flalign}
	Note that movement to neighbouring ladders may change what leg you are on. This is due to half of neighbouring ladders being side-by-side neighbours and the other half are front-and-back neighbours.
	
	\subsection*{Effective Hubbard ladder Hamiltonian}
	When $U$ is strongly repulsive and density is close to unit filling $H_{\mathrm{HL}}$ and thus also the full set of ladders have a spectrum which contains clusters of energy eigenstates separated by large gaps. Thus, analogous to \cref{eff_ham_nufhc} it is possible to derive an effective Hamiltonian
	\begin{equation}
	H_{\mathrm{3D}}^{\mathrm{eff}} = \sum_{kl}H_{\mathrm{HL}}(\textbf{R}_{kl}) - \frac{t_\perp^2}{\Delta E_p}P_0H_\perp^2P_0,
	\end{equation}
	where 
	\begin{equation}
	\Delta E_p = 2E(N+1,1/2) - E(N,0) - E(N+2,0),
	\end{equation}
	is the pairing energy for a single ladder at particle number $N$ and $E(N, S)$ is the energy of a ladder at particle number $N$ and spin $S$. The operator $P_0$ is a projector to the lowest energy manifold of the total system. As in \cref{Sec::Model/sub::pert_theo}, this removes certain terms within $H_\perp^2$, such as two particles moving to two separate ladders.
	
	Expanding $P_0H_\perp^2P_0$ yields a new operator which is quartic and acts like an effective interaction. Each interaction involves particles on two different ladders, e.g., moving two particles from one ladder to an adjacent one.
	
	\subsection*{Mean-field Hamiltonian}
	With a quartic interaction where half the operators involve one ladder and the other half involve the other we can make an ansatz of quasi-free states:
	\begin{equation}
	\braket{c^\dagger_ic^\dagger_jc_kc_l} = \braket{c^\dagger_ic^\dagger_j}\braket{c_kc_l} + \braket{c^\dagger_ic_l}\braket{c^\dagger_jc_k} - \braket{c^\dagger_ic_k}\braket{c^\dagger_jc_l}.
	\end{equation}
	This allows us to create a mean-field Hamiltonian which produces expectation values of this form. In the process we assume that expectation values involving operators on different ladders are of negligible size and ignore them. This leads to the mean-field Hamiltonian (for one ladder)
	\begin{equation}
	H_{\mathrm{MF}} = H_{\mathrm{HL}} - H_{\mathrm{pair}} - H_{PH},
	\end{equation}
	where
	\begin{align}
	H_{\mathrm{pair}} = \sum_{ii',jj'}\alpha_{ii',jj'}\left(c^\dagger_{ij\downarrow} c^\dagger_{i'j'\uparrow} + c_{i'j'\uparrow} c_{ij\downarrow} \right), \\
	\label{lad_hph}H_{PH} = -\sum_{ii',jj',\sigma} \beta_{ii',jj',\sigma}c^\dagger_{ij\sigma}c_{i'j'\sigma},
	\end{align}
	and the pairing amplitudes are given by
	\begin{gather}
	\alpha_{ii',00} = \frac{2t_\perp^2}{E_p}\left(\braket{c_{i'1\uparrow}c_{i1\downarrow}} + 2\braket{c_{i'0\uparrow}c_{i0\downarrow}}\right) \\
	\alpha_{ii',11} = \frac{2t_\perp^2}{E_p}\left(\braket{c_{i'0\uparrow}c_{i0\downarrow}} + 2\braket{c_{i'1\uparrow}c_{i1\downarrow}}\right) \\
	\alpha_{ii',10} = \frac{4t_\perp^2}{\Delta E_p}\braket{c_{i'0\uparrow}c_{i1\downarrow}} \\
	\alpha_{ii',01} = \frac{4t_\perp^2}{\Delta E_p}\braket{c_{i'1\uparrow}c_{i0\downarrow}},
	\end{gather}
	whereas the exchange terms are given by
	\begin{gather}
	\beta_{ii',00,\sigma} = \frac{2t_\perp^2}{E_p}\left(\braket{c^\dagger_{i1\sigma}c_{i'1\sigma}} + 2\braket{c^\dagger_{i0\sigma}c_{i'0\sigma}}\right), \\
	\beta_{ii',11,\sigma} = \frac{2t_\perp^2}{E_p}\left(\braket{c^\dagger_{i0\sigma}c_{i'0\sigma}} + 2\braket{c^\dagger_{i1\sigma}c_{i'1\sigma}}\right), \\
	\beta_{ii',10,\sigma} = \frac{4t_\perp^2}{E_p}\braket{c^\dagger_{i'0\sigma}c_{i1\sigma}}, \\
	\beta_{ii',01,\sigma} =\frac{4t_\perp^2}{E_p}\braket{c^\dagger_{i'1\sigma}c_{i0\sigma}}.
	\end{gather}
	Note that the hermiticity of \cref{lad_hph} is not apparent from the expression but is hidden in the sum.

    \section{Analytical $T_c$ and gap
    $\Delta_\rho(T=0)$\label{App:critical_temperature_and_charge_gap_analitycal}}
	
	We give here details on how to explicitly compute the value of the ratio between $T_c$ and the charge gap $\Delta_\rho(T=0)$ when there is only one massless sector. For $T_c$, we study how the mean-field order parameter approaches zero above the critical temperature $T>T_c$. For the gap, the effective Hamiltonian is sine-Gordon-like and its spectrum has been largely studied in the literature. We thus rely on the exact solution of the gap and the perturbative one of $T_c$.
	
	The critical temperature can be estimated by noticing that close to $T \sim T_c$ the order parameter approaches zero. As stated in \cite{Cazalilla2006a}, we can expand the right-hand side of the self-consistent condition, \cref{order_parameter_analytical}, in powers of the (real) order parameter itself $\Psi_c = \langle \Psi^\dagger \rangle =  \langle \psi ^\dagger_{ \uparrow} (x) \psi ^\dagger_{\downarrow} (x) \rangle$. In the path-integral formalism, the expansion in power of $\Psi_c(T \simeq T_c) \ll 1$ of the average reads
\begin{equation}
    \begin{aligned}
        \Psi_c & = \frac{1}{\mathcal{Z}_{\theta_\rho}}\int \, \mathcal{D}\theta_\rho \, e^{-\mathcal{S}_{\theta_\rho}^{(0)} + \Bar{t}_\perp \Psi_c \int d\textbf{r} (\Psi(\textbf{r}) + \Psi^{\dagger}(\textbf{r}))} \Psi(\textbf{r}') \\ 
        & \simeq \frac{\int \mathcal{D} \theta_\rho \, e^{-\mathcal{S}_{\theta_\rho}^{(0)} }e^{\Bar{t}_\perp \Psi_c \int d\textbf{r} (\Psi(\textbf{r}) + \Psi^{\dagger}(\textbf{r}))} \Psi(\textbf{r}') }{ \mathcal{Z}_{\theta_\rho}^{(0)} \left (1+ \mathcal{O}(\Psi_c\right)) }  \\
        & \simeq \langle e^{\Bar{t}_\perp \Psi_c \int d\textbf{r} (\Psi(\textbf{r}) + \Psi^{\dagger}(\textbf{r}))} \Psi(\textbf{r}') \rangle_0   \\
        & = \Bar{t}_\perp \Psi_c \int d\textbf{r} \, \langle \Psi^{\dagger}(\textbf{r}) \Psi(0, 0) \rangle_0 + \mathcal{O}(\Psi_c^2)
    \end{aligned}
\end{equation}
	where the integration over $\mathcal{D}\theta_\rho$ stands for averaging over all possible configurations and $\textbf{r}=(x, \tau)$, which means that $\int d\textbf{r} = \int_0^L dx \, \int _0^\beta d\tau$, with $\tau$ the imaginary time. The letter $\mathcal{S}$ denotes the action and $\mathcal{Z}$ the partition function while the superscript $"0"$ refers to the quadratic component of the mean-field Hamiltonian. The effective coupling is $\Bar{t}_\perp =\frac{ t_{\perp}^2}{\Delta_\sigma} z_c$.
	If we neglect second order terms, the resulting equation reads
	\begin{equation}
	    1 + \Bar{t}_\perp g_1^R(k=\omega=0,T \simeq T_c)=0
	\end{equation}
    with $g_1^R$ being the zero component Fourier transform of the retarded correlation function. It is defined as
    \begin{equation}
        \begin{aligned}
            g_1^R(\textbf{r},T) &=  -\langle \mathcal{T}_\tau \Psi^{ }(\textbf{r}) \Psi^\dagger(0,0)\rangle_0\\
            &= - (A_F \rho_0 C)^2 \langle \mathcal{T}_\tau e^{i\sqrt{2}\theta_\rho (\textbf{r})} e^{-i\sqrt{2}\theta_\rho (0)}\rangle_0\\
            &= - (A_F \rho_0 C)^2 e^{-\langle[\theta_\rho(\textbf{r})-\theta(0)]^2\rangle _0}
        \end{aligned}
    \end{equation}
	where we write explicitly the imaginary time ($\tau$) ordering operator $\mathcal{T}_\tau$. In the thermodynamical limit, such averages $\langle \dots \rangle_0$ are well known quantities and can be exactly computed in 1D systems for quadratic Hamiltonians \cite{Giamarchi2003}. By analytically continuing to real time $\tau = it+\epsilon(t)$, with $\epsilon(t)=\text{sgn}(t)\epsilon$, the critical temperature is
	\begin{widetext}
    \begin{equation}\label{scaling_eq_Tc}
    T_c = \left [\frac{t_{\perp}^2}{2\Delta_\sigma} C^2 z_c \frac{(\rho_0 A_F)^2}{u_\rho} \sin \left ( \frac{\pi}{2 K_\rho} \right )  \left (\frac{2 \pi \alpha }{u_\rho }\right)^{ \frac{1}{K_\rho}} \left (  \frac{u_\rho}{2 \pi}  \right)^2 B^2 \left (\frac{1}{4K_\rho}, 1 - \frac{1}{2K_\rho} \right ) \right]^{\frac{K_\rho}{ 2K_\rho-1}}
    \end{equation}
	\end{widetext} 
	with $B(x,y)= \Gamma(x)\Gamma(y)/\Gamma(x+y)$ the beta function and $A_F$ a prefactor that depends on the specific microscopic model. The choice of the cut-off $\alpha$ is arbitrary but should be sufficiently small so that the spectrum can be linearized. Moreover, the non-universal constant $A_F$ is such that the final result is cut-off independent.
	
    The charge gap is instead computed by noticing that the effective model is expressed by a sine-Gordon Hamiltonian. By using variational method, thus approximating $\cos \theta \sim 1 - \frac{1}{2}\theta^2$ and computing the action \cite{Giamarchi2003}, we have that the spectrum is gapped and behaves as $E(k) =\sqrt{(uk)^2 + \Delta_{\text{var}}^2}$. Even if the variational method gives the right scaling for the gap $\Delta_\rho ( t_\perp)$, we use here the exact formula \cite{Lukyanov_gap_sine_gordon} so we have also the exact prefactors
	\begin{widetext}
    	\begin{equation}\label{scaling_eq_charge_gap}
        \Delta_\rho(T=0) = u_\rho \left [ (\rho_0 A_F)^2 \alpha^{\frac{1}{K_\rho}} \frac{K_\rho / 2}{\kappa^2(K_\rho/2)(4K_\rho-1)} \tan{\left ( \frac{\pi}{2} \frac{1}{4K_\rho-1}\right )} \frac{z_c}{u_\rho} \frac{ t_{\perp}^2 C^2}{2\Delta_\sigma} \right ]^{\frac{K_\rho}{2K_\rho-1}} \sin \left ( \frac{\pi}{4K_\rho-1}\right )
        \end{equation}
    \end{widetext}
	with $\kappa (K)$ a combination of gamma functions $\Gamma(K)$
    \begin{equation} \label{combination_gamma_functions_ratio}
    \kappa (K) = \frac{1}{2 \sqrt{\pi}}\frac{\Gamma \left ( \frac{1}{8K}\right ) \Gamma \left ( \frac{1}{2} \frac{8K}{8K-1}\right ) }{\Gamma \left (1-  \frac{1}{8K}\right )\Gamma \left ( \frac{1}{2}\frac{1}{8K-1}\right )}
    \end{equation}
	Notably, the unknown constants cancel out if we consider the ratio \cref{analytic_fermionic_ratio}. Moreover, this value depends only on the interaction $U$, which modifies the Luttinger parameter $K_\rho$. We recall that this result is valid as long as $t_\perp$ is the smallest energy scale of the problem.
	
	\section{\label{App::Renormalization_group}Renormalization group theory }
	
	In this section, we give details on the Renormalization Group (RG) procedure implemented to describe an effective 1D Hubbard chains when interactions are small and the pair size is much larger than the microscopic cutoff (``lattice spacing''). The idea is to integrate out all degrees of freedom corresponding to energies between the bandwidth $W$ (corresponding to microscopic cutoff $\alpha_0$) and the spin gap $\Delta_\sigma$ (corresponding to the pair size $\xi_\sigma = u_\sigma / \Delta_\sigma$).  It is important to underline that the RG treatment is only valid when $\Delta_\sigma < W$. 
    
    \subsection{First step RG: renomalization of charge and spins}	
	In order to find the RG equations, we compute perturbatively (in the couplings) the correlation $ \langle \psi^\dagger (x)\psi(x)\rangle $ \cite{Jose_RG_equations_from_correlations_perturbaive,Giamarchi2003}. Let us start from the Luttinger parameter $K_\sigma$ and the dimensionless interaction $g = \frac{U}{\pi v_F}$, with $v_F$ the Fermi velocity
	\begin{equation}\label{RG_eq_K_sigma_vs_interaction}
	    \begin{aligned}
	         \frac{dK_\sigma(l)}{dl} &= - \frac{K_\sigma^2(l) g^2(l)}{2} \\
	         \frac{dg(l)}{dl} &= (2-2K_\sigma(l))g(l)
	    \end{aligned}
	\end{equation}
	Because the system is spin-isotropic, the equations are equivalent to the ones from the XY problem \cite{Kosterlitz_XYmodel_critical}. For $U<0$, we flow towards larger $g$ and smaller $K_\sigma$ and we need to stop the flow when $g \propto \mathcal{O}(1)$, say at the RG length $l_1$. This fictitious length is defined from $\alpha(l) = \alpha(l=0) e^l$ with $\alpha (l=0)$ the original cut-off (lattice spacing).
	
	In presence of interchain tunnelling we need to complete the above equations by the ones generated by the interchain tunnelling:
	\begin{equation}\label{RG_eq_tperp_pairhopping}
	    \begin{aligned}
	        \frac{dt_\perp(l)}{dl} &= \left (2- \frac{\tilde{K}_\rho + \tilde{K}_\sigma(l)}{4}  \right) t_\perp(l), \\
	        \frac{d\tilde{J}(l)}{dl} &= \left(2- \frac{1}{K_\rho}-K_\sigma(l) \right)\tilde{J}(l) + \tilde{J}_s(l),
	    \end{aligned}
	\end{equation}
	where $\tilde{K}_\nu = K_\nu + {K_\nu}^{-1}$.  The dimensionless couplings are defined as
    \begin{align}
        \tilde{J} = \frac{\pi \alpha^2}{4 u_\rho}  ( \rho_0 A_F)^2    J  \quad \text{ and }\quad   \tilde{J}_s =  \frac{\alpha^2}{2 u_\rho^2} t^2_\perp
    \end{align}
	and the subscript $s$ stands for source term. Note that the transverse hopping is also, in principle, contributing to the renormalization of the other parameters \cref{RG_eq_K_sigma_vs_interaction} and of $K_\rho$. In practice, because we consider that the interchain tunnelling is the smallest quantity in the problem and in particular that $t_\perp \ll \Delta_\sigma$, we will neglect such renormalization in the first step of the RG. In particular $K_\rho$ can be considered as essentially constant in the first step of the RG. Note also that the combination of interchain hopping and interactions lead to a modification of the naive scaling of the interchain tunnelling. In addition to its own renormalization, the single particle interchain tunnelling also generates via RG the \emph{pair} tunnelling. This is due to the fact that pairs of electrons that hop within a distance $|r_1-r_2|<\alpha(l)$ are to be considered local. Moreover, we need to enforce the condition $\tilde{J}(l=0)=0$ because the original Hamiltonian \cref{fullham} has only single-particle hopping, not pair hopping. It is clear that, as we renormalize, the pair-hopping term is generated and eventually will be the relevant coupling. It easy to see that if from \cref{RG_eq_K_sigma_vs_interaction} we flow towards smaller $K_\sigma$, then the $1/K_\sigma$ term in \cref{RG_eq_tperp_pairhopping} makes $t_\perp$ an irrelevant coupling.
	
	We have to stop this first step of the flow when the coupling constant in \cref{RG_eq_K_sigma_vs_interaction} is of order one. At that scale ($l=l_1$) the microscopic cutoff is of the order of the pair size. Another estimation of $l_1$, is the pair size $\alpha(l_1) \sim u_\sigma/\Delta_\sigma$ where $\Delta_\sigma$ is the spin gap and $u_\sigma$ is the velocity of the spin sector, before pairs become local. 
	At that scale the single particle tunnelling is suppressed because of the gap in the spin sector which can be formally seen in the RG equations \cref{RG_eq_tperp_pairhopping} by the fact that for small $K_\sigma$, $t_\perp$ is formally irrelevant. 
	
	\subsection{Second step RG: pair hopping and dimensional crossover}
	In the second RG step $l>l_1$, we have only pair-hopping and spin excitations and single particle hopping  are suppressed. The spin sector is out of the picture, and the RG equation expressing the pair-hopping coupling becomes
	\begin{equation} \label{RG_second_step_J}
	    \frac{d\tilde{J}(l)}{dl} = \left(2- \frac{1}{K_\rho}\right)\tilde{J}(l)
	\end{equation}
	
	We thus see that at the scale $l_1$ we are now left with only the charge sector as a massless sector and an effective Josephson coupling between the chains. The situation is thus similar to the one we had in the large spin gap limit but with a different Josephson coupling than the strong coupling limit $J \sim \frac{t^2_\perp}{\Delta_\sigma}$. This has consequences for the absolute values of the $T_c$ and charge gap at zero temperature but the ratio is unchanged compared to \cref{analytic_fermionic_ratio}.
	
	The absolute values of the $T_c$ or the charge gap can simply be computed by continuing the flow of \cref{RG_second_step_J} until the Josephson coupling itself become of order one. This defines a second RG length $l^*$. The condition is set by $\tilde{J}(l^*) \propto \mathcal{O}(1)$, which means that for $l>l^*$ the coupling is so large that we are back to a 2D/3D system. From this length, we can define either the critical temperature or the charge gap, because their ratio is fixed from \cref{analytic_fermionic_ratio}. By integrating \cref{RG_second_step_J},, we find that the RG length $l^*$ at which the DC occurs is
	\begin{equation}
        \alpha(l^*) = \alpha(l_1) \left ( \frac{1}{\tilde{J} (l_1)} \right )^{\frac{1}{2- 1/K_\rho}}
    \end{equation}
	with $\alpha(l_1)$ the pair size.
	Moreover, from dimensional analysis we know that energies scale as $\tilde{\Delta}_\rho(0) = \tilde{\Delta}_\rho(l)e^{-l}$. We observe that $\tilde{\Delta}_\rho(l^*) \propto \tilde{J}(l^*) \sim 1$ and
	we find that the original (dimensionless) charge gap $\tilde{\Delta}_\rho(0)$ reads
	\begin{equation}
	    \tilde{\Delta}_\rho(0) =  \tilde{J} (l_1) ^{\frac{1}{2- 1/K_\rho}} \frac{\Delta_\sigma}{u_\sigma } \frac{v_F}{W}
	\end{equation}
	where we define the lattice spacing from $\alpha(l=0) = v_F/W$, the bandwidth as $W=2t$ and the Fermi velocity $v_F=2t \sin{(\frac{\pi \rho_0}{2})}$, with $\rho_0$ the unperturbed density (1 at half-filling). 
	
	Finally let us note that if we look directly at the pair operator we have 
	\begin{equation}
	    \psi^\dagger_{R,\uparrow} \psi^\dagger_{L,\downarrow} \propto e^{i\sqrt2\theta_\rho} \cos(\sqrt8 \phi_\sigma)
	\end{equation}
	Averaging over the massive spin sector would lead to 
\begin{equation}
	    \psi^\dagger_{R,\uparrow} \psi^\dagger_{L,\downarrow} \propto C e^{i\sqrt2\theta_\rho} 
	\end{equation}	
where the prefactor $C$ could be related to 
\begin{equation}
    C \propto \langle \cos(\sqrt8\phi_\sigma) \rangle_{H_\sigma}
\end{equation}
Because of the gap in the spin sector this average is nonzero.  
	
	\bibliography{nufhc_ref,analyt_ref,gunnar_ref_bib}
\end{document}